\documentclass[preprint]{elsarticle}

\usepackage{lineno,hyperref}
\usepackage{float}

\usepackage{amssymb}
\usepackage{amsmath}
\usepackage[numbers]{natbib}

\usepackage{graphicx}
\usepackage{geometry}
\usepackage{subcaption} 

\usepackage{bm}
\usepackage{siunitx}
\usepackage{booktabs}

\def\mathbi#1{\textbf{\em #1}}

\modulolinenumbers[5]	

\journal{preprint for arXiv}

\begin{document} 

\begin{frontmatter}

\title{A novel phase-field based cohesive zone model for modeling interfacial failure in composites}


\author[1]{Pei-Liang Bian}

\author[1]{Hai Qing\corref{mycorrespondingauthor}}
\cortext[mycorrespondingauthor]{Corresponding author, ordid: 0000-0002-1022-7917}
\ead{qinghai@nuaa.edu.cn}

\address[1]{State Key Laboratory of Mechanics and Control of Mechanical Structures, Nanjing University of Aeronautics and Astronautics, Nanjing, 210016, P.R. China}

\begin{abstract}
The interface plays a critical role in the mechanical properties of composites. In the present work, a novel phase-field based cohesive zone model (CZM) is proposed for the cracking simulation. The competition and interaction between the bulk and interfacial cracking are taken into consideration directly in both displacement- and phase-field. A modified family of degradation functions is utilized to describe traction-separation law in the CZM. Finite element implementation of the present CZM was carried out with a completely staggered algorithm. Several numerical examples, including a single bar tension test, a double cantilever beam test, a three-point bending test, and a single fiber-reinforced composite test, are carried out to validate the present model by comparison with existing numerical and experimental results.The present model shows its advantage on modeling interaction between bulk and interfacial cracking.
\end{abstract}



\end{frontmatter}

\linenumbers


%
%
%
\section{Introduction}
The interface plays a role in the mechanical properties of micro-structures. The cohesive zone models (CZM) have been widely used for simulating interfacial cracking propagation in materials. To simulate interfacial cracking, many approaches of the CZM have been developed. Apart from the extended finite element method (XFEM) \cite{Mos.1999,Fries.2010} and the embedded finite element method (EFEM) \cite{Simo.1993,Oliver.2006}, interface finite elements \cite{Xu.1994,Ortiz.1999} is the most popular approach \citep{Simo.1993,Oliver.2006,Linder.2007,Song.2006}. 


The phase-field method (PFM) is derived from Griffith's theory, in which crack propagation is described as the competition between the elastic and free surface energy \cite{Griffith.1921}. 
Recent two decades, the PFM has been widely used to solve the problem of crack propagation. 
It has been used to deal with brittle fracture in epoxy\cite{Xie.2016}, rubber\cite{Miehe.2014}, polycrystals in alloy\cite{Schneider.2016}, and composites \cite{EspadasEscalante.2019}. Besides, ductile and fatigue fracture can also be simulated with the PFM \cite{Miehe.2014,Russ.2020}.
Apart from linear elastic fracture mechanics (LEFM), some cohesive fracture models were also proposed with PFM.
Verhoosel and Borst proposed a novel phase-field based cohesive model and investigated cracking in materials by approximating the discrete displacement jump with an auxiliary field. \cite{Verhoosel.2013}. 
Wu proposed a unified phase-field theory for the mechanics of damage and quasi-brittle failure, which bridges the damage and fracture mechanics for quasi-brittle failure in solids \cite{Wu.2017,Wu.2018}.

On the other hand, the PFM is used not only for investigating the cracking in the bulk region but also for crack propagation at interfaces.
Nguyen et al. presented a phase-field model to simulate the displacement and traction jump across interfaces in materials \cite{Nguyen.2016}. A level-set method was proposed to describe diffuse displacement jump at interfaces, a level-set method has been proposed.
In addition, they also presented a phase-field model to simulate the displacement and traction jump across interfaces in materials \cite{Nguyen.2019}. In this case, the unilateral contact condition was adapted to distinguish the different fracture behaviors in tension/compression for the bulk cracking and in normal/tangential direction for the interfacial cracking.
Xia et al. used a similar concept to analyze hydraulic fracturing with interfacial damage \cite{Xia.2017}.
In addition, Hansen-D\"{o}rra et al. proposed a phase-field interfacial model by modifying the fracture toughness of the interfacial zone \cite{HansenDorr.2019}. The relationship between the interfacial zone width and convergence was also investigated in the research.
However, these models need regions with finite width to model the interfacial properties, which limit the usage of models for composites with high filler volume fraction. 
Besides, interfacial elements based on the phase-field with nearly zero thickness have also been developed.
Paggi and Reinoso developed a modeling framework that combines the phase-field model for brittle fracture in the bulk and interfaces for a pre-existing interface \cite{Paggi.2017}. 
Meanwhile, this algorithm was also extended to 3D cases \cite{Carollo.2017} and used to model complex crack paths in ceramic laminates \cite{Carollo.2018}.
Quintanas-Corominas et al. developed a PF-CZM approach for modeling the interaction between the delamination and interlaminar damage in long fiber composite materials \cite{QuintanasCorominas.2020}.
Nevertheless, most of the system equations of these schemes are solved by the monolithic algorithm, which means that the unknown displacement and phase-field need to be solved simultaneously. However, the robustness of the algorithm can not be guaranteed for unstable crack propagation, which is commonly solved with the staggered algorithm. Under the staggered algorithm, the displacement and phase-field values are solved separately. This method is widely used for PFM due to its robustness. 

In the present work, a phenomenological cohesive element is proposed to simulate the interfacial debonding in materials. The cohesive fracture in interfaces is simulated with interface elements. The main idea of the work is a combination of Paggi's zero thickness element and Hansen-D\"{o}rra's replacement of the fracture toughness. The main features of the present model are summarized as follows:
\begin{itemize}
    \item The cohesive zone is described as interfacial cohesive elements, which is similar to the approach in \cite{Paggi.2017}.
    \item The staggered algorithm can be used to solve the coupled problem between displacement- and phase-field, which can guarantee the convergence of iteration.
    \item The phase-field value on each node is shared by bulk and interfacial elements. No additional variable is necessary for the present model.
    \item Both mode I and II fracture in the interfaces can be simulated under the present framework, the stiffness, strength ,and critical energy release rate in the normal and tangential direction can be given individually.
    \item Closure of the interfacial cracks is also able to be simulated with the present scheme.
\end{itemize}

The manuscript is structured as follows: in section \ref{secBulkPF}, we revisit the concept of the classical phase-field method for brittle fracture.
In section \ref{secCohesivePF}, a new type of the phase-field based cohesive element is proposed, which follows the process for derivation of the bulk cracking. In section \ref{secDegradationFunc}, a family of modified degradation functions is proposed to describe the stiffness degradation in the interfacial region. Some numerical examples are listed in section \ref{secNumericalExample}.

\section{Revisiting on the phase-field approach to fracture in the bulk region}
\label{secBulkPF}
\subsection{Regularized frame work of crack in the bulk region}
\begin{figure}
    \centering
    \includegraphics[width=0.5\textwidth]{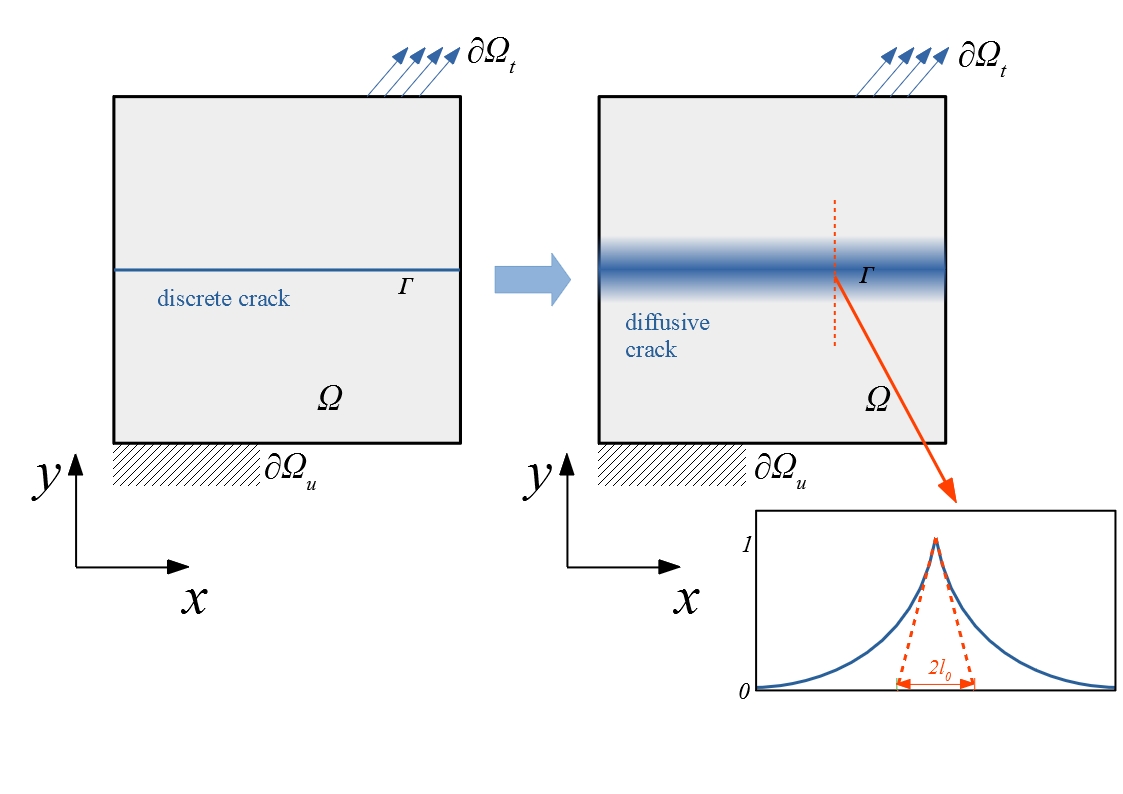}
    \caption{Schematic representation of diffusive crack in a bulk region.}
    \label{figdiffcrack}
\end{figure}

Let $\varOmega$ be a region, which has boundary $\partial\varOmega$. In addition, the region $\varOmega$ is constrained with displacement boundary $\partial\varOmega_u$ and traction $\partial\varOmega_t$ and $\partial\varOmega_u\cup\partial\varOmega_t=\partial\varOmega$. As illustrated in Fig. \ref{figdiffcrack}, the basic idea of the regularized crack is that a discrete crack located in a bulk region $\varOmega$ can be modeled with a scale field $\phi$. The scale field here is also called as the diffusive crack. The value of $\phi \in [0,1]$ is known as the phase-field value on the material point. $\phi=0$ indicates the unbroken state and $\phi=1$ means total failure state at the material point. Then the free surface energy of the crack here can be rewritten as:
\begin{equation}
    \label{eqcrackarea}
    \varGamma(\phi) = \int_{\varOmega}\gamma^{\rm bulk}(\phi)\mathrm{d}\varOmega
\end{equation}
where the $\gamma^{\rm bulk}(\phi)$ is the density of crack and can be defined as \cite{Miehe.2010}:
\begin{equation}
    \label{eqdencrack}
    \gamma^{\rm bulk}(\phi,\nabla\phi) = \frac{1}{2\ell_0}\left\{\phi^2+ \ell_0^2\nabla\phi\cdot\nabla\phi\right\}
\end{equation}
where $\ell_0$ is a regularization length controlling the process region of the damage diffusion. According to the minimization principle, the scalar field $\phi$ can be obtained:
\begin{equation}
    \phi(\bm{x}) = \arg\left\{\inf_{\phi\in\mathcal{S}_{\phi}}\varGamma(\phi)\right\}
\end{equation}
By solving the Euler-Lagrangian equation, the analytical solution of the phase-field can be obtained as:
\begin{equation}
    \phi(x) = \phi_0\mathrm{e}^{-|\frac{x}{\ell_0}|}
    \label{eq_phsol}
\end{equation}
where the $\phi_0$ is the phase-field value at the origin. Eq. (\ref{eq_phsol}) is also known as the approximated Dirac function \cite{Verhoosel.2013}.
\subsection{Governing equation of the phase-field model for the brittle fracture in the bulk region}
Following \cite{Miehe.2010}, the discrete crack located in $\varGamma$ can be replaced by diffusive crack.
The variational energy $\varPsi$ of the bulk region $\varOmega$ can be defined as:
\begin{equation}
    \varPsi_{\rm bulk}(\bm{u},\phi) = \int_{\varOmega}\psi_{\rm bulk}^{\rm el}(\bm{\varepsilon},\phi)+\mathcal{G}^{\rm bulk}_c\gamma^{\rm bulk}(\phi,\nabla\phi)\mathrm{d}\varOmega-\int_{\partial\Omega^t}\bm{\bar{t}}\cdot\bm{u}\mathrm{d}S
    \label{eq_bulkenergy}
\end{equation}
where $\psi^{\rm el}$ is the elastic energy density. $\mathcal{G}_c^{\rm bulk}$ is the critical energy release rate of the bulk material. The $\bm{\bar{t}}$ is the traction on the traction boundary $\partial\Omega^t$. 
For linear elastic material, the elastic energy density $\psi^{\rm el}$ is defined as: 
\begin{equation}
    \psi_{\rm bulk}^{\rm el}(\bm{\varepsilon},\phi) = \omega^{\rm bulk}(\phi)\psi_{\rm bulk}^{\rm el+}(\bm{\varepsilon})+\psi_{\rm bulk}^{\rm el-}(\bm{\varepsilon})
\end{equation}
where $\psi_{\rm bulk}^{\rm el+}$ and $\psi_{\rm bulk}^{\rm el-}$ are the positive and negative parts of elastic energy density, respectively. Under this decomposition algorithm, the crack closure can be modeled. 
$\omega^{\rm bulk}(d)$ is the degradation function of the material here. The form of the degradation function is discussed in detail later. 
Then the governing equations and boundary conditions of the crack propagation can be obtained by optimization of the total variational energy $\varPsi_{\rm bulk}$
\begin{equation}
    \left\{\bm{u(x)},\phi(x)\right\} = \arg\left\{\inf\varPsi_{\rm bulk}(\bm{u},\phi)\right\}
\end{equation}
Based on the Euler-Lagrangian equation, the governing equations and boundary conditions can be obtained as follows: 
\begin{subequations}
    \begin{align}
        \nabla\cdot\bm{\sigma} = \bm{0}\ \text{in}\ \varOmega
    \end{align}
    \begin{align}
        \label{eqPF}
        \frac{\mathcal{G}^{\rm bulk}_c}{\ell_0}\left[\phi-\ell_0^2\Delta\phi\right]+\frac{\partial\omega^{\rm bulk}(\phi)}{\partial\phi}\mathcal{H}=0\ \text{in} \ \varOmega
    \end{align}
    \begin{align}
        \nabla\cdot\bm{\sigma}=\overline{\bm{t}}\ \text{on}\ \partial\varOmega_t
    \end{align}
    \begin{align}
        \bm{u} = \overline{\bm{u}}\ \text{on}\ \partial\varOmega_u
    \end{align}
    \begin{align}
        \nabla\cdot\phi=0\ \text{on}\ \partial\varOmega 
    \end{align}
\end{subequations}
The $\mathcal{H}$ is introduced to guarantee the irreversibility of the crack for the phase-field, which is defined as \cite{Miehe.2010}:
\begin{equation}
    \mathcal{H}(\bm{x},t) = \max_{\tau\in[0,t]}\psi_{\rm bulk}^{\rm el+}(\bm{\varepsilon},\tau)
\end{equation}


\section{Mechanics of the phase-field based cohesive model}
\label{secCohesivePF}
In the present model, a new PFM based cohesive element is developed to simulate the interfacial fracture. To this end, a derivation process similar to the PFM in the bulk region is used here. 
\subsection{Definition of crack density at the interface}
\begin{figure}
    \centering
    \begin{subfigure}[b]{0.49\textwidth}
        \centering
        \includegraphics[width=\textwidth]{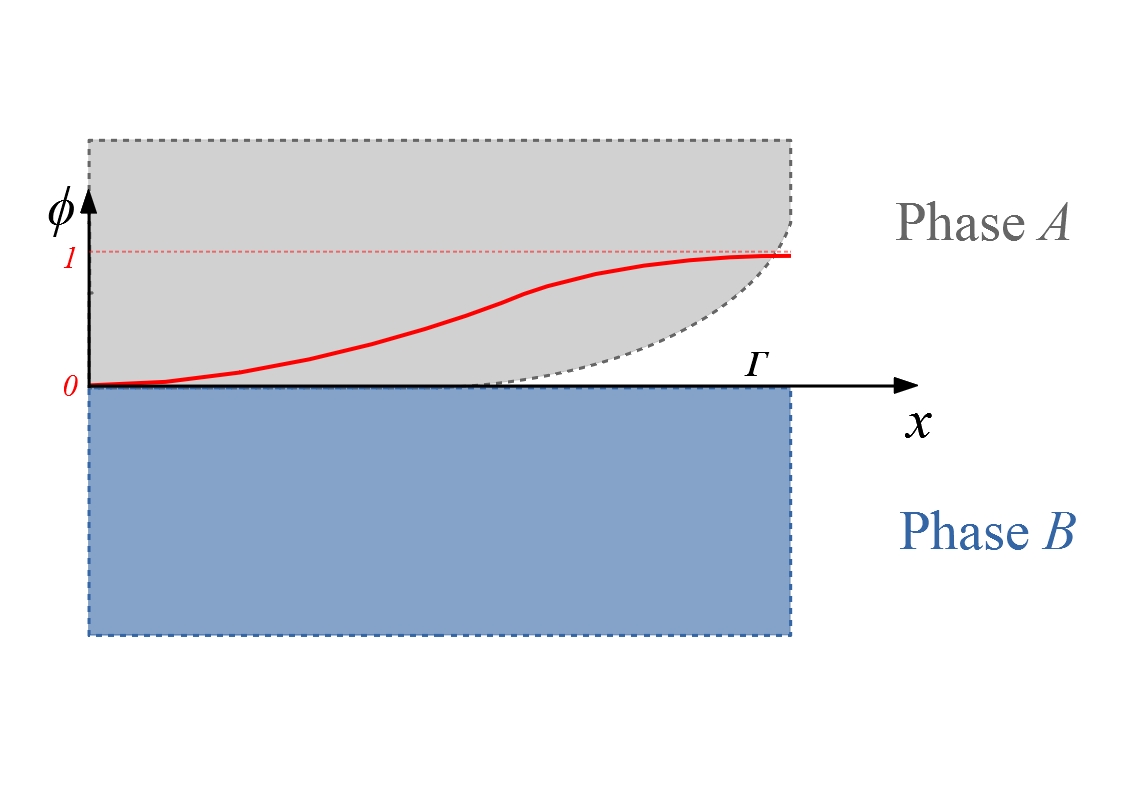}    
        \caption{}
        \label{figintPF}
    \end{subfigure}
    \hfill
    \begin{subfigure}[b]{0.49\textwidth}
        \centering
        \includegraphics[width=\textwidth]{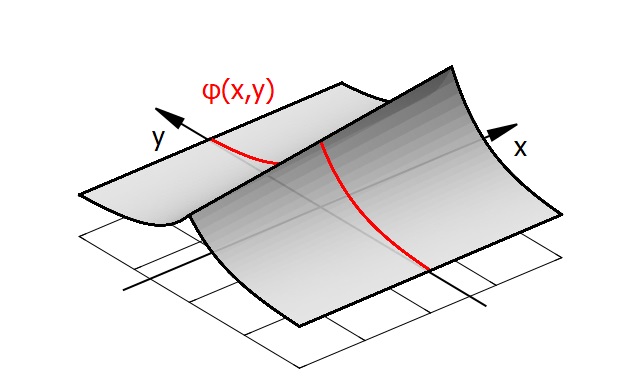}    
        \caption{}
        \label{figPFdistribution}
    \end{subfigure}
    \caption{Phase-field value at the interface between phase A and B:(a)distribution of phase-field value at interface and (b)assumed phase-field distribution}
\end{figure}
As illustrated in Fig. \ref{figintPF}, the phase-field value $\phi$ is also used to describe the separation of an interfacial crack $\varGamma$. Same as in the bulk region, $\phi=0$ represents no separation, and $\phi=1$ represents total separation. Different from the density of crack defined in Eq. (\ref{eqdencrack}), a linear density is needed to be defined to describe the area of the interfacial crack. To this end, let the crack density be distributed as illustrated in Fig. \ref{figPFdistribution}. An interfacial crack is located along the x-axis and has not been ultimately separated. Then the distribution of the crack density can be assumed as:
\begin{equation}
    \phi(x,y)=\phi(x,0)\mathrm{e}^{-|\frac{y}{\ell_0}|}
    \label{eqPFdistribution}
\end{equation}
Taking Eq. (\ref{eqPFdistribution}) into Eqs. (\ref{eqcrackarea}) and (\ref{eqdencrack}), we can rewrite the area of diffusive crack as follows:
\begin{equation}
    \varGamma(\phi) = \int_{-\infty}^{\infty}\int_{-\infty}^{\infty}\gamma^{\rm bulk}(\phi) \mathrm{d}x\mathrm{d}y = \int_{-\infty}^{\infty}\gamma^{\rm int} \mathrm{d}x
\end{equation}
where $\gamma^{\rm int}$ is the linear density of the crack, which can be obtained by integrating $\gamma^{\rm bulk}(\phi)$ across y-axis:
\begin{equation}
    \gamma^{\rm int} = \int_{-\infty}^{\infty} \gamma^{\rm bulk}(\phi) \mathrm{d}y= \phi^2+\frac{\ell_0^2}{2}\nabla_s\phi\cdot\nabla_s\phi
\end{equation}
where the $\nabla_s\cdot$ is the gradient operator along the x-axis.
\subsection{Interfacial cohesive fracture}
The separation of cohesive interfaces is defined as the displacement jump between both sides of the interfaces here \cite{scheider2001cohesive}. 
\begin{equation}
    \bm{\delta}=\left[\!\left[\bm{u}\right]\!\right] =\mathbf{u}^+-\mathbf{u}^-
\end{equation}
The $\mathbf{u}^+$ and $\mathbf{u}^-$ here are the displacement of the opposite sides of a crack, respectively. $\left[\!\left[\bm{u}\right]\!\right]$ represents the jump of the displacement field across the crack here. $\bm{\delta}$ is the separation displacement of the crack. For two-dimension problem, the $\bm{\delta}$ can be described in a local coordination with normal component $\delta_{n}$ and tangential component $\delta_{t}$. 

Once the crack opens, the damage will degrade the stiffness of the interfacial material. The separation-traction law (TSL) for the cohesive model can be described as follows:
\begin{subequations}
    \begin{align}
        \sigma_t =\omega^{\rm int}_t(\phi)k_{t,0} \delta_t
    \end{align}
    \begin{align}
        \sigma_n = 
        \begin{cases}
            \omega^{\rm int}_n(\phi)k_{n,0} \delta_n \ \  \text{$\delta_n\geq0$} \\
            k_{n,0} \delta_n  \qquad \ \   \text{$\delta_n<0$}
        \end{cases}
    \end{align}
    \label{eq:TSL}
\end{subequations}
Where $\sigma_i$, $k_{i,0}$ and $\delta_i$ are the traction, initial interfacial stiffness and separation displacement, respectively. $i=t,n$ represent tangential and normal direction, respectively. The $\omega_t^{\rm int}(\phi)$ and $\omega_n^{\rm int}(\phi)$ here are the degradation function in the tangential and normal directions, respectively.
The closure effect in the normal direction is also taken into consideration here. Then the interfacial deformation energy in the normal and tangential directions are defined as:
\begin{subequations}
    \begin{align}
        \psi_{{\rm int},t}^{\rm el} = \frac{\omega^{\rm int}_t(\phi) k_{t,0}\delta_t^2}{2}
    \end{align}
    \begin{align}
        \psi_{{\rm int},n}^{\rm el} = \psi_{{\rm int},n}^{\rm el+} + \psi_{{\rm int},n}^{\rm el-}  = \frac{H(\delta_n)\omega^{\rm int}_n(\phi) k_{n,0}\delta_n^2}{2} + \frac{H(-\delta_n)k_{n,0}\delta_n^2}{2}
    \end{align}  
\end{subequations}
where $H(x)=\max(x,0)$ is the step function. Under the present decomposition scheme, the negative part of the energy will not contribute to fracture during loading.

\subsection{Governing equation for the present phase-field based cohesive fracture}
Similar to the PFM in the bulk region above-mentioned in section \ref{secBulkPF}, the variational energy of the interface $\varPsi_{int}$ can be written as follows:
\begin{equation}
    \varPsi_{\rm int}(\mathbf{u},\phi)=\int_{\varGamma}\psi_{\rm int}^{\rm el}(\bm{\delta},\phi)+\mathcal{G}^{\rm int}_c \gamma^{\rm int}(\phi,\nabla_s\phi) \mathrm{d}\varGamma-\int_{\varGamma} \bm{t}\cdot\bm{\delta} \mathrm{d}\varGamma
    \label{eq_intenergy}
\end{equation}
With the Euler-Lagrangian equation, the optimization of the $\varPsi_{int}$ is equivalent to the TSL listed in  Eq.(\ref{eq:TSL})  and following equation of the phase-field:
\begin{equation}
    \mathcal{G}_c^{\rm int}[2\phi-\ell_0^2\Delta_s\phi]+\frac{\partial\omega_n^{\rm int}(\phi)}{\partial\phi}\mathcal{H}_n+\frac{\partial\omega_t^{\rm int}(\phi)}{\partial\phi}\mathcal{H}_t = 0
    \label{eq_intpf}
\end{equation}
where state variables $\mathcal{H}_t$ and $\mathcal{H}_n$ are defined as:
\begin{subequations}
    \begin{align}
        \mathcal{H}_t(\bm{x},t) = \max_{\tau\in[0,t]}\psi_{{\rm int},t}^{\rm el}(\bm{\delta},\tau)
    \end{align}
    \begin{align}
        \mathcal{H}_n(\bm{x},t) = \max_{\tau\in[0,t]}\psi_{{\rm int},n}^{\rm el+}(\bm{\delta},\tau)
    \end{align}
\end{subequations}
The $\mathcal{G}_c^{\rm int}$ is the mixed critical energy release rate for the interfacial crack, which is defined as
\begin{equation}
    \mathcal{G}_c^{\rm int}=\frac{\mathcal{G}_{\mathrm{I}}^{\rm int}}{\mathcal{G}_{\mathrm{I}}^{\rm int}+\mathcal{G}_{\mathrm{II}}^{\rm int}}\mathcal{G}_{c,\mathrm{I}}^{\rm int}+\frac{\mathcal{G}_{\mathrm{II}}^{\rm int}}{\mathcal{G}_{\mathrm{I}}^{\rm int}+\mathcal{G}_{\mathrm{II}}^{\rm int}}\mathcal{G}_{c,\mathrm{II}}^{\rm int}
\end{equation}
where $\mathcal{G}_{\mathrm{I}}^{\rm int}$ and $\mathcal{G}_{\mathrm{II}}^{\rm int}$ are mode I and II energy release rate, respectively. $\mathcal{G}_{c,\mathrm{I}}^{\rm int}$ and $\mathcal{G}_{c,\mathrm{II}}^{\rm int}$ are the critical mode I and II energy release rate, respectively. It is worth mentioning that the $\mathcal{G}_{\mathrm{I}}^{\rm int}$ and $\mathcal{G}_{\mathrm{II}}^{\rm int}$ here are functions of the displacement field, which means that they can be treated as constants when solving the phase-field.
    
\subsection{Scheme of surface energy equivalence between interfaces and the bulk region}
\begin{figure}
    \centering
    \begin{subfigure}[b]{0.49\textwidth}
        \centering
        \includegraphics[width=\textwidth]{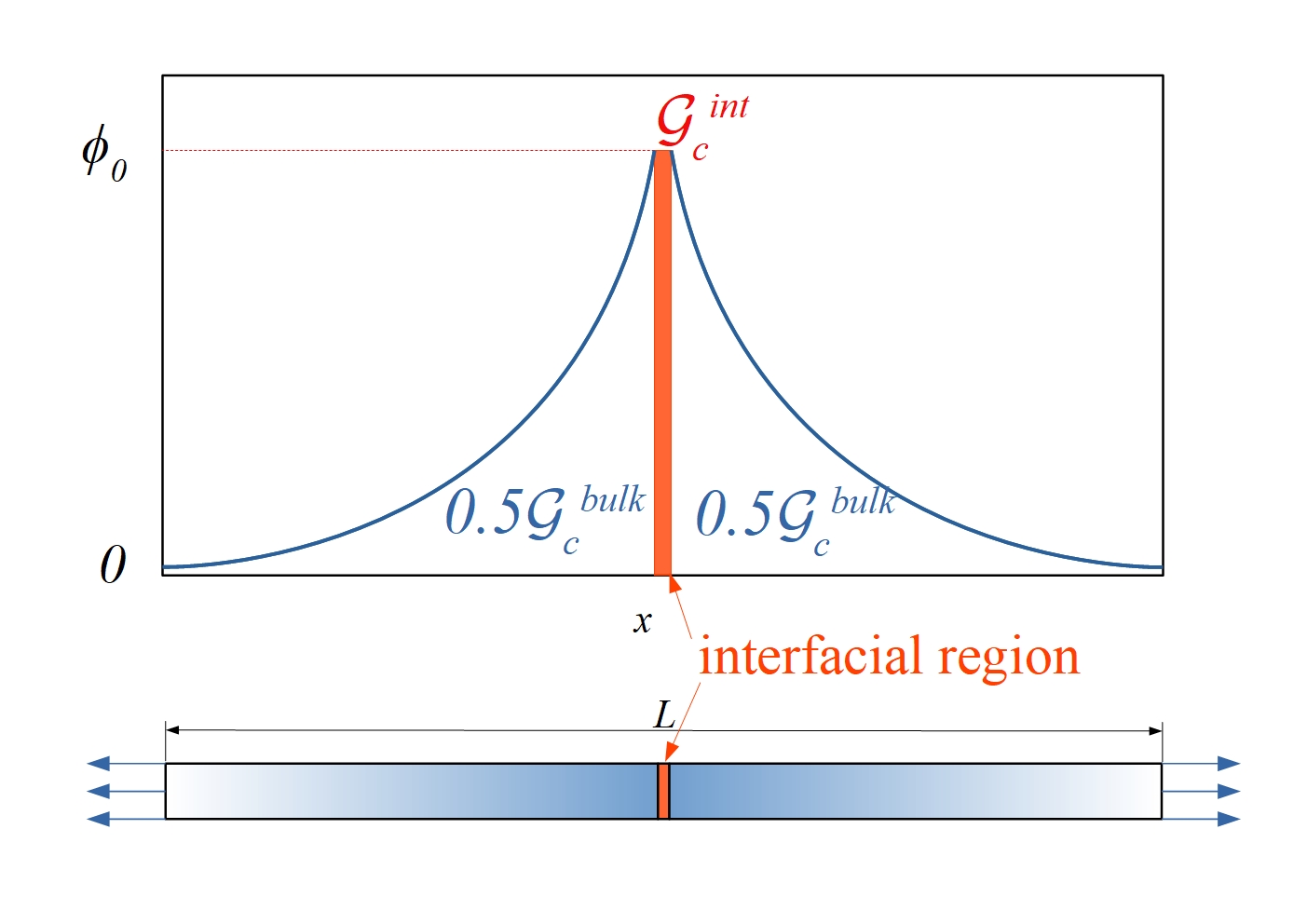}    
        \caption{}
        \label{fig_singlebar}
    \end{subfigure}
    \hfill
    \begin{subfigure}[b]{0.49\textwidth}
        \centering
        \includegraphics[width=\textwidth]{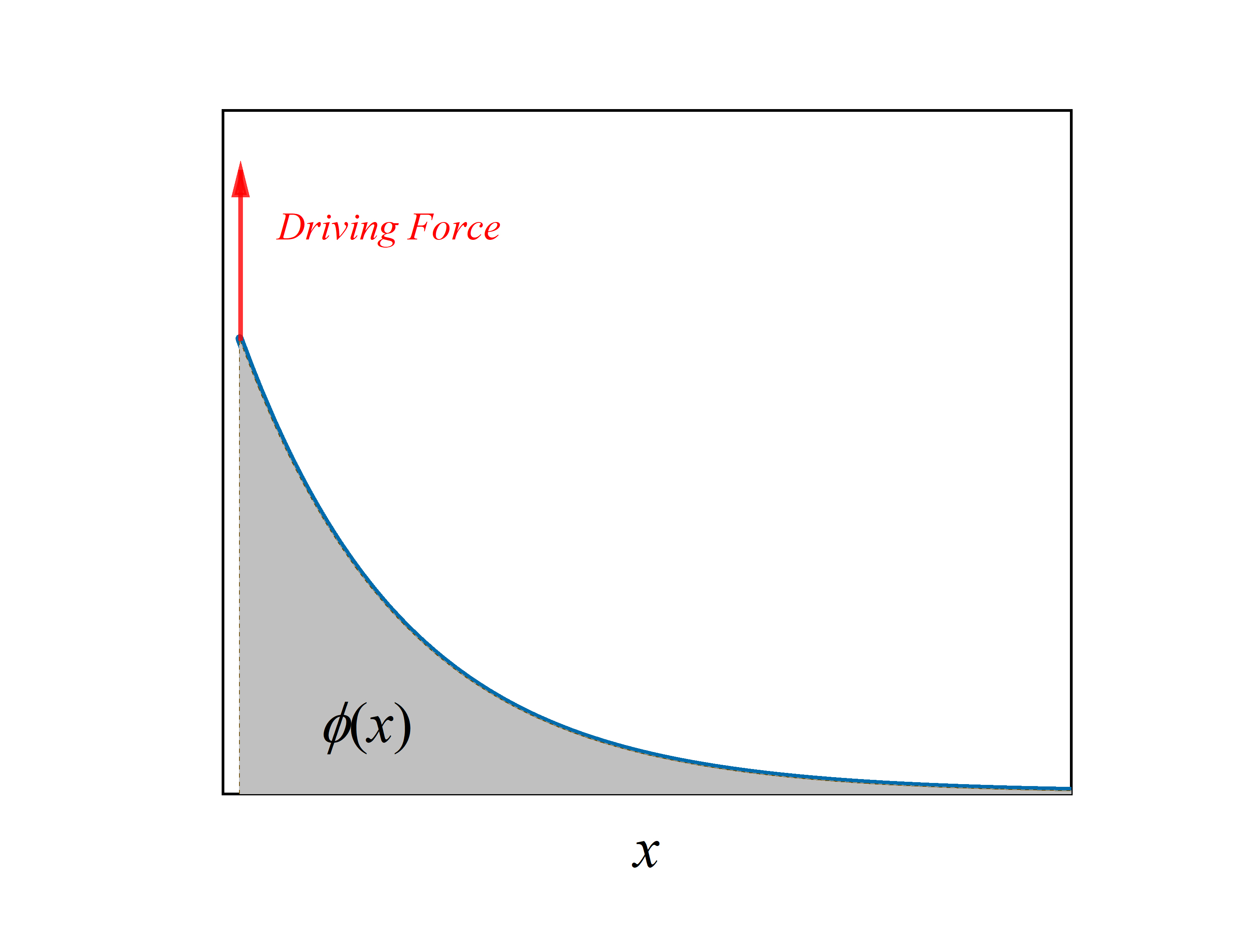}   
        \caption{}
        \label{fig_drivingforce}
    \end{subfigure}
    
    \caption{Driving force acting on the diffusive crack in the matrix: (a) the interface embedded in a single bar and (b) the concept of driving force.}
    
\end{figure}

Till now, the governing equations for the interfacial cracking has been derived. Combining the bulk and interfacial  variational energy in Eqs. (\ref{eq_bulkenergy}) and (\ref{eq_intenergy}), the global variational energy can be expressed as the sum of the bulk and interfacial energy\cite{Paggi.2017}:
\begin{equation}
    \varPsi(\bm{u},\phi) = \int_{\varOmega}\psi_{\rm bulk}^{\rm el}(\bm{\varepsilon},\phi)+\mathcal{G}^{\rm bulk}_c\gamma^{\rm bulk}(\phi,\nabla\phi)\mathrm{d}\varOmega+\int_{\varGamma}\psi_{\rm int}^{\rm el}(\bm{\delta},\phi)+\mathcal{G}^{\rm int}_c \gamma^{\rm int}(\phi,\nabla_s\phi) \mathrm{d}\varGamma-\int_{\partial\Omega^t}\bm{\bar{t}}\cdot\bm{u}\mathrm{d}S
\end{equation}
The external traction term of the Eq. (\ref{eq_intenergy}) is removed because the interface region is embedded in the bulk materials. Then the displacement and phase-field in both bulk region and interfaces can be solved by optimizing the variational function $\varPsi$. However, there is still a paradox under the present framework. Let us consider an interfacial region embedded in the center of a bar as illustrated in Fig. \ref{fig_singlebar}. Once the crack is completely open, the phase-field will distribute as Eq. (\ref{eq_phsol}) and the central phase-field $\phi_0=1$, then the free surface energy of the diffusive crack can be obtained as:
\begin{equation}
    \varPi = \int_{\varOmega}\gamma^{\rm bulk}\mathrm{d}\varOmega+\int_{\varGamma}\gamma^{\rm int}\mathrm{d}\varGamma = (\mathcal{G}_c^{\rm bulk}+\mathcal{G}_c^{\rm int})A
\end{equation}
where $A$ is the cross-section area of the bar. Here we notice that the overall $\mathcal{G}_c$ is the sum of the $\mathcal{G}_c^{\rm bulk}$ and $\mathcal{G}_c^{\rm int}$, which indicates that the interfacial cracking could be suppressed or prevented by the bulk region. To dealing with this problem, an external driving force is introduced to drive cracking in the bulk region when the interfacial debonding occurs. To this end, let us consider a half crack illustrated in Fig. \ref{fig_drivingforce}. With divergence theorem, the derivation of the free surface energy is:
\begin{equation}
    \delta \left( \mathcal{G}^{\rm bulk}_c\int_{\varOmega^{\rm half}}\gamma^{\rm bulk}\mathrm{d}\varOmega \right) =
    \frac{\mathcal{G}^{\rm bulk}_c}{\ell_0}\int_{\varOmega^{\rm half}}\{2\phi-\ell_0^2\Delta\phi\}\delta\phi\mathrm{d}\varOmega+
    \mathcal{G}^{\rm bulk}_c\ell_0\int_{\partial\varOmega^{\rm half}}\nabla\phi\cdot\boldsymbol{n}\delta\phi\mathrm{d}S
\end{equation}
The last term is the traction of the phase-field. Considering that $\phi(0)=\phi_0$ and $\nabla\phi(+\infty)=0$, the traction $\mathcal{F}$ of the phase-field on the $x=0$ boundary to drive this diffusive crack is
\begin{equation}
    \mathcal{F} = \mathcal{G}^{\rm bulk}_c\ell_0\nabla\phi = -\mathcal{G}_c^{\rm bulk}\phi_0
\end{equation}
For the case that both sides of the interface connected to the bulk materials, the $\mathcal{F}$ will be doubled to drive the crack initiation on both sides. The $\mathcal{F}$ is added at the interface during the calculation. Under the present compensation scheme, the $\mathcal{G}_c$ can be adjusted to the value of that of interface $\mathcal{G}_c^{int}$. Meanwhile, we can define the penalty stiffness, strength ,and the critical energy release rate independently on the normal and tangential direction.

\subsection{Finite element implementation of the cohesive zone model}
\begin{figure}[]
    \centering
    \includegraphics[width=0.3\textwidth]{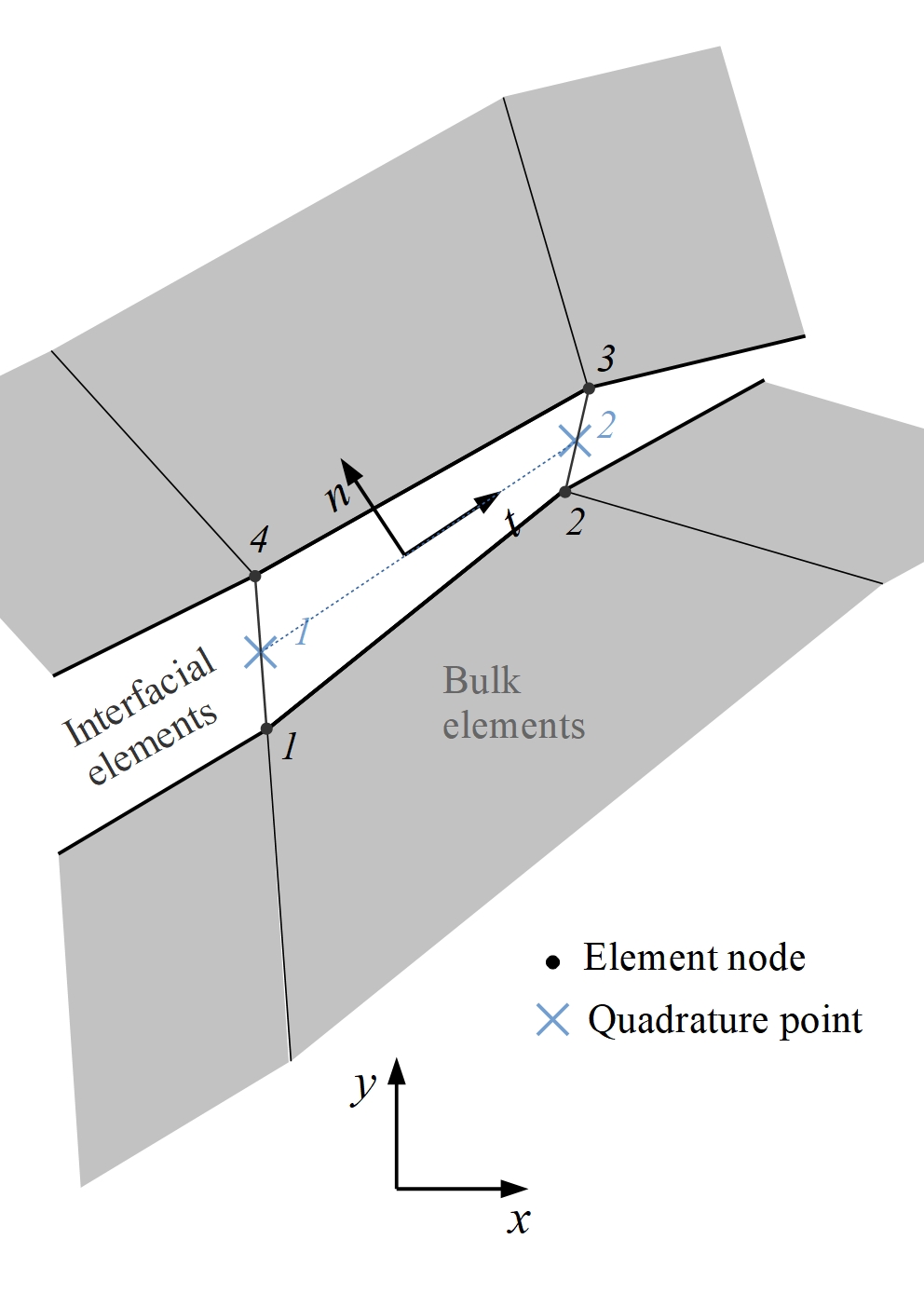}	
	\caption{The interface element with node numbering and quadrature points.}
	\label{fig:elementshape}
\end{figure}

Once the governing equations have been derived through the above-mentioned derivation process, the finite elements are used for the spatial discretization. The shape of the interface element is illustrated in Fig. \ref{fig:elementshape}. Nodes on both sides of the element are connecting to the bulk elements. The thickness of the element is zero when no load is applied to it.
Once all the interfacial elements are inserted into the model, the displacement and phase-field can be solved numerically. Due to the poly-convexity of the displacement and phase-field, a standard staggered algorithm is used here, which means that the unknown displacement and phase-field are solved separately in an increment. The process will repeat until both fields converge to desired tolerance. Details of the staggered algorithm can be seen in \cite{Miehe.2010}. The discretization process of the bulk region also follows the same way with \cite{Miehe.2010}. For the unknown interfacial nodal displacement $\mathbi{u}$ and interfacial nodal phase-field values $\boldsymbol{\phi}$, the internal residual vector and Jacobi matrix of displacement and phase-field can be expressed as:
\begin{equation}
    \mathbi{r}_{uu} = \int_{\varGamma}\mathbi{B}_{u}^\mathrm{T}\mathbi{R}^\mathrm{T}\left[\begin{array}{ccc}
        \sigma_t \\
        \sigma_n \\
    \end{array}\right]\mathrm{d}\varGamma
    \label{eq:ru}
\end{equation}
\begin{equation}
    \mathbi{K}_{uu} = \int_{\varGamma}\mathbi{B}_{u}^\mathrm{T}\mathbi{R}^\mathrm{T}\left[\begin{array}{ccc}
        \frac{\partial\sigma_t}{\partial\delta_t} & 0 \\
        0 & \frac{\partial\sigma_n}{\partial\delta_n}
    \end{array}\right]\mathbi{R}\mathbi{B}_{u} \mathrm{d}\varGamma
    \label{eq:Kuu}
\end{equation}
\begin{equation}
    \mathbi{r}_{\phi}=\int_{\varGamma}\mathcal{G}^{\rm int}_c \ell_0^2 \mathbi{B}_{\phi,s}^\mathrm{T}\nabla\phi + \left(2\mathcal{G}_c^{\rm int}\phi+\frac{\partial\omega_t^{\rm int}(\phi)}{\partial\phi}\mathcal{H}_t+\frac{\partial\omega_n^{\rm int}(\phi)}{\partial\phi}\mathcal{H}_n\right)\mathbi{N}_{\phi}^\mathrm{T}-\alpha\mathcal{G}_c^{\rm bulk}\mathbi{N}_{\phi}^\mathrm{T}\phi\mathrm{d}\varGamma
    \label{eq:rp}
\end{equation}
\begin{equation}
    \mathbi{K}_{\phi\phi}=\int_{\varGamma}\mathcal{G}_c^{\rm int} \ell_0^2 \mathbi{B}_{\phi,s}^\mathrm{T}\mathbi{B}_{\phi,s} +\left(2\mathcal{G}_c^{\rm int}+\frac{\partial^2\omega_t^{\rm int}(\phi)}{\partial\phi^2}\mathcal{H}_t+\frac{\partial^2\omega_n^{\rm int}(\phi)}{\partial\phi^2}\mathcal{H}_n\right)\mathbi{N}_{\phi}^\mathrm{T}\mathbi{N}_{\phi}-\alpha\mathcal{G}_c^{\rm bulk}\mathbi{N}_{\phi}^\mathrm{T}\mathbi{N}_{\phi}\mathrm{d}\varGamma
    \label{eq:Kpp}
\end{equation}
 The details of the finite element implementation of the interface can be found in Appendix \ref{sec:appendix}.
The parameter $\alpha$ here is the switch that controls the surface energy equivalence. The $\alpha=1$ represents the single side of the interface is embedded in phase-field and $\alpha=2$ means both sides are embedded in phase-field. Then the Newton-Raphson method is used to solve the unknown displacement and phase-field values on each node:
\begin{subequations}
    \begin{align}
        \Delta\boldsymbol{\phi} = \mathbi{K}_{\phi\phi}^{-1}\mathbi{r}_{\phi}
    \end{align}
    \begin{align}
        \Delta\mathbi{u} = \mathbi{K}_{uu}^{-1}\mathbi{r}_{u}
    \end{align}
\end{subequations}
Carrying out the above-mentioned staggered algorithm, the unknown nodal displacement and phase-field can be solved. The present finite-element scheme is carried out in Julia, which is a newborn language designed from the beginning for high-performance \cite{Julia-2017}. 

\section{Choice of the degradation function}
\label{secDegradationFunc}
\subsection{The degradation function in the classical phase-field method for brittle fracture}
In the classical framework of the phase-field models for brittle fracture, the energy degradation function $\omega(\phi)$ is defined as a square function:
\begin{equation}
    g_2(\phi) = (1-\phi)^2
\end{equation}
To simplify the discussion in this section, let $k=k_{n,0}=k_{t,0}$ and $\mathcal{G}_c=\mathcal{G}_{c,{\rm I}}^{\rm int}=\mathcal{G}_{c,{\rm II}}^{\rm int}$. Meanwhile, only uniaxial loading is considered in the section. For the classical phase-field framework of LEFM, there is a paradox between the critical energy release rate $\mathcal{G}_c$ and the ultimate stress $\sigma_c$. For given regularized length $\ell_0$ and module $E_m$, $\sigma_c$ can be determined by the $\mathcal{G}_c$. Considering the interfacial phase-field equation listed in Eq. (\ref{eq_intpf}), the ultimate traction $\sigma_c$ is
\begin{equation}
    \sigma_c = \frac{3}{8}\sqrt{\frac{3}{2}k\mathcal{G}_c}
\end{equation}
For the traditional interfacial cohesive element, the penalty stiffness $k$ is always set to be a great value to prevent displacement discontinuity before interfacial failure. Meanwhile, $k$ also plays a critical role in the algorithm that prevents crack closure. The above-mentioned reasons make the $g_2(\phi)$ to be greatly improper for the interfacial cracking simulation.
To overcome this problem, it is necessary to use other functions to replace the $g_2(\phi)$ while simulating the cohesive fracture. 
\subsection{The modified degradation function family and determination the parameter}
According to \cite{Sargado.2018}, any candidate of the degradation function has to satisfy the following constraints:
\begin{enumerate}[(a)]
    \item $\omega'(\phi)\leq0,\ \ \phi\in (0,1)$
    \item $\omega(0)=1$ and $\omega(1)=0$
    \item $\omega'(0)<0$ and $\omega'(1)=0$
\end{enumerate}
The first condition indicates that $\omega(\phi)$ should be a monotonic decreasing function. The second one suggests that $\phi=0$ is the unbroken state and $\phi=1$ is a completely broken state. The last constrain is the start and stop condition of cracking.
Inspired by Wu's work \cite{Wu.2017,Wu.2018}, a modified family of degradation function is used here:
\begin{equation}
    \omega_p = \frac{(1-\phi)^p}{(1-\phi)^p+a\phi},\ p \in \mathrm{N}_{+}
\end{equation}

It can be proved that for any integer greater than one, the above-mentioned conditions can be satisfied. For the given $\mathcal{G}_c$ and $\sigma_c$, the unknown parameter $a$ can be determined. The $\omega(\phi)$ here can be implemented either in bulk or interfacial cracking. For the interfacial case, the derivative of traction is zero when it reaches the ultimate value. According to \cite{Sargado.2018}, it can be written as:
\begin{equation}
    \frac{\mathrm{d} \sigma}{\mathrm{d} \delta} = \frac{2\phi\left[\frac{\partial\omega_p(\phi)}{\partial\phi}\right]^2+\omega_p(\phi)\left[\frac{\partial\omega_p(\phi)}{\partial\phi}-\phi\frac{\partial^2\omega_p(\phi)}{\partial\phi^2} \right]}{\frac{\partial\omega_p(\phi)}{\partial\phi}-\phi\frac{\partial^2\omega_p(\phi)}{\partial\phi^2}}k
\end{equation}
By solving the square equation $\frac{\partial \sigma}{\partial\delta} = 0$, the phase-field value corresponding to the ultimate stress can be obtained by Eq. (\ref{eq_intpf}):
\begin{equation}
    \phi_c = \frac{-p+\sqrt{5 p+4} \sqrt{p}-2}{2
    \left(p^2-1\right)}
\end{equation}
It is worth noting that $\phi_c$ here is independent on the material parameters, i.e. $k$ or $\mathcal{G}_c$. Then the corresponding separation displacement $\delta_c$ can be determined as:
\begin{equation}
    \delta_c = \sqrt{-\frac{4\mathcal{G}_c\phi_c}{\frac{\partial\omega_p(\phi_c)}{\partial\phi}k}}
\end{equation}
Taking $\delta_c$ into the TSL listed in Eq. (\ref{eq:TSL}), the parameter $a$ now can be expressed as:
\begin{equation}
    a = \frac{4\mathcal{G}_ck}{\sigma_c^2}\frac{\phi_c(1-\phi_c)^{p+1}}{1+(p-1)\phi_c}
\end{equation}
The $\omega_p(\phi)$ can also be used in the bulk elements for simulating cohesive fracture. For the case that $\omega_p(\phi)$ was used as $\omega^{\rm bulk}(\phi)$ in the bulk region, the above-mentioned derivation can still be used by replacing $\mathcal{G}_c$ and $k$ with $\frac{\mathcal{G}_c^{\rm bulk}}{2\ell_0}$ and $E$, respectively. 
\subsection{The TSL of the interfacial model with the new degradation function}
\begin{figure}[]
    \begin{subfigure}{0.49\textwidth}
        \includegraphics[width=\textwidth]{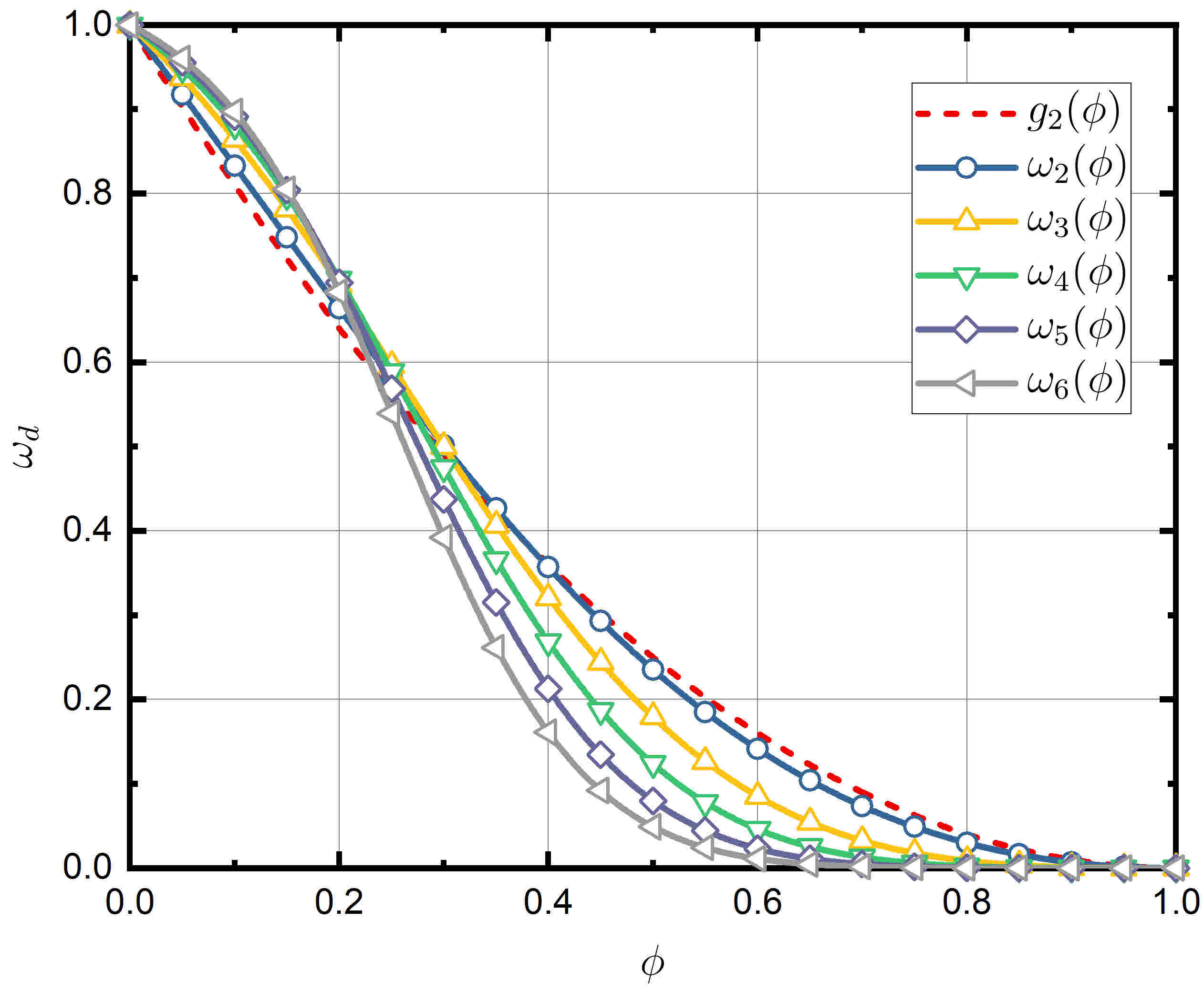}
        \caption{}
    \end{subfigure}
    \begin{subfigure}{0.49\textwidth}
        \includegraphics[width=\textwidth]{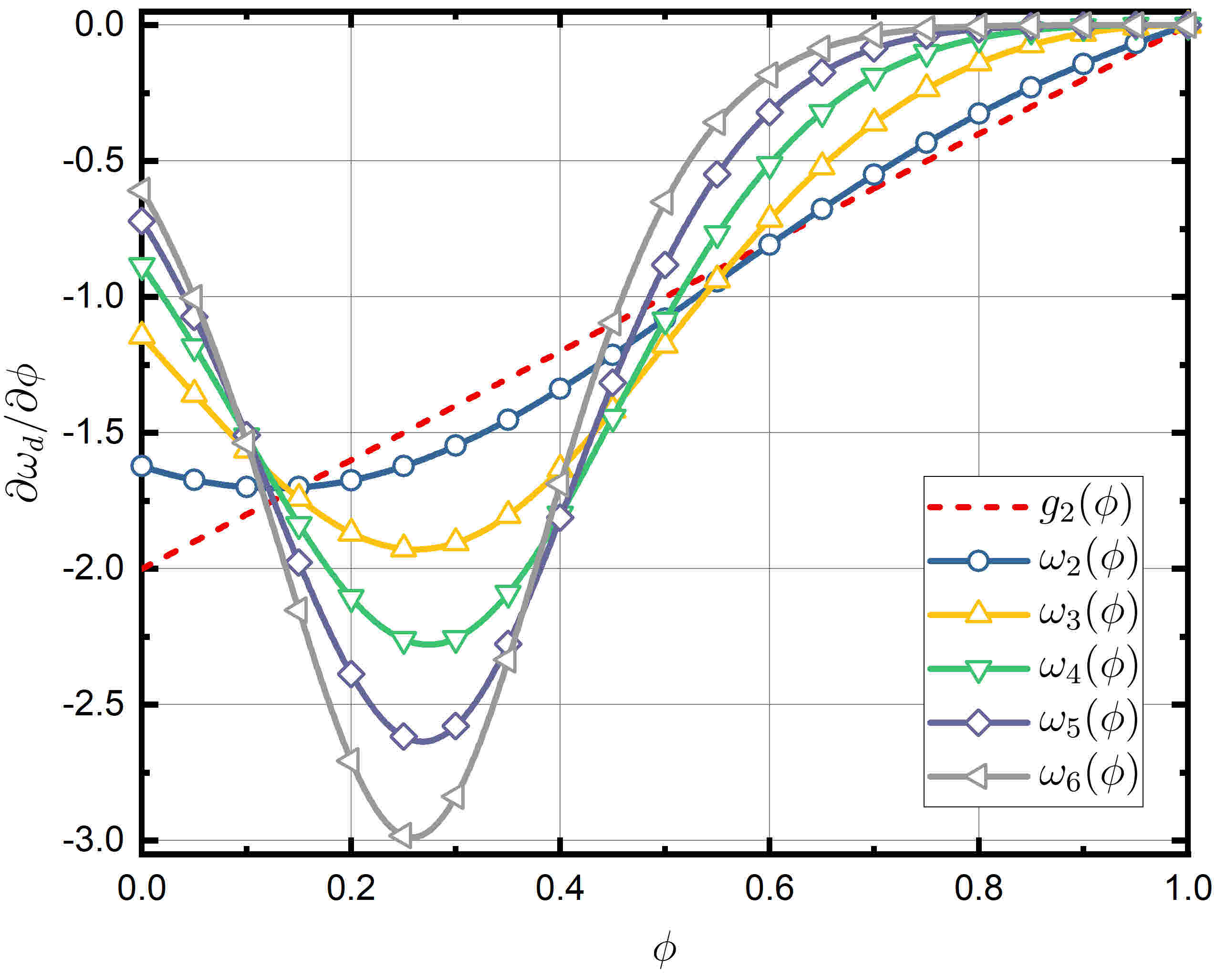}
        \caption{}
    \end{subfigure}
    \caption{Comparison between classical $g_2(\phi)$ and $\omega_p(\phi)$ with the identical mechanical parameters: (a) degradation functions and (b) the first order derivations.}
    
    \label{figG2andWd}
\end{figure}

Once the unknown parameter $a$ is determined by the above-mentioned process, the degradation function $\omega_p(\phi)$ can be finally determined. The comparison between the $g_2(\phi)$ and $\omega_p(\phi)$ are illustrated in Fig. \ref{figG2andWd}. Both the functions and their derivation are illustrated. The $\mathcal{G}_c$ are identical for all the degradation function. The $\sigma_c$ of the $\omega_p(\phi)$ is set to be $\sqrt{27k/128\mathcal{G}_c}$, which is the same value of $g_2(\phi)$. It can be seen that the difference between $g_2(\phi)$ and $\omega_p(\phi)$ increases with increasing $p$ value.
Meanwhile, the absolute value of $\partial\omega_p(\phi)/\partial\phi$ is lower than $\partial g_2(\phi)/\partial\phi$ when $\phi$ is close to zero, which suggests that the $\omega_p$ are slower for crack initiation and material can have more quasi-elastic stage.
In addition, $\omega_p(\phi)$ are flatter when $\phi$ approaches to one, which indicates that the degradation process of residual stiffness with the present $\omega_p(\phi)$ is softer than that with $g_2(\phi)$.

\begin{figure}[]
    \centering
    \begin{subfigure}{0.49\textwidth}
        \includegraphics[width=\textwidth]{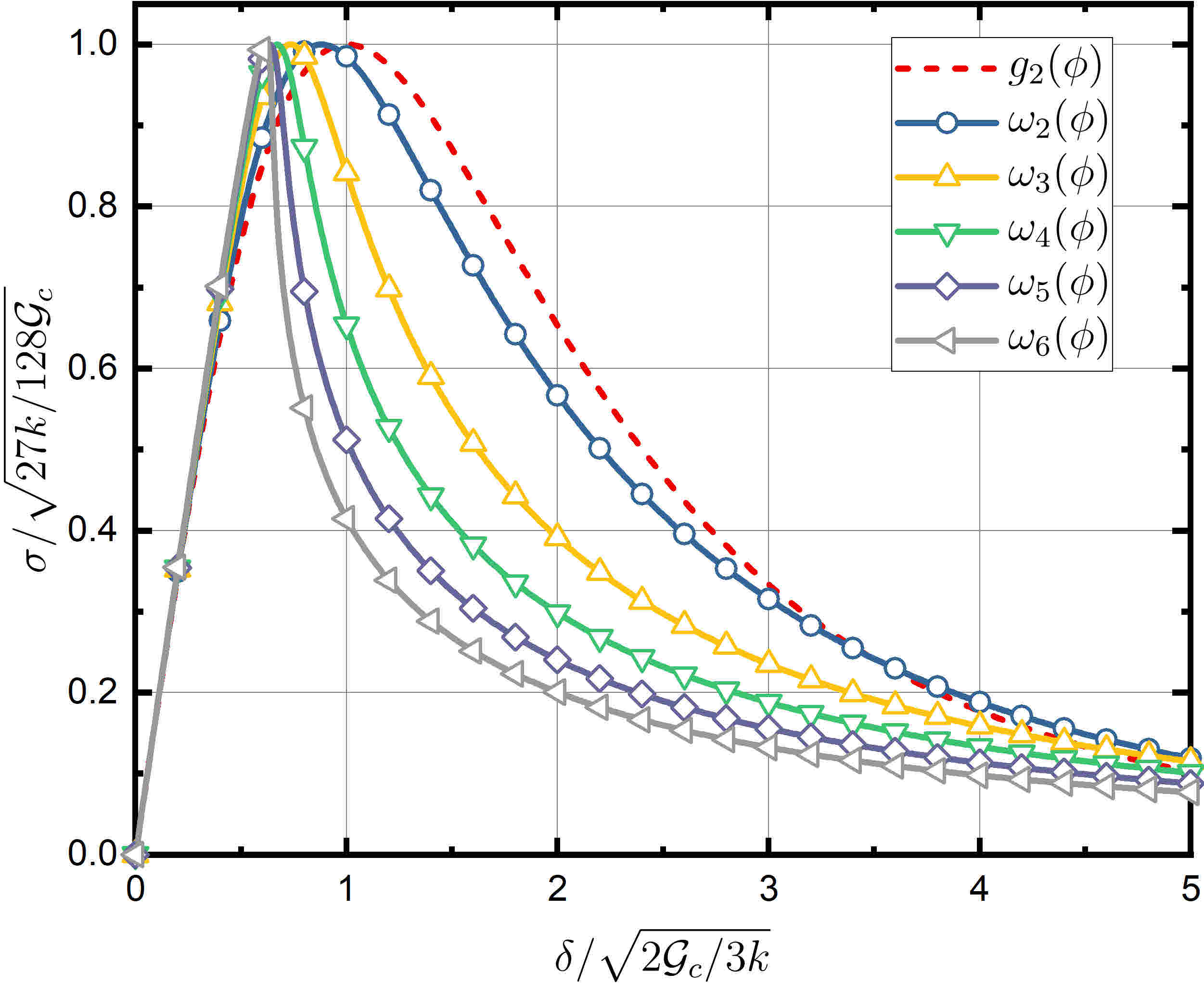}
        \caption{}
        \label{fig_TSL}
    \end{subfigure}
    \begin{subfigure}{0.49\textwidth}
        \includegraphics[width=\textwidth]{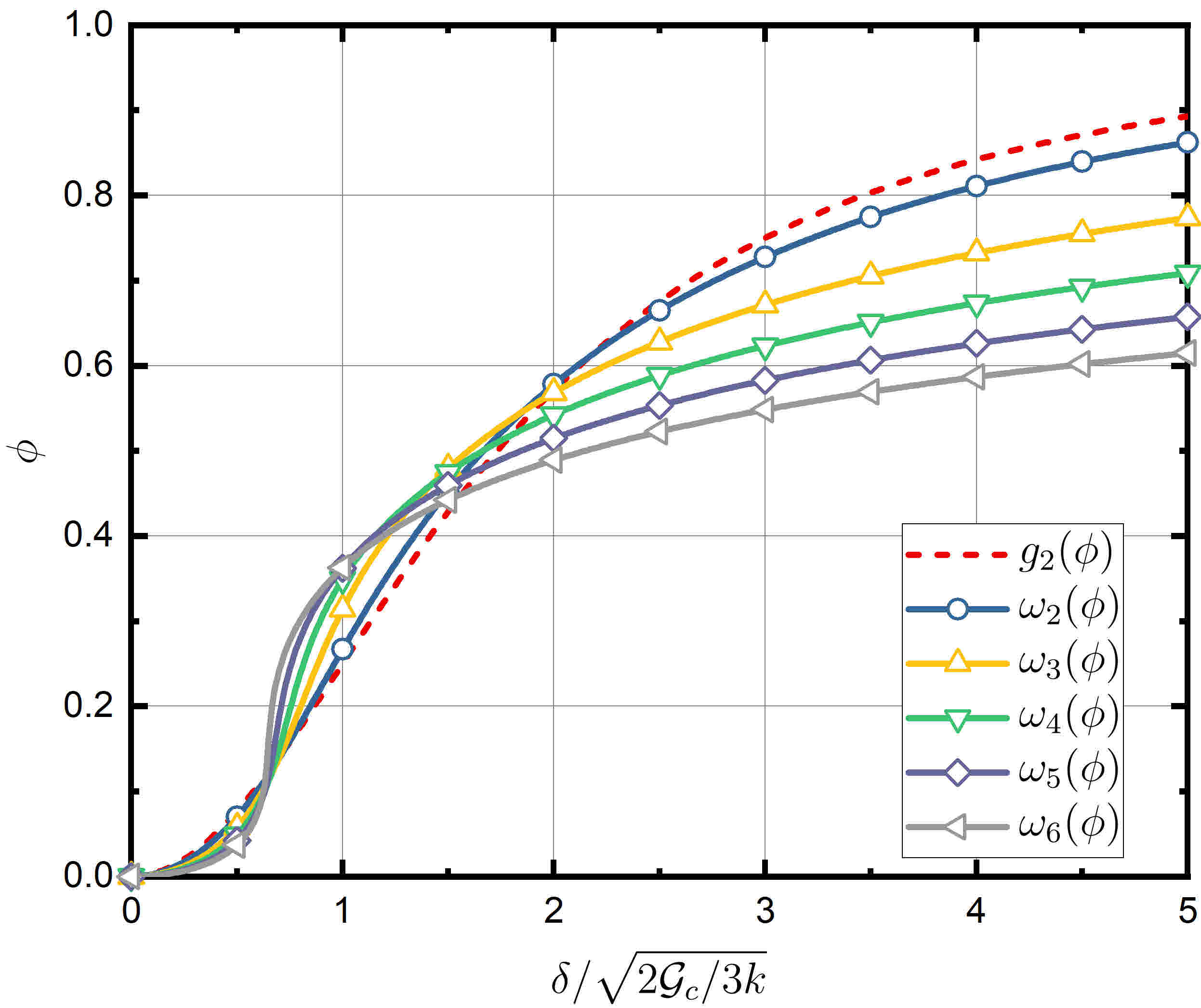}
        \caption{}
        \label{fig_phi2delat}
    \end{subfigure}
    \caption{Traction-separation law of the cohesive zone model with different degradation functions: (a) relationship between normal separated and traction and (b) relationship between separation and the phase-field under uniaxial load.}
    \label{fig_TSLall}
\end{figure}

In Fig. \ref{fig_TSL}, the normalized traction-separation relationship determined by the Eq. (\ref{eq:TSL}) is illustrated. For the high order $\omega_p(\phi)$, there is a longer quasi-elastic stage when the interface under uniaxial loading. Once reaching the ultimate stress, the stiffness for the interface with high order $\omega_p(\phi)$ degrades steeply compared with $g_2(\phi)$ and $\omega_2(\phi)$. However, the degrading speed of the high order $\omega_p(\phi)$ decreases with further loading, which agrees with the morphology of the degradation function. 

The relationship between $\phi$ and normalized displacement is illustrated in Fig. \ref{fig_phi2delat}. The $g_2(\phi)$ and $\omega_2(\phi)$ have a quite similar shape. On the other hand, high order $\omega_p(\phi)$ shows an obvious character of three-stage. The $\phi$ increases slowly when it stays near zero, which can explain the reason for the long quasi-elastic range. In the second stage, the $\phi$ increases rapidly and slows down with the increasing load. In addition, the corresponding separation displacement also decreases with increasing $p$ value.

In next, differences between the $g_2(\phi)$ and $\omega_p(\phi)$ with the identical $\mathcal{G}^{\rm int}_c$ and different $\sigma_c$ are also investigated for $p=2$. The ultimate values $\sigma_{max}$of the traction in the cohesive model with $\omega_2(\phi)$ are set to be 0.25, 0.5, 1, 2, and 4 times of the value of $g_2(\phi)$, which is marked as $\sigma_c$ here. 

\begin{figure}[]
    \centering
    \begin{subfigure}{0.49\textwidth}
        \includegraphics[width=\textwidth]{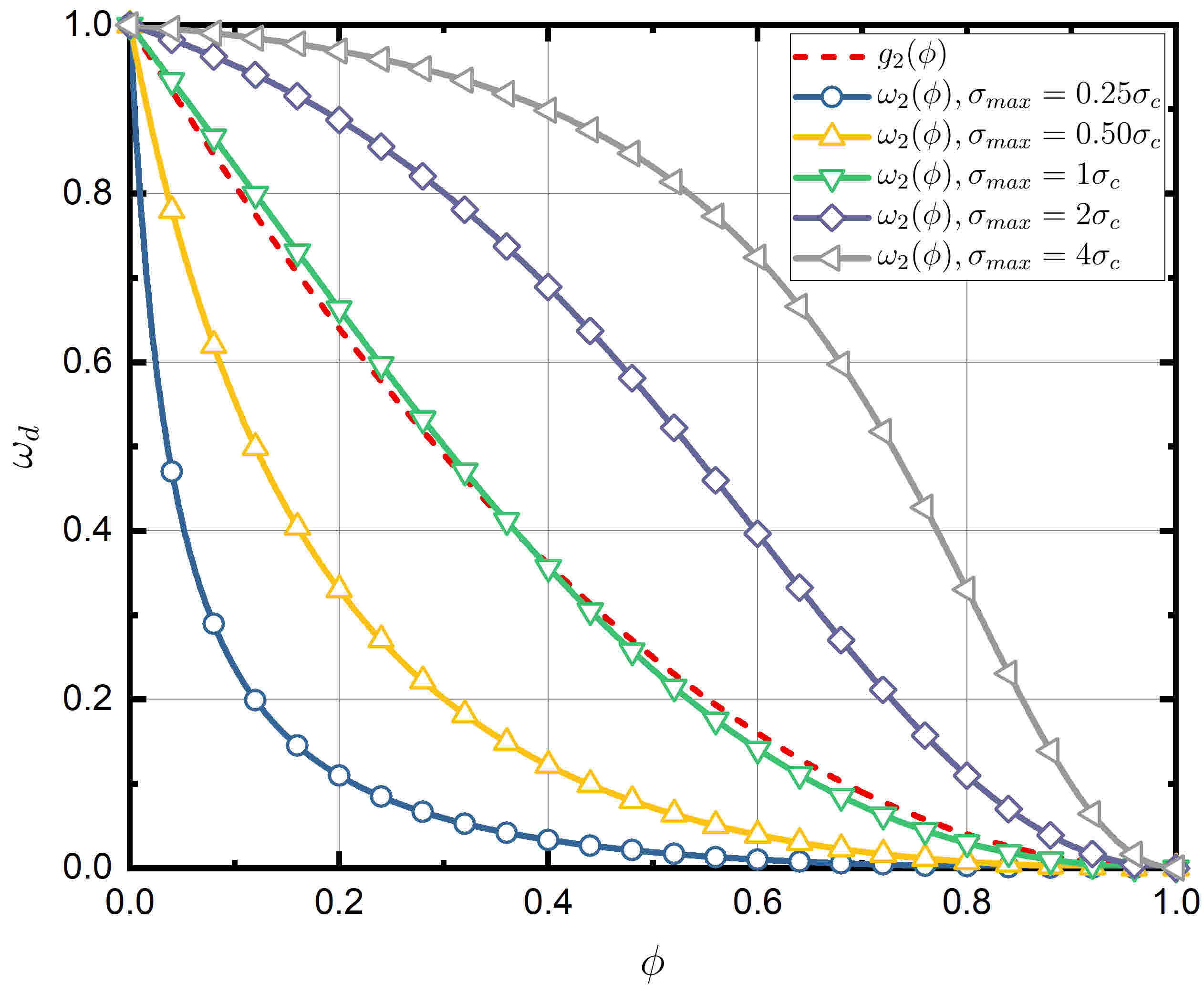}
        \caption{}
        \label{fig_g2andWd_sigma}
    \end{subfigure}
    \begin{subfigure}{0.49\textwidth}
        \includegraphics[width=\textwidth]{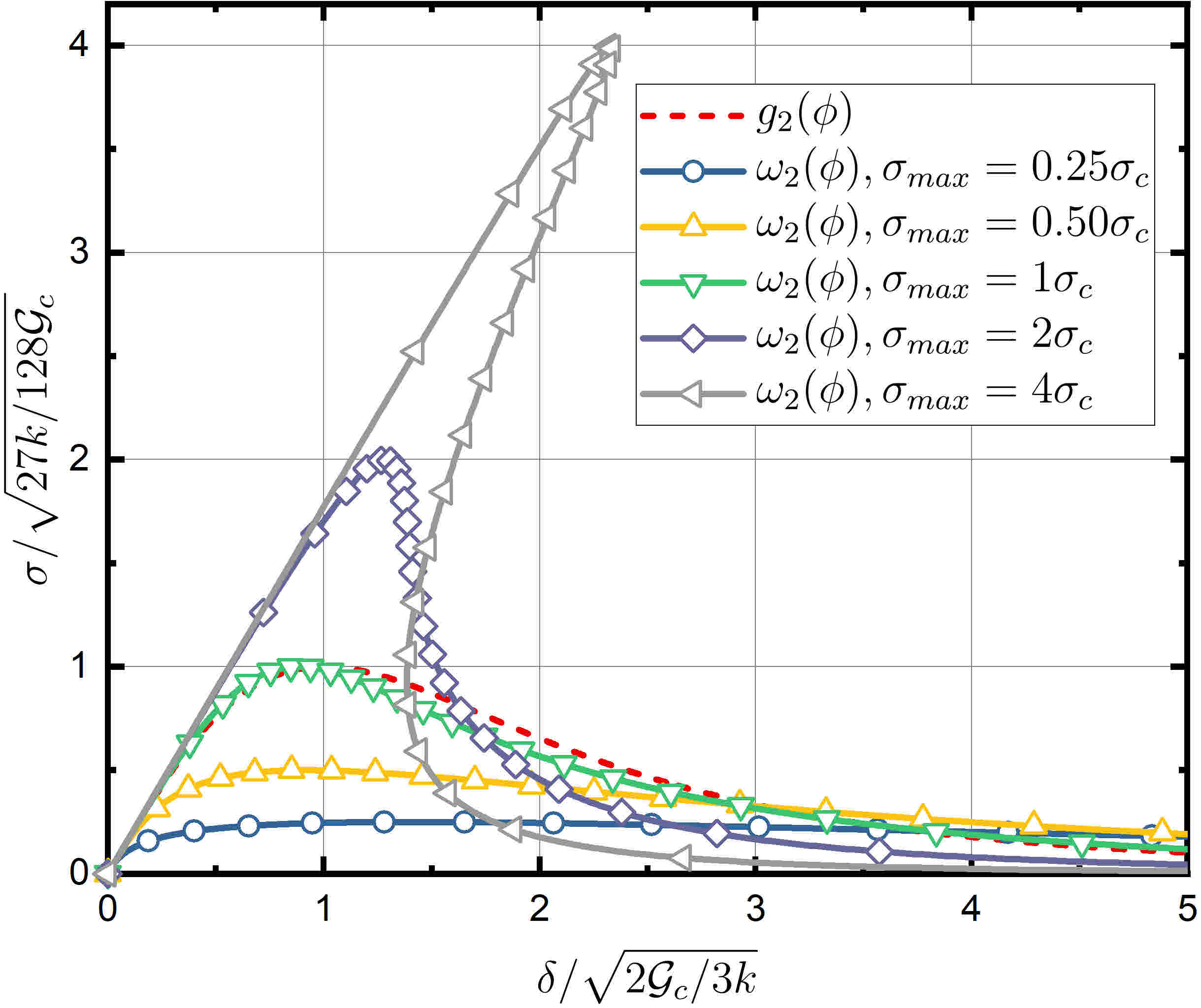}
        \caption{}
        \label{fig_TSL_sigma}
    \end{subfigure}

    \begin{subfigure}{0.49\textwidth}
        \includegraphics[width=\textwidth]{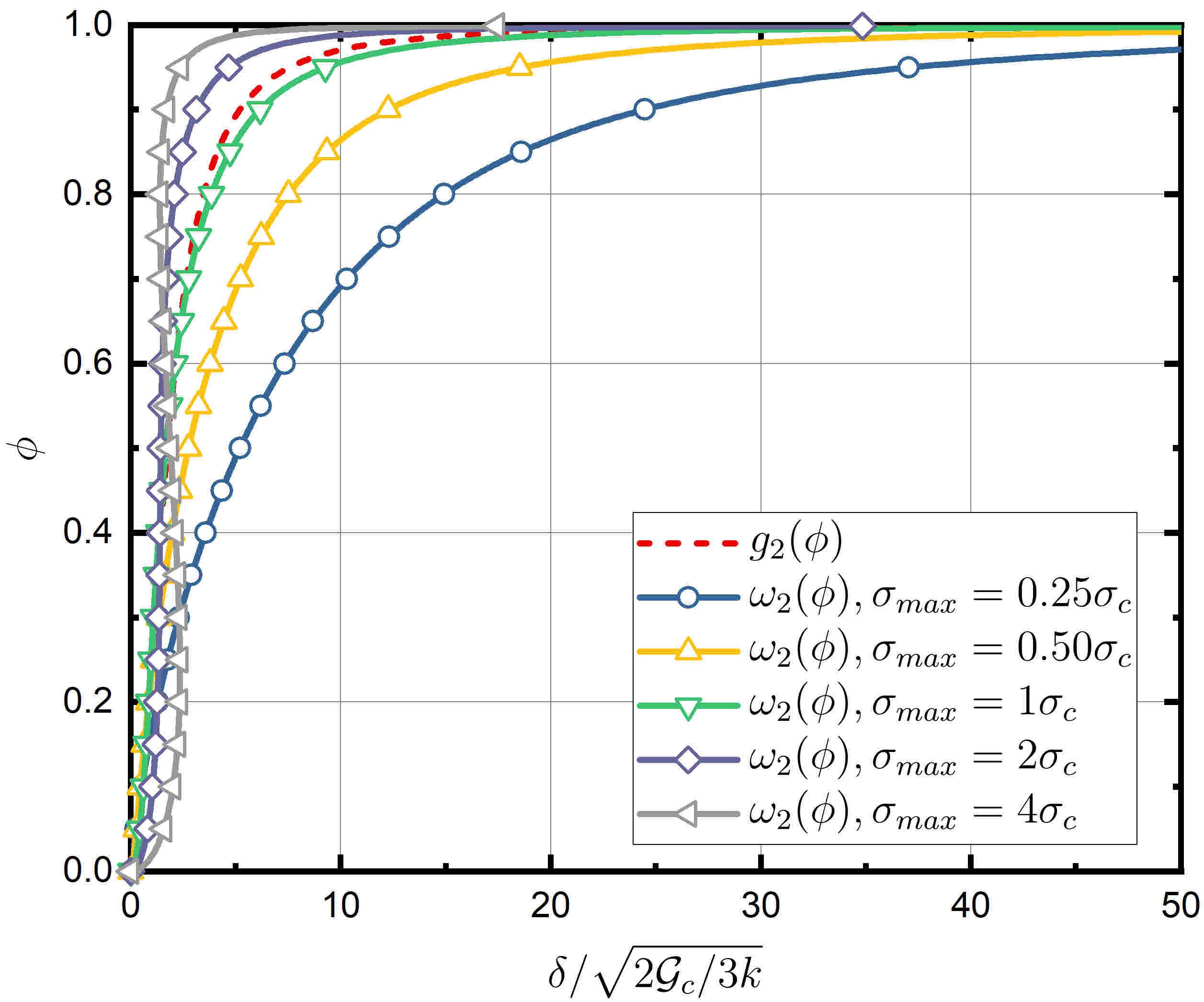}
        \caption{}
        \label{fig_phi2delat_sigma}
    \end{subfigure}
    \caption{Comparison between the classical $g_2(\phi)$ and $\omega_2(\phi)$ with the identical $\mathcal{G}_c$ and different $\sigma_c$: (a) degradation function, (b) traction-separation relationship and (c) relationship between separation and phase-field.}
    \label{fig_TSL_simga}
\end{figure}

The shapes of the degradation functions are illustrated in Fig. \ref{fig_g2andWd_sigma}. The cases for the $\sigma_{max}>\sigma_c$, the degradation functions are concave. On the other hand, the degradation functions are convex for the cases that $\sigma_{max}\le\sigma_c$. It can also be seen that the stiffness degrades steeply at the initial stage for the case that $\sigma_{max}\le\sigma_c$. In contrast, when $\phi$ is close to 1, the function $\omega_2(\phi)$ decreases gently for the cases that $\sigma_{max}\le\sigma_c$. The cases where $\sigma_{max}>\sigma_c$ is just the opposite.

The traction-separation curves for different $\sigma_{\rm max}$ are illustrated in Fig. \ref{fig_TSL_sigma}. It can be seen that for the cases that $\sigma_{\rm max}=4\sigma_c$, there is an obviously unstable range during the loading, which is known as snap-back behavior\cite{Sargado.2018}. The phenomenon suggests that the excessive strength is not suitable for the degradation function here. 

The relationship between normalized separation displacement and phase-field value is illustrated in Fig. \ref{fig_phi2delat_sigma}. The curve becomes steep with increasing $\sigma_{\rm max}$. Meanwhile, the unstable unloading stage of the case that $\sigma_{\rm max}=4\sigma_c $ can be seen obviously here.
\section{Numerical examples}
In this section, several numerical examples are presented to validate the present CZM under the assumption of two-dimensional plane-strain.
\label{secNumericalExample}
\subsection{Preliminary test on a single bar with a weak interface}
Considering an interfacial crack embedded in a single bar illustrated in Fig. \ref{fig_singlebar}, the total length of the bar $L_{\rm bar}=1\ {\rm mm}$. The $\omega^{\rm bulk}(\phi)$ is the classical $g_2(\phi)$ and the $\omega^{\rm int}(\phi)=\omega_t^{\rm int}(\phi)=\omega_n^{\rm int}(\phi)$ is $\omega_p(\phi)$. The properties of the interfacial and bulk materials are listed in table \ref{tb:matSB}. 
\begin{table}[htbp]
    \caption{Properties of the interfacial and bulk materials in a single bar.}\label{tb:matSB}
    \centering
    \begin{tabular}{cc}
        \toprule
        Material properties & Values \\
        \midrule
        Bulk material Young's modulus & $E_m = 4000\ \mathrm{MPa}$ \\
        Bulk material Poisson's ratio & $\upsilon_m = 0.4  $ \\
        Bulk material critical energy release rate & $\mathcal{G}_c^{\rm bulk} = 0.25\ \mathrm{N/mm}$ \\
        Normalized length & $\ell_0 = 0.02\ \mathrm{mm}$\\
        Interfacial penalty stiffness &  $k_t=k_n=100,000\ \mathrm{MPa/mm}$\\
        Interfacial ultimate stress & $\sigma_t=\sigma_n = 10\ \mathrm{MPa}$ \\
        Interfacial critical energy release rate & $\mathcal{G}_c^{\rm int} = 0.05\ \mathrm{N/mm}$ \\
        \bottomrule
    \end{tabular}
\end{table}

\begin{figure}[]
    \centering
    \begin{subfigure}{0.49\textwidth}
        \centering
        \includegraphics[width=\textwidth]{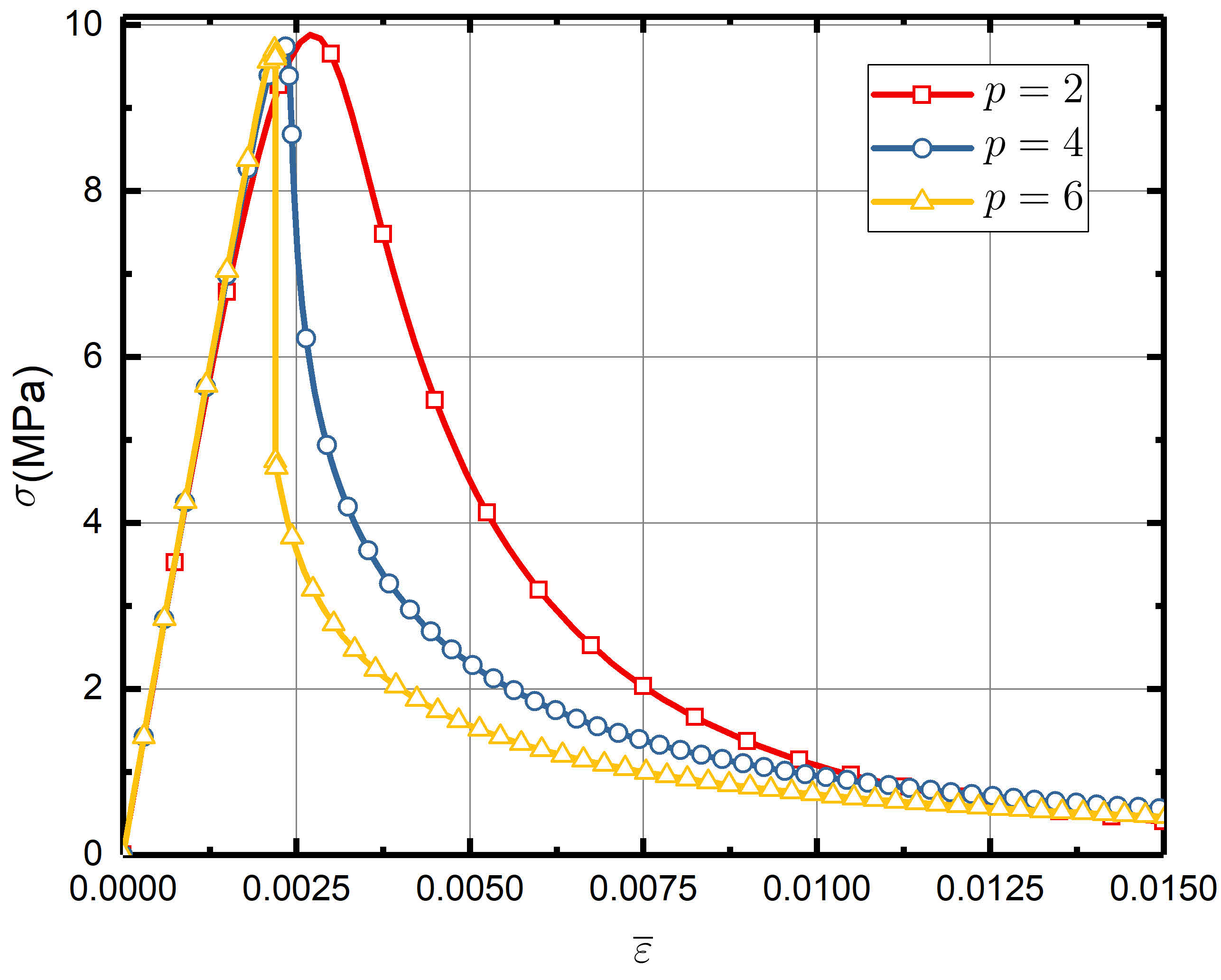}
        \caption{}
        \label{fig_singlebarp}
    \end{subfigure}
    \begin{subfigure}{0.49\textwidth}
        \centering
        \includegraphics[width=\textwidth]{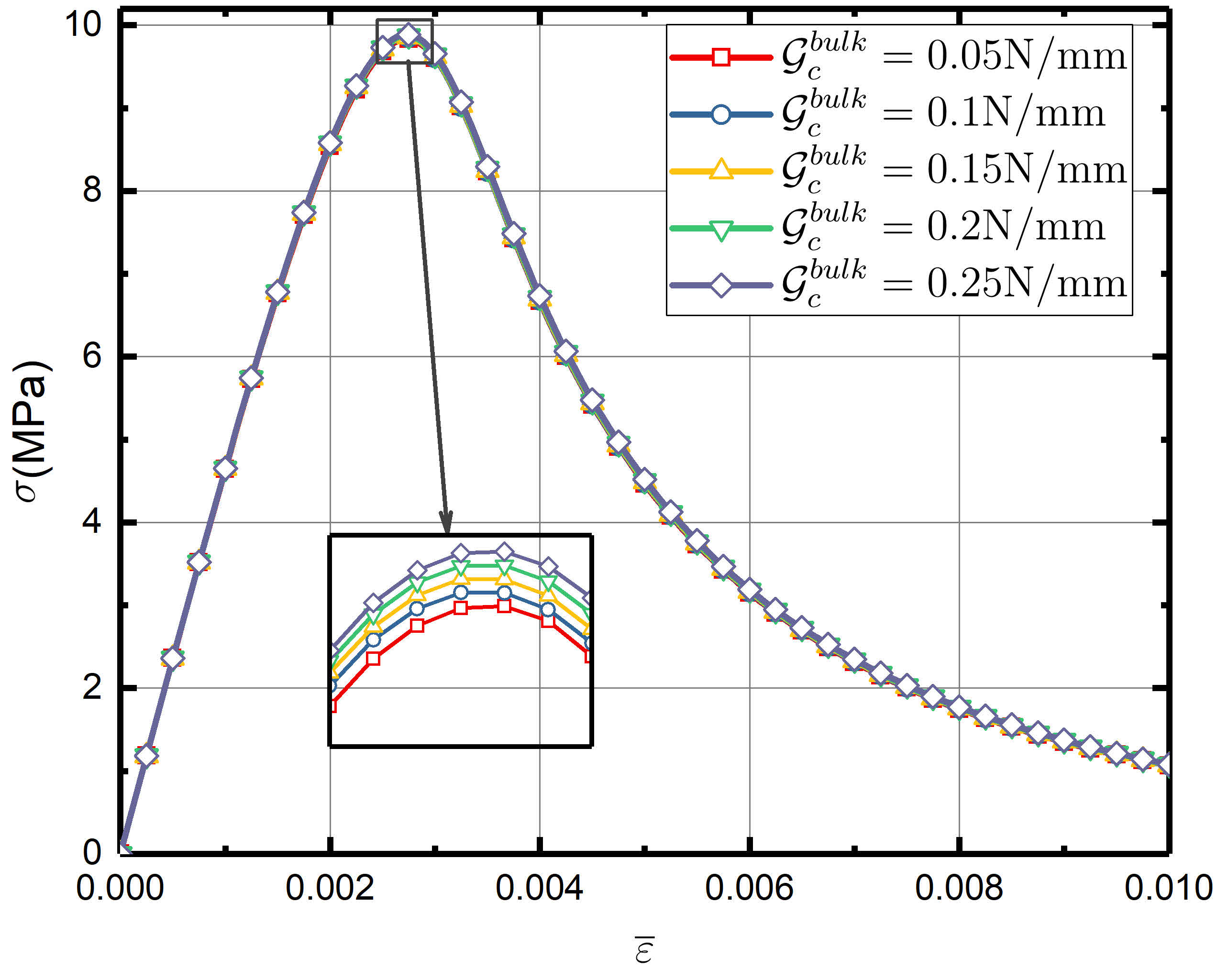}
        \caption{}
        \label{fig_singlebarGc}
    \end{subfigure}
    \caption{Single bar test: (a) displacement-force curves with different $p$ values and (b) displacement-force curves for $p=2$ with different $\mathcal{G}_c^{\rm bulk}$.}
    \label{}
\end{figure}

Displacement-force relationships are illustrated in Fig. \ref{fig_singlebarp}. The $p$ value of $\omega^{\rm int}(\phi)$ is 2, 4 and 6 here. It can be seen that the ultimate stress is identical for different $p$. Meanwhile, the softening curve is similar to that in TSL, in which the bulk region is not activated. However, the ultimate stress is lower than $\sigma_c$. The main reason is that the damage also initializes in the bulk region, which decreases the overall strength.

The comparison between different $\mathcal{G}_c^{\rm bulk}$ is illustrated in Fig. \ref{fig_singlebarGc}. In the test, the $p=2$ is fixed for the interfacial element. It can be seen that the $\mathcal{G}_c^{\rm bulk}$ plays a quite minor role in the overall mechanical properties. In addition, the ultimate stress approaches the $\sigma_c$ with increasing $\mathcal{G}_c^{\rm bulk}$, which also indicates that the present algorithm is suitable for the case that $\mathcal{G}_c^{\rm bulk}\gg \mathcal{G}_c^{\rm int}$.

\subsection{Double cantilever beam test}
\begin{figure}[]
     \centering
    \begin{subfigure}{0.48\textwidth}
        \centering
        \includegraphics[width=\textwidth]{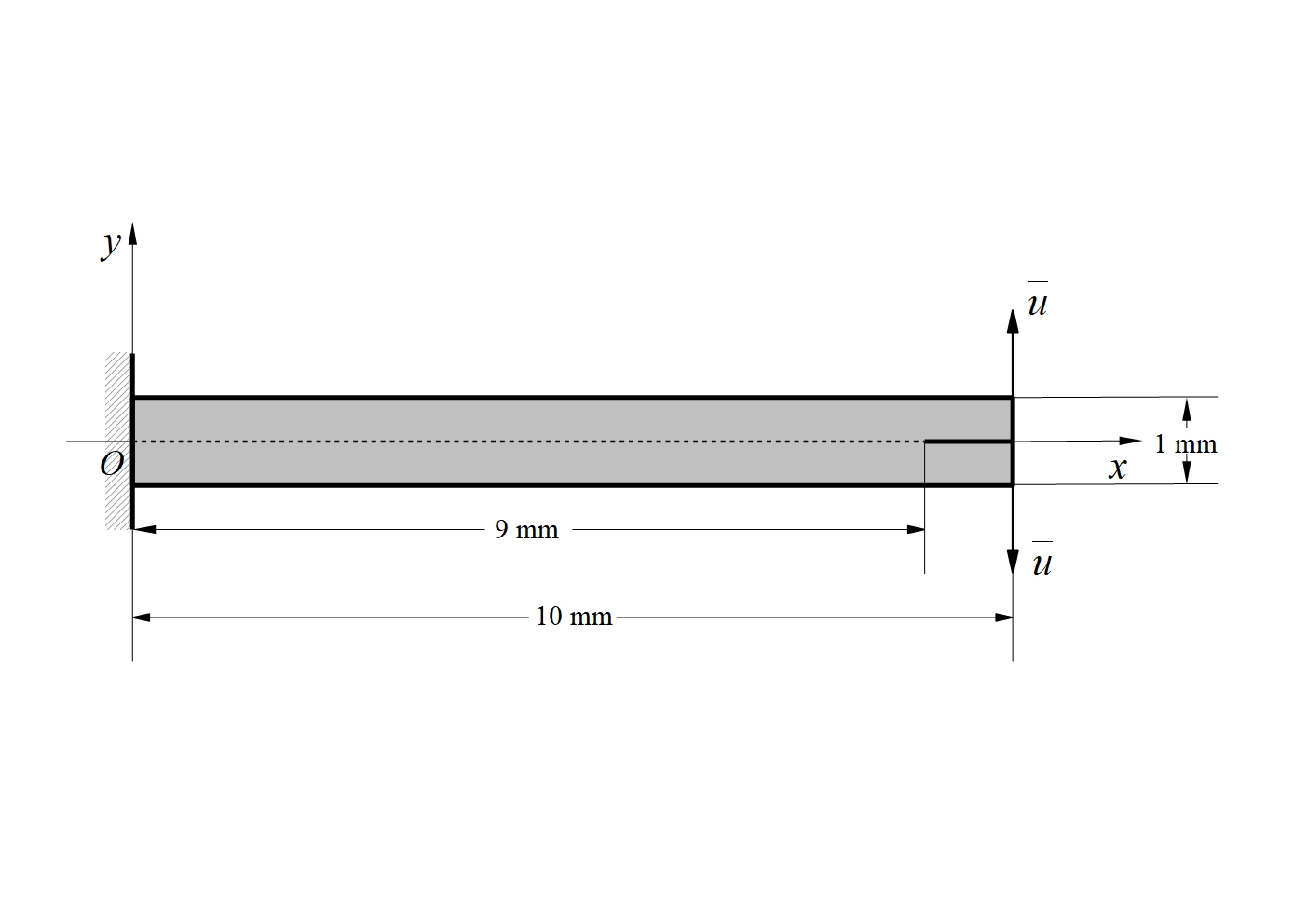}
        \caption{}
        \label{figDCBspecimen}
    \end{subfigure}
    \begin{subfigure}{0.48\textwidth}
        \centering
        \includegraphics[width=\textwidth]{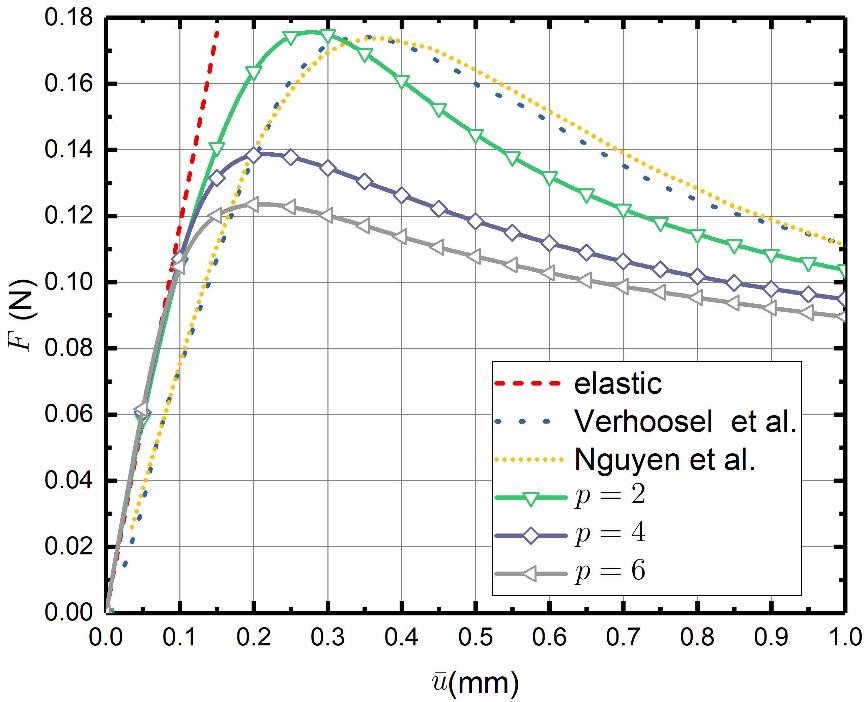}
        \caption{}
        \label{figDCBDFcurve}
    \end{subfigure}
    \caption{The DCB test: (a) the size of the specimen and (b) displacement-force curve with different cohesive models.}
\end{figure}

In the subsection, a double cantilever beam (DCB) test is carried out to validate the performance of the present cohesive model. The size of the specimen here is illustrated in Fig. \ref{figDCBspecimen}. A predefined crack is located at the right side and interfacial elements are inserted in the middle layer of the specimen. A pair of concentrated forces are applied on two nodes on the right side of the beam. Meanwhile, the nodes on the left side are fixed during the loading process. The values of material properties are listed in table \ref{tb:matDCB}. The main difference between the single bar test and DCB is that the gradient term, $\nabla\phi$, is not zero during the crack propagation.
\begin{table}[htbp]
    \caption{Properties of the interfacial and bulk materials in the DCB test.\cite{Nguyen.2016}}\label{tb:matDCB}
    \centering
    \begin{tabular}{cc}
        \toprule
        Material properties & Values \\
        \midrule
        Bulk material Young's modulus & $E_m = 100\ \mathrm{MPa}$ \\
        Bulk material Poisson's ratio & $\upsilon_m = 0.3  $ \\
        Bulk material critical energy release rate & $\mathcal{G}_c^{\rm bulk} = 1.0\ \mathrm{N/mm}$ \\
        Normalized length & $\ell_0 = 0.05\ \mathrm{mm}$\\
        Interfacial penalty stiffness &  $k_t=k_n=100,000\ \mathrm{MPa/mm}$\\
        Interfacial ultimate stress & $\sigma_t=\sigma_n = 1.0\ \mathrm{MPa}$ \\
        Interfacial critical energy release rate & $\mathcal{G}_c^{\rm int} = 0.1\ \mathrm{N/mm}$ \\
        \bottomrule
    \end{tabular}
\end{table}

The displacement-load curve is illustrated in Fig. \ref{figDCBDFcurve}. Compared with the result from \cite{Verhoosel.2013} and \cite{Nguyen.2016}, the main character of the present model is the long quasi-elastic stage. Apart from different TSL, the different ways to evaluate the displacement jump can also lead to the difference. The stiffness starts to degrade when force is about to reach the peak value. Meanwhile, the ultimate force decreases with increasing $p$ value. The main reason is that the stiffness of interfacial elements degrades steeply once the ultimate $\sigma_c$ is reached. This phenomenon is not obvious in the single bar test due to the lack of damage gradient.

\begin{figure}[]
   \centering
   \begin{subfigure}[t]{0.3\textwidth}
       \centering
       \includegraphics[width=\textwidth]{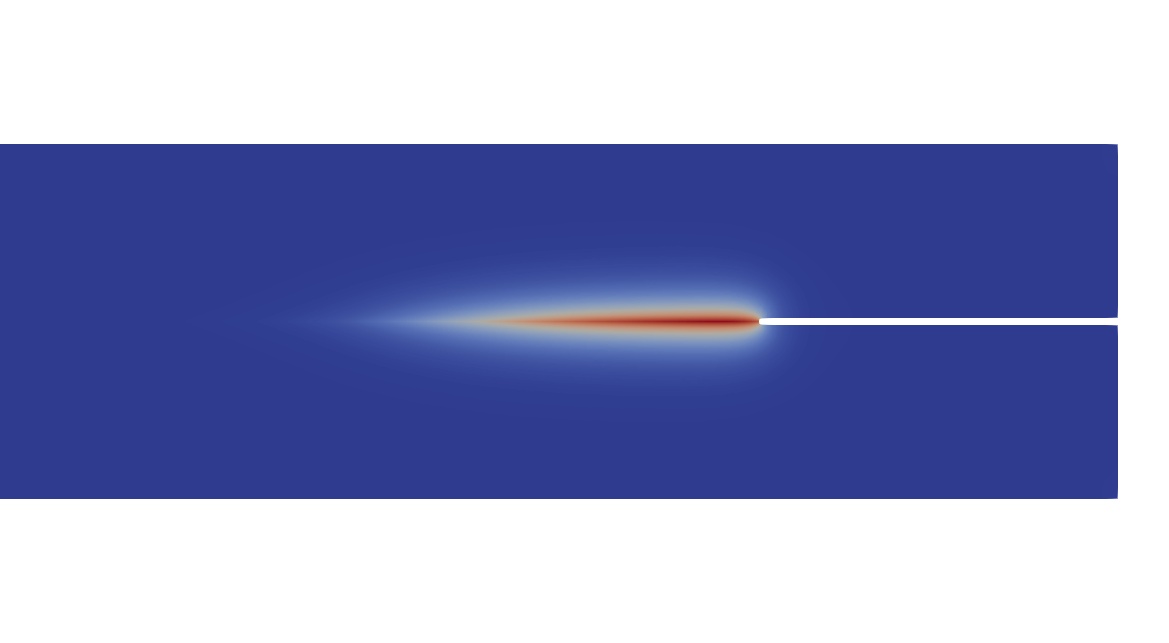}
   \end{subfigure}
   \begin{subfigure}[t]{0.3\textwidth}
        \centering
        \includegraphics[width=\textwidth]{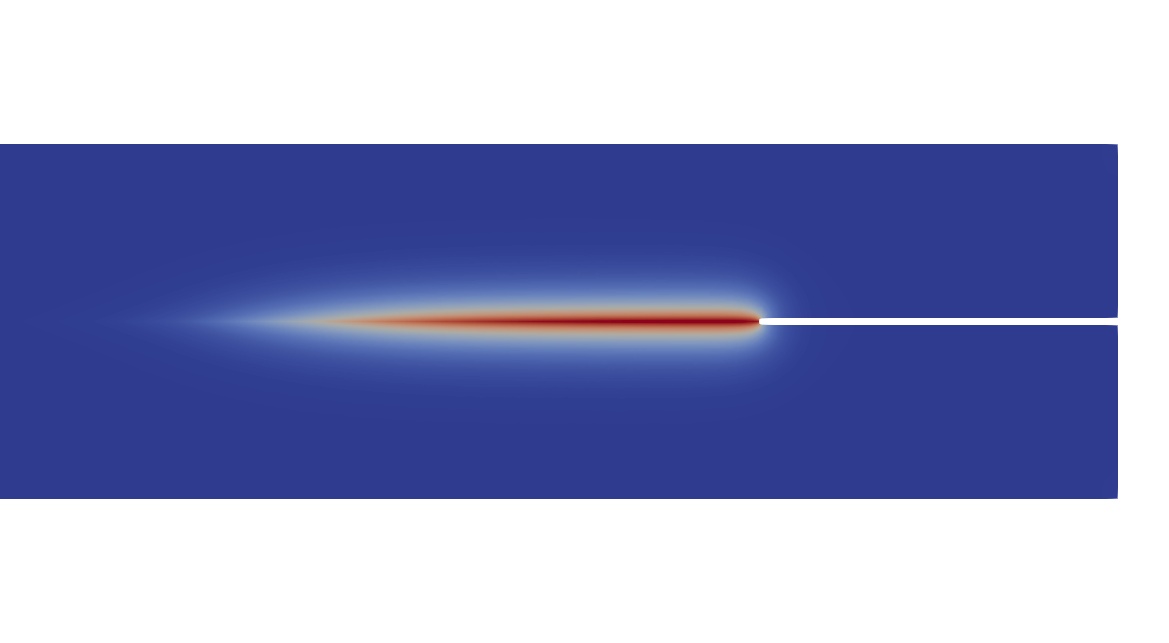}
        \caption{}
    \end{subfigure}
    \begin{subfigure}[t]{0.3\textwidth}
        \centering
        \includegraphics[width=\textwidth]{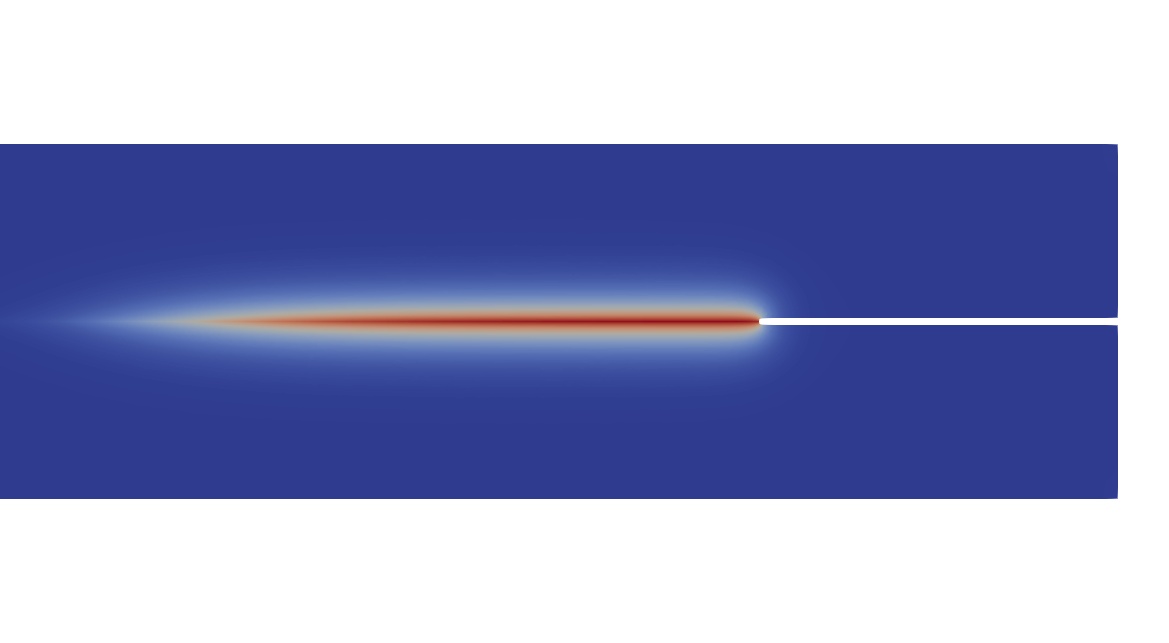}
    \end{subfigure}

    \begin{subfigure}[t]{0.3\textwidth}
        \centering
        \includegraphics[width=\textwidth]{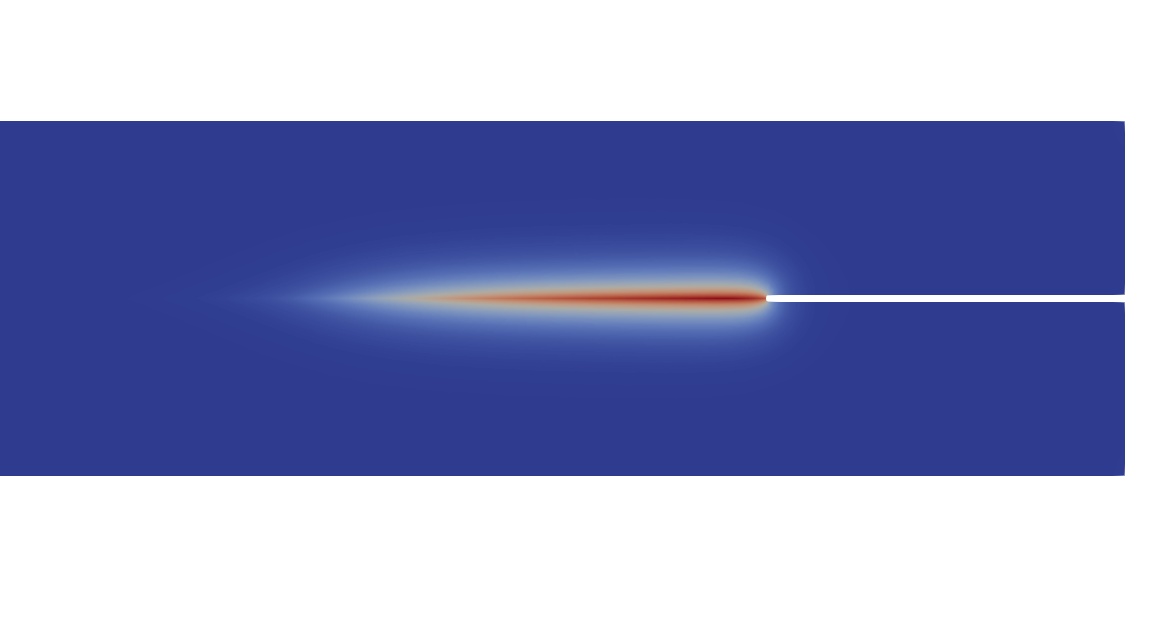}
    \end{subfigure}
    \begin{subfigure}[t]{0.3\textwidth}
         \centering
         \includegraphics[width=\textwidth]{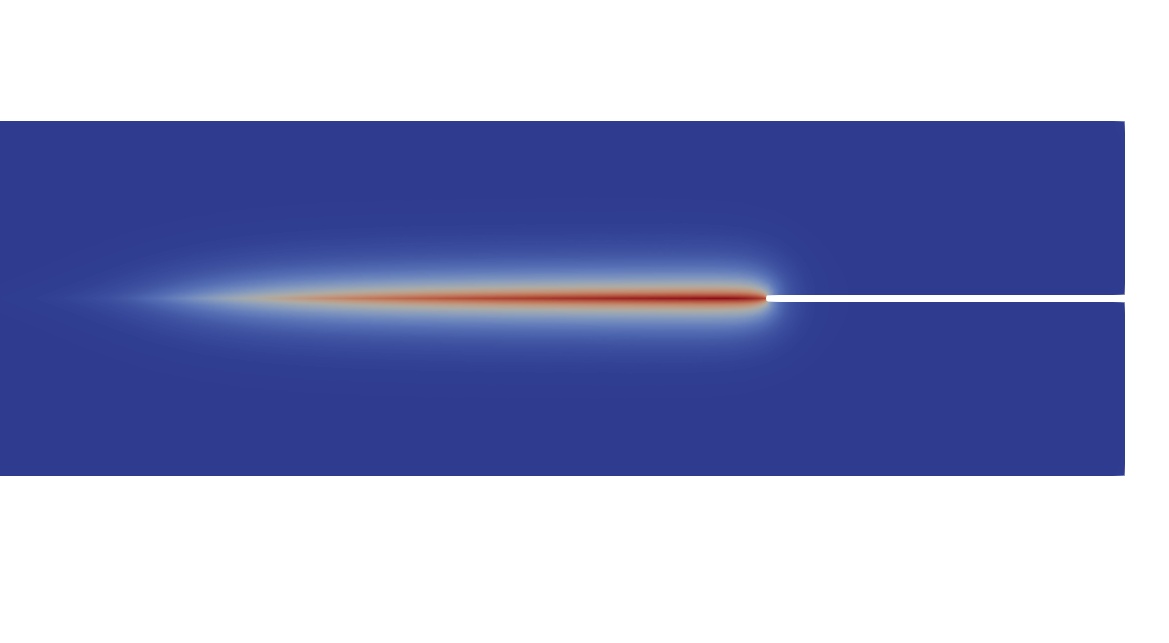}
         \caption{}
     \end{subfigure}
     \begin{subfigure}[t]{0.3\textwidth}
         \centering
         \includegraphics[width=\textwidth]{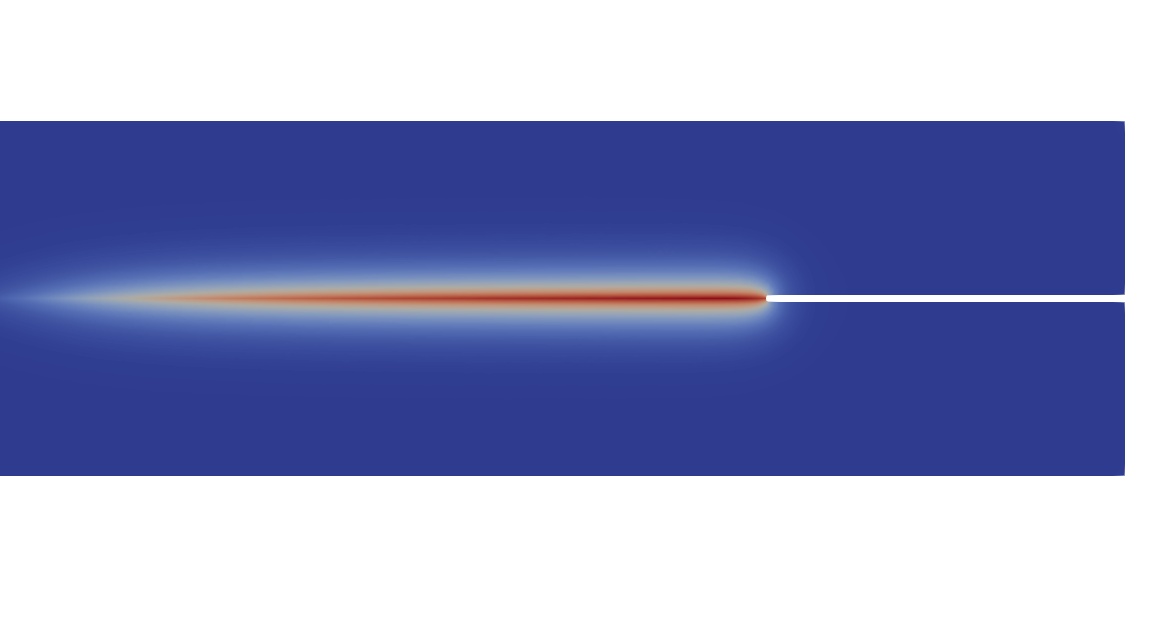}
     \end{subfigure}
     
    \begin{subfigure}[t]{0.3\textwidth}
        \centering
        \includegraphics[width=\textwidth]{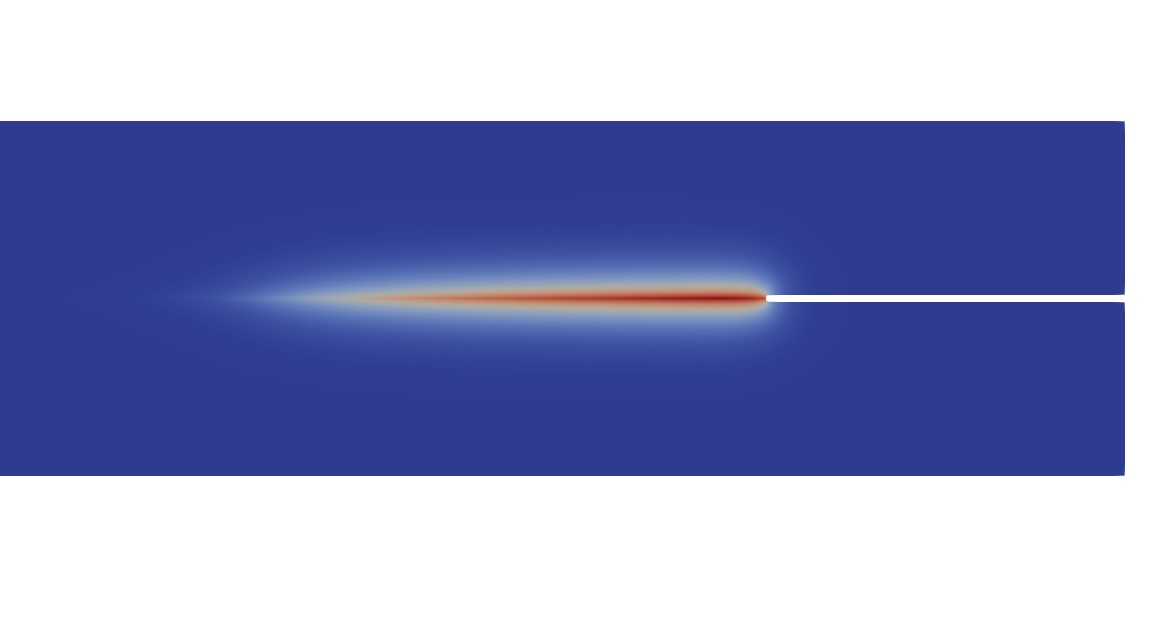}
    \end{subfigure}
    \begin{subfigure}[t]{0.3\textwidth}
         \centering
         \includegraphics[width=\textwidth]{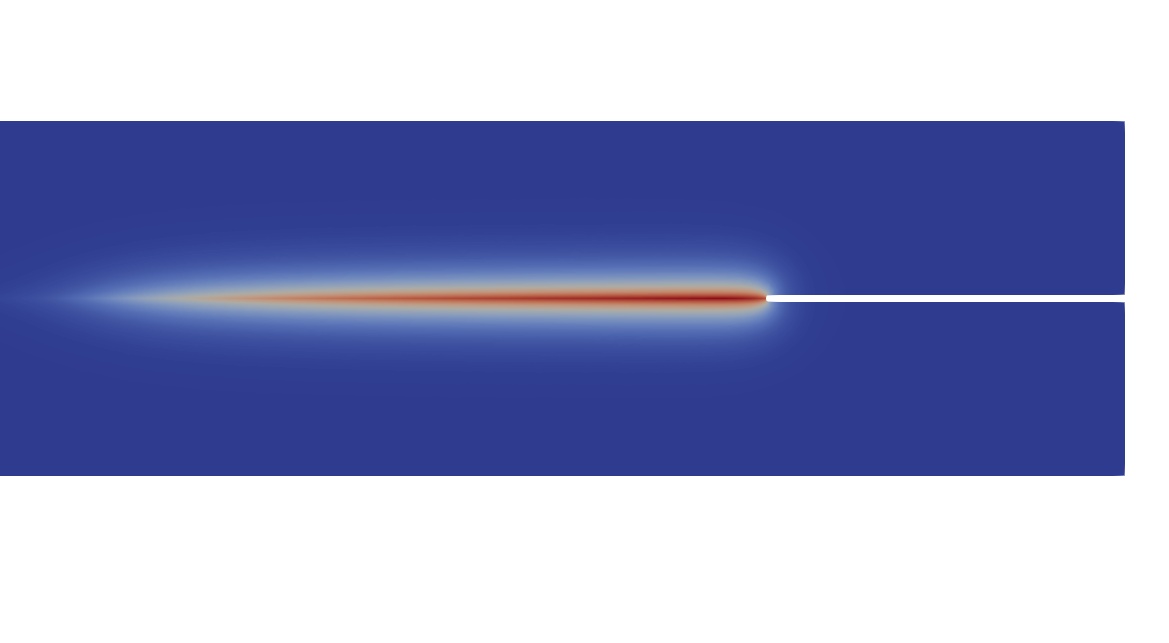}
         \caption{}
     \end{subfigure}
     \begin{subfigure}[t]{0.3\textwidth}
         \centering
         \includegraphics[width=\textwidth]{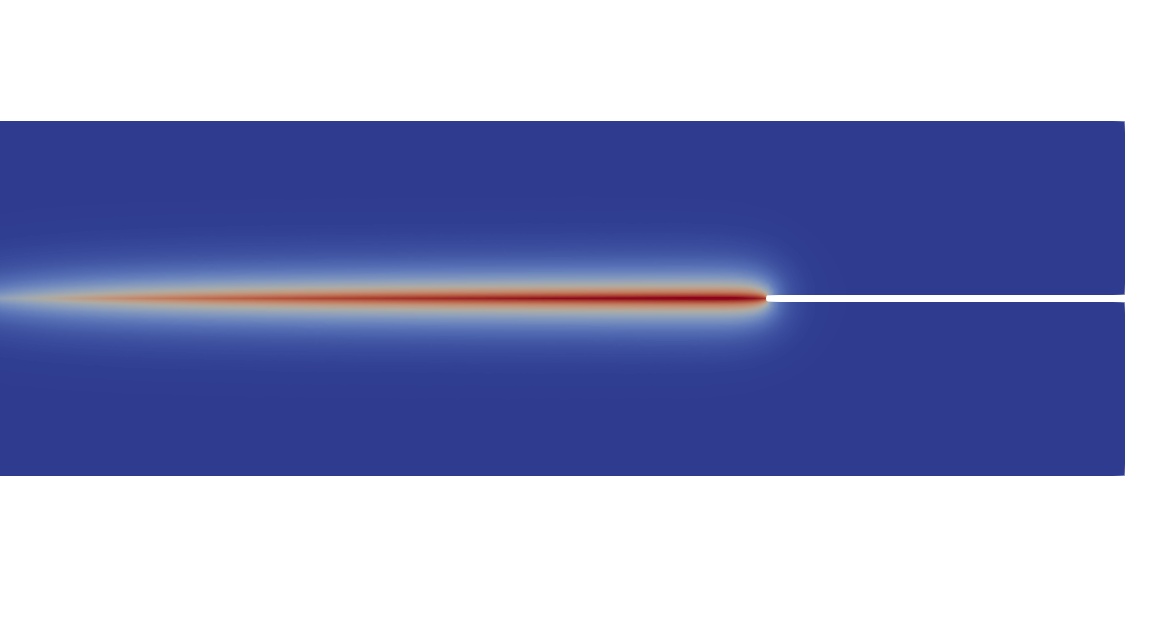}
     \end{subfigure}
     \caption{The debonding region in the DCB test: (a) $p=2$, (b) $p=4$ and $p=6$ for $\overline{u}=0.5\mathrm{mm},\ 0.75\mathrm{mm}\ \text{and}\ 1.0\mathrm{mm}$.} 
     \label{figCrackinDCB}

\end{figure}

Cracks in the beam under different loads are illustrated in Fig. \ref{figCrackinDCB}. The length of crack increases with increasing $p$, which also indicates that high-order $\omega_p(\phi)$ is softer under the identical external load.
\subsection{Three-point bending test}
\begin{figure}[]
    \centering
    \begin{subfigure}{0.48\textwidth}
        \centering
        \includegraphics[width=\textwidth]{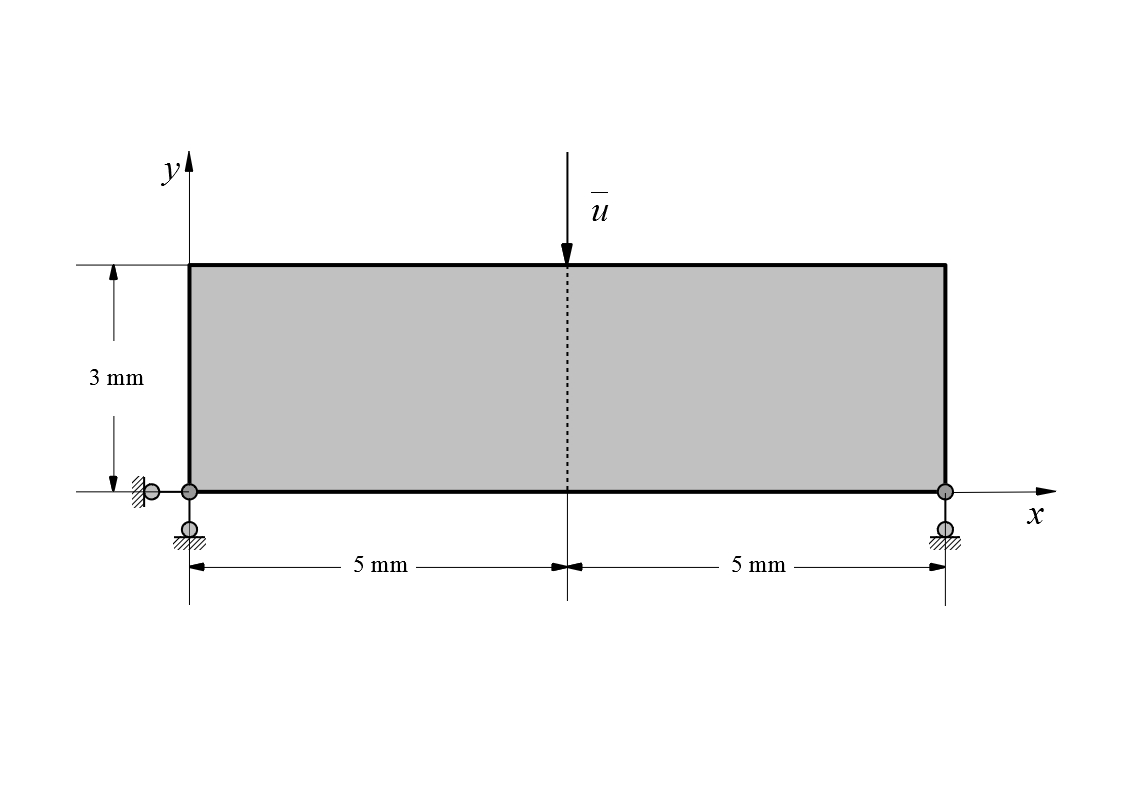}
        \caption{}
        \label{figTPBspecimen}
    \end{subfigure}
    \begin{subfigure}{0.48\textwidth}
        \centering
        \includegraphics[width=\textwidth]{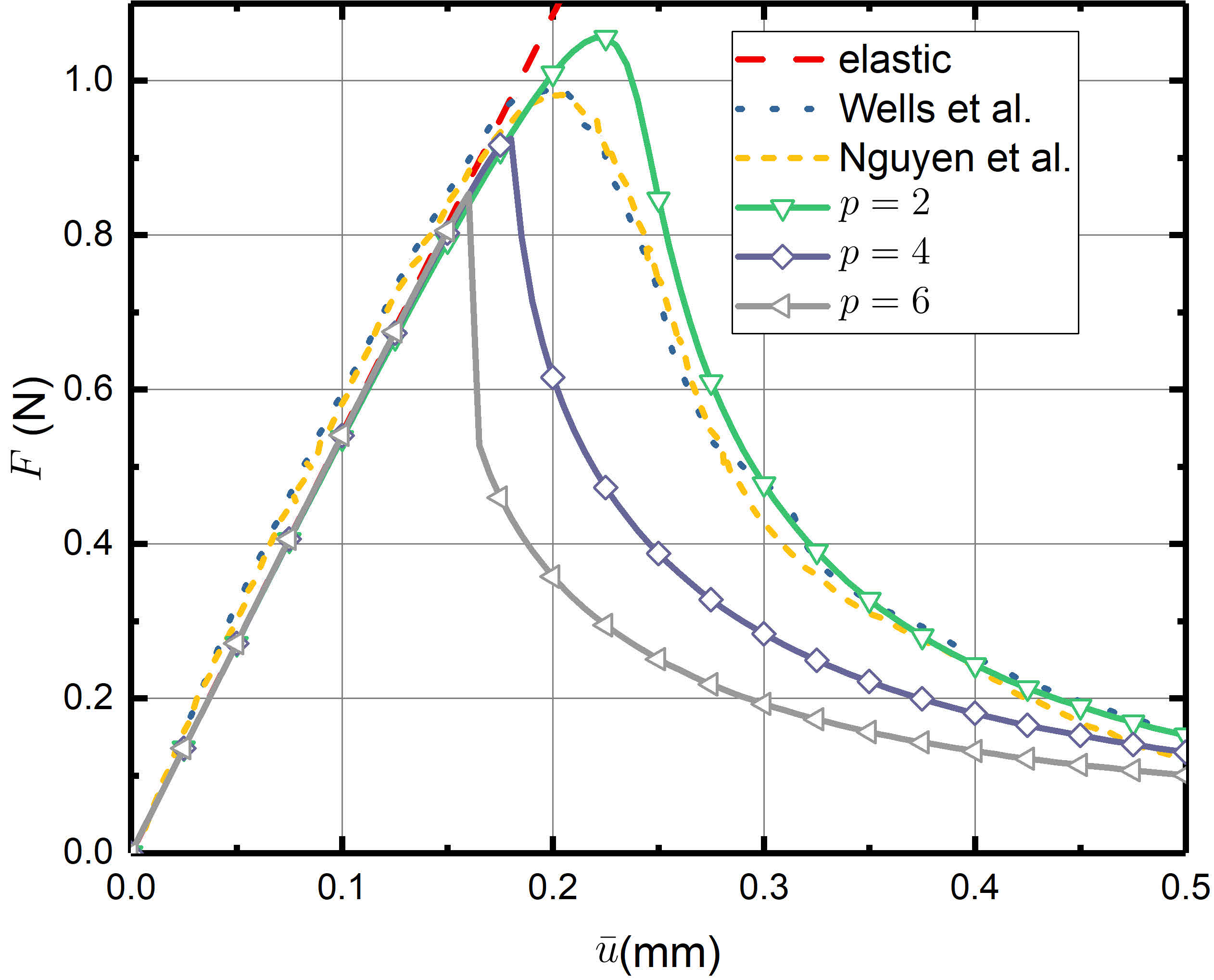}
        \caption{}
        \label{figTPBDFcurve}
    \end{subfigure}
    \caption{The three point bending-test: (a) the size of the beam and (b) displacement-force curve with different cohesive models.}
\end{figure}

In the subsection, a three-point bending (TPB) test is carried out. The load point is located at the middle point of a simply supported beam top surface. The cohesive element is set in the middle of the beam. Other details of the specimen are illustrated in Fig. \ref{figTPBspecimen}. The values of material properties are listed in table \ref{tb:matTPB}.

\begin{table}[htbp]
    \caption{Properties of the interfacial and bulk materials in the TPB test.\cite{Nguyen.2016}}\label{tb:matTPB}
    \centering
    \begin{tabular}{cc}
        \toprule
        Material properties & Values \\
        \midrule
        Bulk material Young's modulus & $E_m = 100\ \mathrm{MPa}$ \\
        Bulk material Poisson's ratio & $\upsilon_m = 0.0  $ \\
        Bulk material critical energy release rate & $\mathcal{G}_c^{\rm bulk} = 1.0\ \mathrm{N/mm}$ \\
        Normalized length & $\ell_0 = 0.15\ \mathrm{mm}$\\
        Interfacial penalty stiffness &  $k_t=k_n=100,000\ \mathrm{MPa/mm}$\\
        Interfacial ultimate stress & $\sigma_t=\sigma_n = 1.0\ \mathrm{MPa}$ \\
        Interfacial critical energy release rate & $\mathcal{G}_c^{\rm int} = 0.1\ \mathrm{N/mm}$ \\
        \bottomrule
    \end{tabular}
\end{table}

The displacement-force curve is illustrated in Fig. \ref{figTPBDFcurve}. Comparison between the present work and results from other models are also presented \cite{Nguyen.2016,Wells.2001}.  The ultimate force a little higher than the others results. On the other hand, the peak value of the load decreases with increasing $p$ value.

\begin{figure}[]
    \centering
    \begin{subfigure}[t]{0.3\textwidth}
        \centering
        \includegraphics[width=\textwidth]{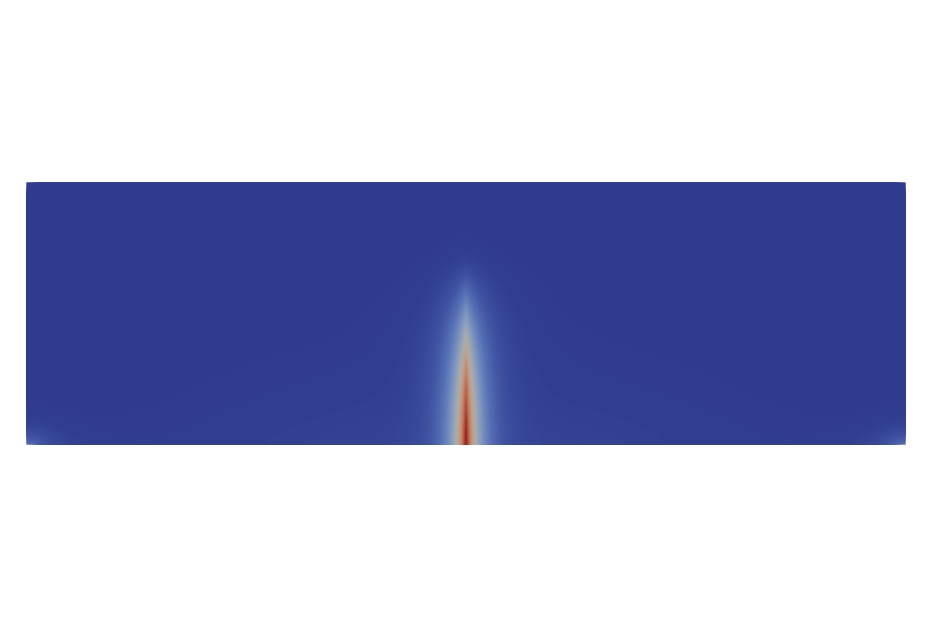}
    \end{subfigure}
    \begin{subfigure}[t]{0.3\textwidth}
         \centering
         \includegraphics[width=\textwidth]{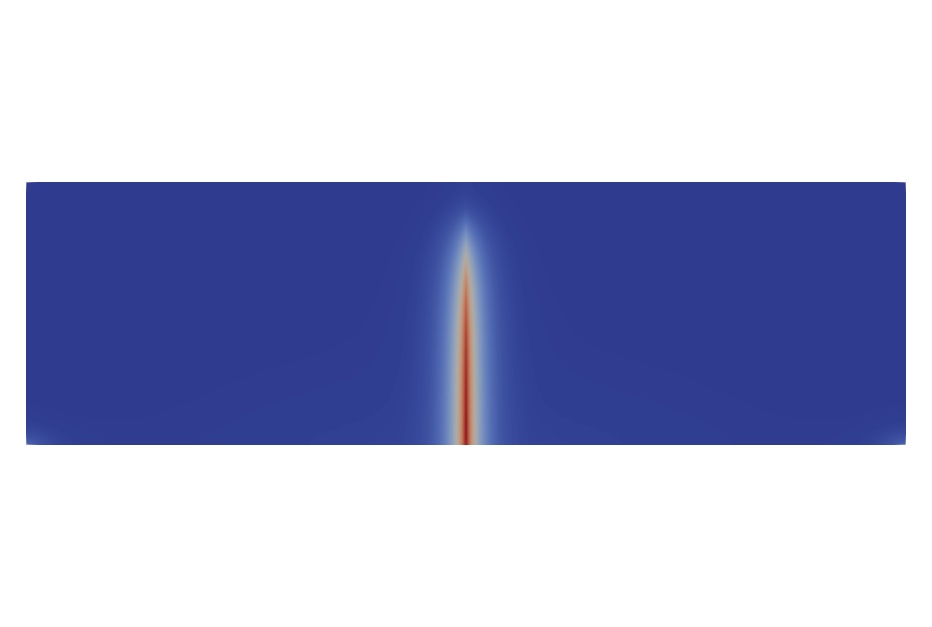}
         \caption{}
     \end{subfigure}
     \begin{subfigure}[t]{0.3\textwidth}
         \centering
         \includegraphics[width=\textwidth]{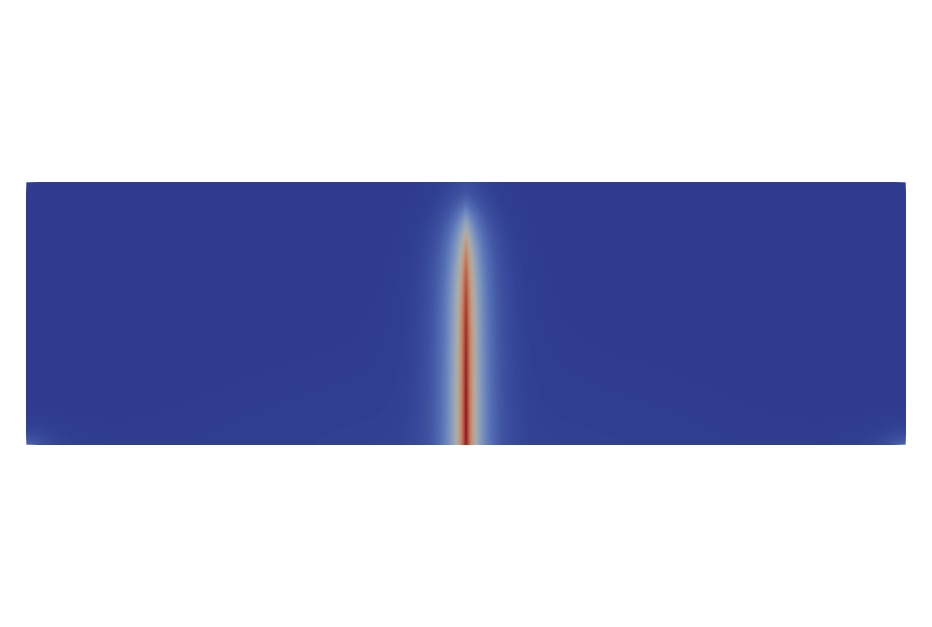}
     \end{subfigure}
 
     \begin{subfigure}[t]{0.3\textwidth}
         \centering
         \includegraphics[width=\textwidth]{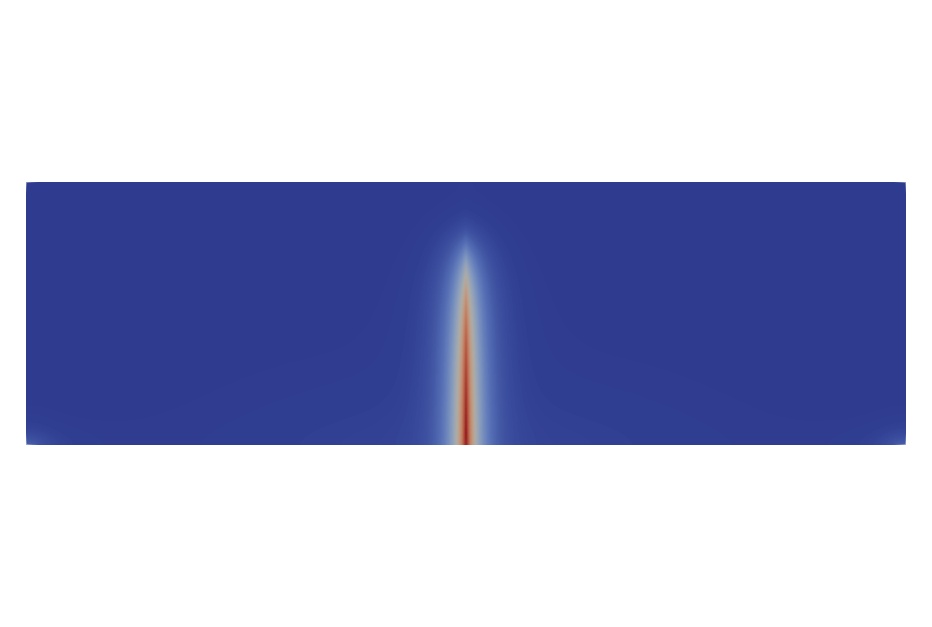}
     \end{subfigure}
     \begin{subfigure}[t]{0.3\textwidth}
          \centering
          \includegraphics[width=\textwidth]{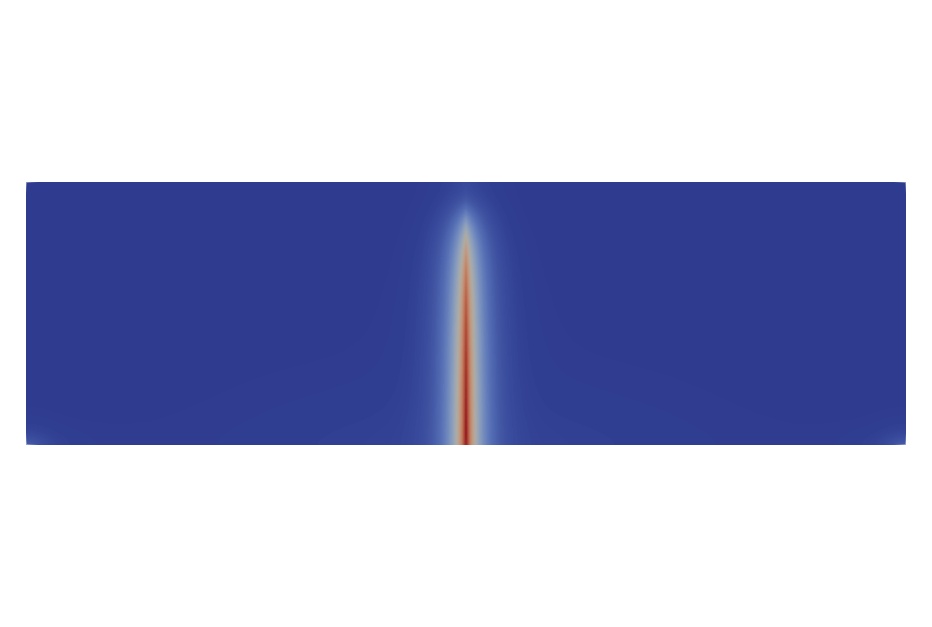}
          \caption{}
      \end{subfigure}
      \begin{subfigure}[t]{0.3\textwidth}
          \centering
          \includegraphics[width=\textwidth]{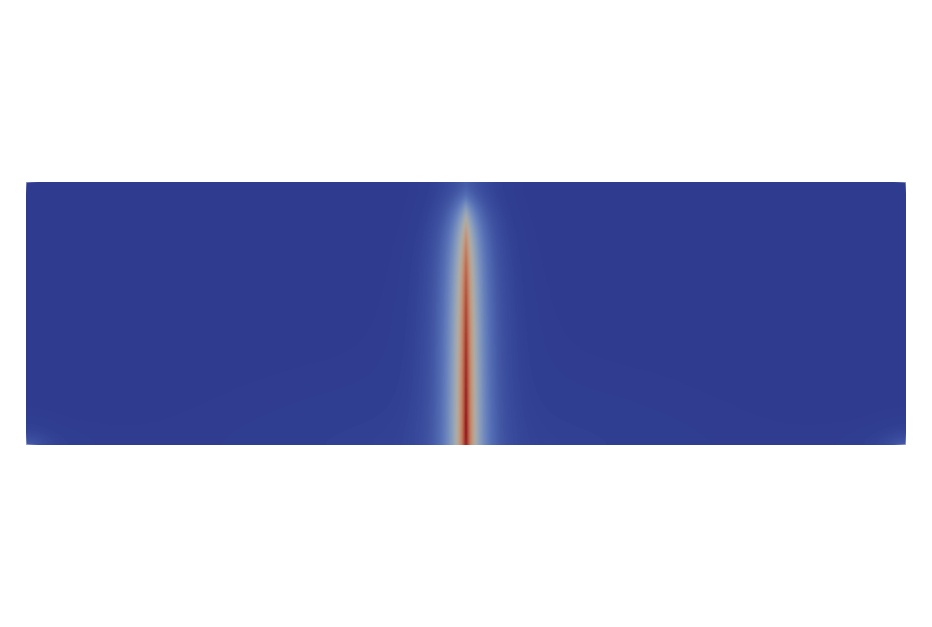}
      \end{subfigure}
      
     \begin{subfigure}[t]{0.3\textwidth}
         \centering
         \includegraphics[width=\textwidth]{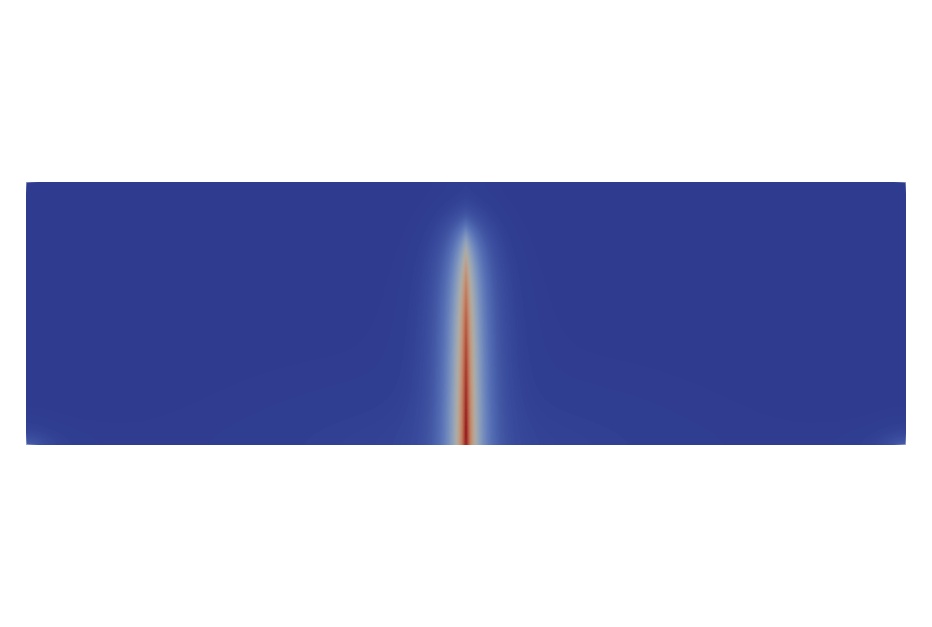}
     \end{subfigure}
     \begin{subfigure}[t]{0.3\textwidth}
          \centering
          \includegraphics[width=\textwidth]{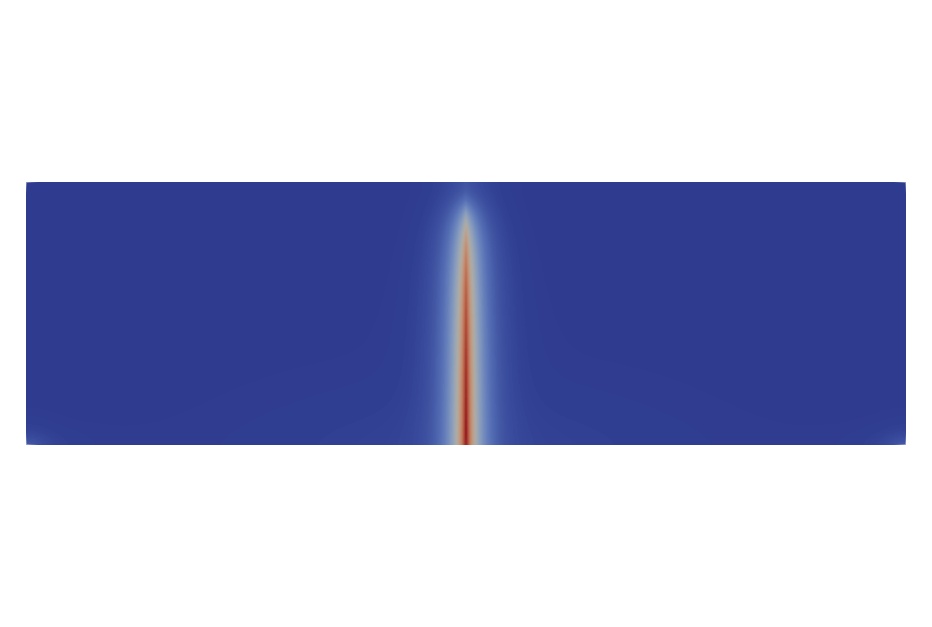}
          \caption{}
      \end{subfigure}
      \begin{subfigure}[t]{0.3\textwidth}
          \centering
          \includegraphics[width=\textwidth]{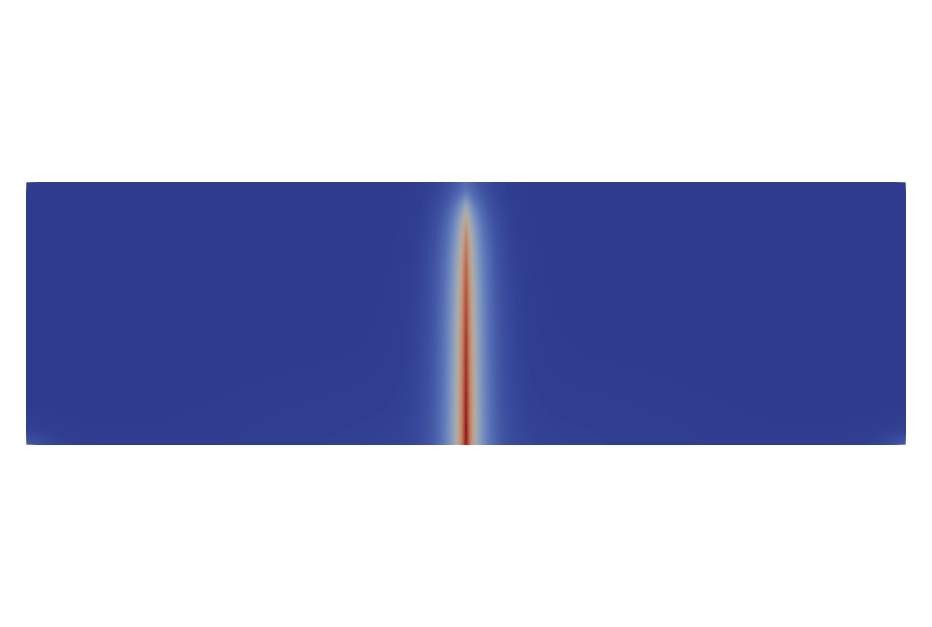}
      \end{subfigure}
      \caption{The debonding region in the TPB test: (a) $p=2$, (b) $p=4$ and $p=6$ for $\overline{u}=0.5\mathrm{mm},\ 0.75\mathrm{mm}\ \text{and}\ 1.0\mathrm{mm}$.} 
      \label{figCrackinTPB}
\end{figure}

\begin{figure}[]
    \centering
    \begin{subfigure}[t]{0.32\textwidth}
        \centering
        \includegraphics[width=\textwidth]{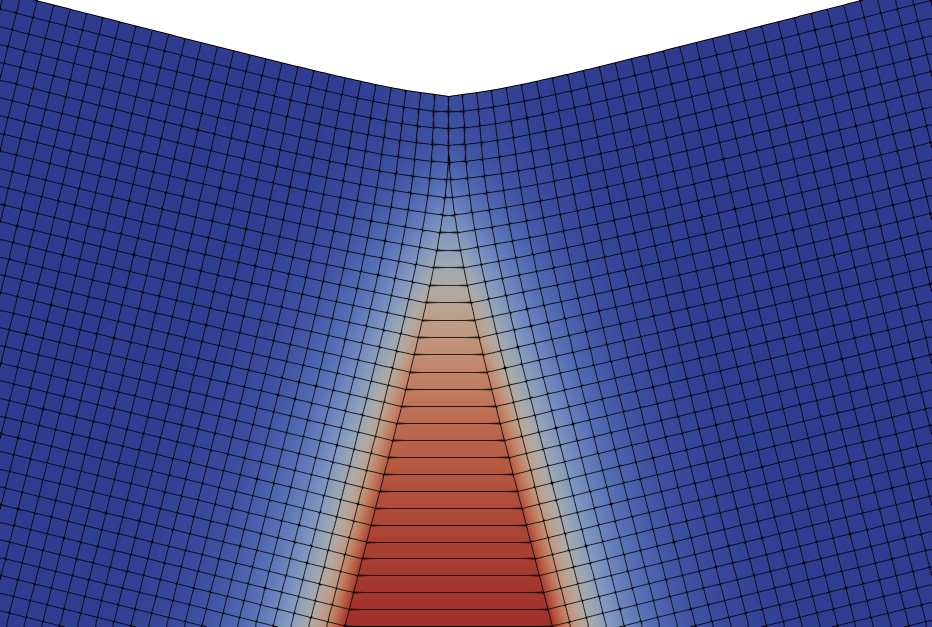}
        \caption{}
    \end{subfigure}
    \begin{subfigure}[t]{0.32\textwidth}
         \centering
         \includegraphics[width=\textwidth]{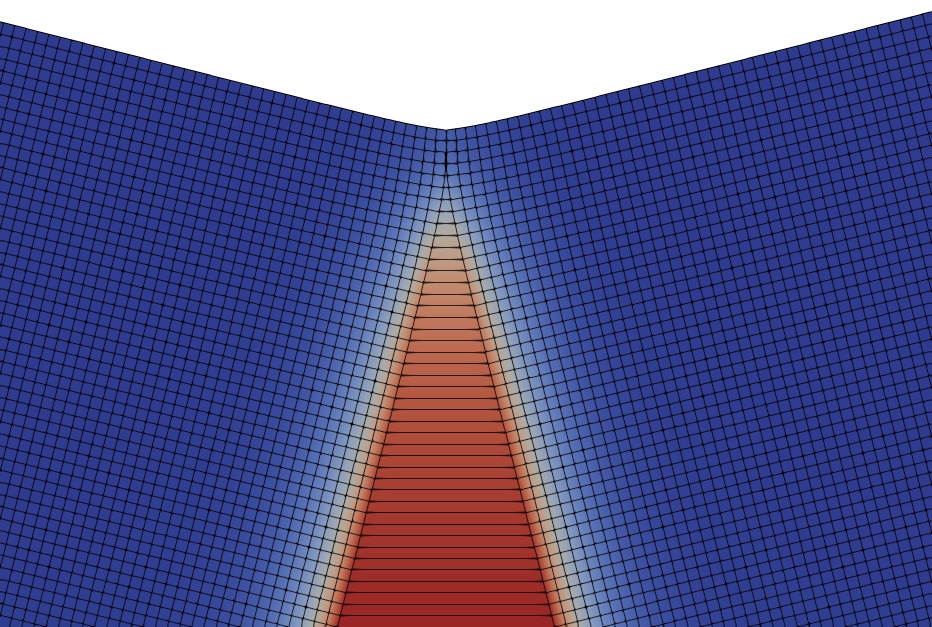}
         \caption{}
     \end{subfigure}
     \begin{subfigure}[t]{0.32\textwidth}
         \centering
         \includegraphics[width=\textwidth]{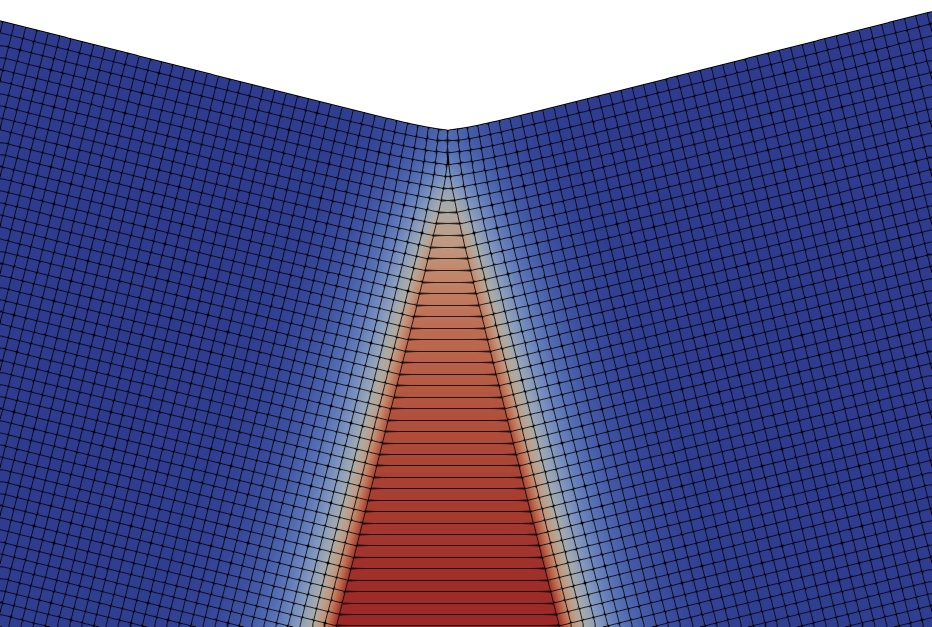}
         \caption{}
     \end{subfigure}
     \caption{Mesh details near the crack tip in three point bending test with different degradation functions $\omega_p(\phi)$: (a) $p=2$, (b) $p=4$ and $p=6$.}
     \label{figDetailsinTPB}
\end{figure}

Cracks in the beam under different loads are illustrated in Fig. \ref{figCrackinTPB}. Under the identical external load, the length of crack increases with increasing $p$. However, the differences on the crack length get smaller with the increasing load value. The details of the mesh near the crack tips are illustrated in Fig. \ref{figDetailsinTPB}. The present cohesive elements separate if they are under tension. Meanwhile, there is no penetration between elements under compression, which means that penalty stiffness can be kept under the present algorithm. 

\subsection{Single-fiber reinforced composites test}
The single fiber test is also carried out to validate the cohesive element properties under a more complicated loading case. A square representative volume element (RVE) is built and periodic boundary conditions (PBC) are used to avoid stress concentration near boundaries of the RVE. The length of RVE is set to be $L = 1\ \mathrm{mm}$ and the diameter of the fiber is $d = 0.5\ \mathrm{mm}$. In the test, the $\omega_p$ will be used not only in the interfaces but also in the bulk region. 

\begin{table}[htbp]
    \caption{Properties of the interfacial and bulk materials in a single fiber reinforced composites test.\cite{Zhang.2020}}\label{tb:matSF}
    \centering
    \begin{tabular}{cc}
        \toprule
        Material properties & Values \\
        \midrule
        Matrix material Young's modulus & $E_m = 4,000\ \mathrm{MPa}$ \\
        Matrix material Poisson's ratio & $\upsilon_m = 0.4  $ \\
        Matrix material critical energy release rate & $\mathcal{G}_c^{bulk} = 0.25\ \mathrm{N/mm}$ \\
        Matrix material strength (only available for $\omega^{\rm bulk}(\phi)=\omega_p(\phi)$)& $\sigma_c = 30\ \mathrm{MPa}$ \\
        Fiber material Young's modulus & $E_f = 40,000\ \mathrm{MPa}$ \\
        Fiber material Poisson's ratio & $\upsilon_f = 0.33\ \mathrm{MPa}$ \\
        Normalized length & $\ell_0 = 0.02\ \mathrm{mm}$\\
        Interfacial penalty stiffness &  $k_t=k_n=100,000\ \mathrm{MPa/mm}$\\
        Interfacial ultimate stress & $\sigma_t=\sigma_n = 10.0\ \mathrm{MPa}$ \\
        Interfacial critical energy release rate & $\mathcal{G}_c^{int} = 0.05\ \mathrm{N/mm}$ \\
        \bottomrule
    \end{tabular}
\end{table}

\begin{figure}[]
    \centering
    \begin{subfigure}[b]{0.48\textwidth}
        \centering
        \includegraphics[width=\textwidth]{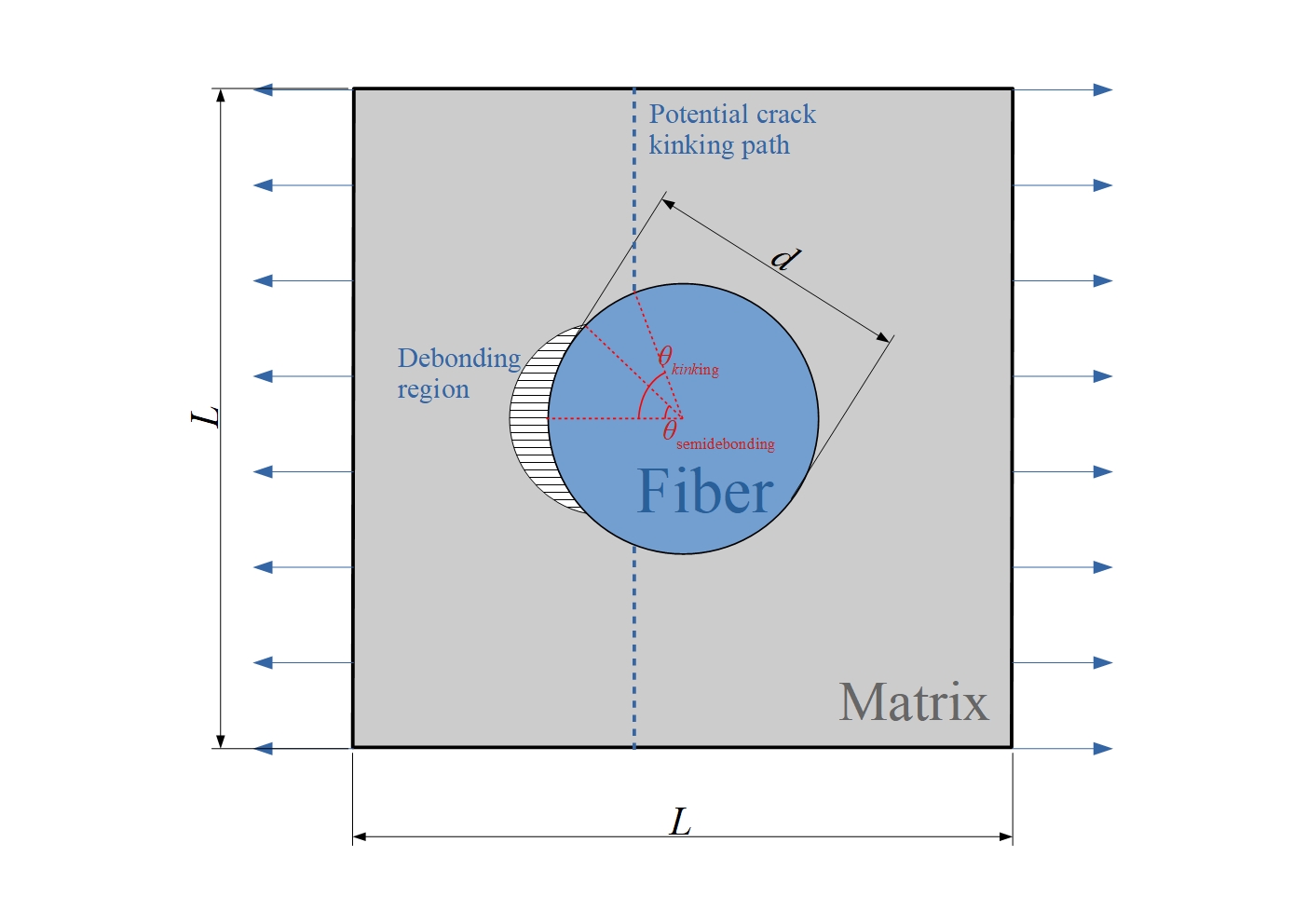}
        \caption{}
    \end{subfigure}
    \begin{subfigure}[b]{0.48\textwidth}
        \centering
        \includegraphics[width=\textwidth]{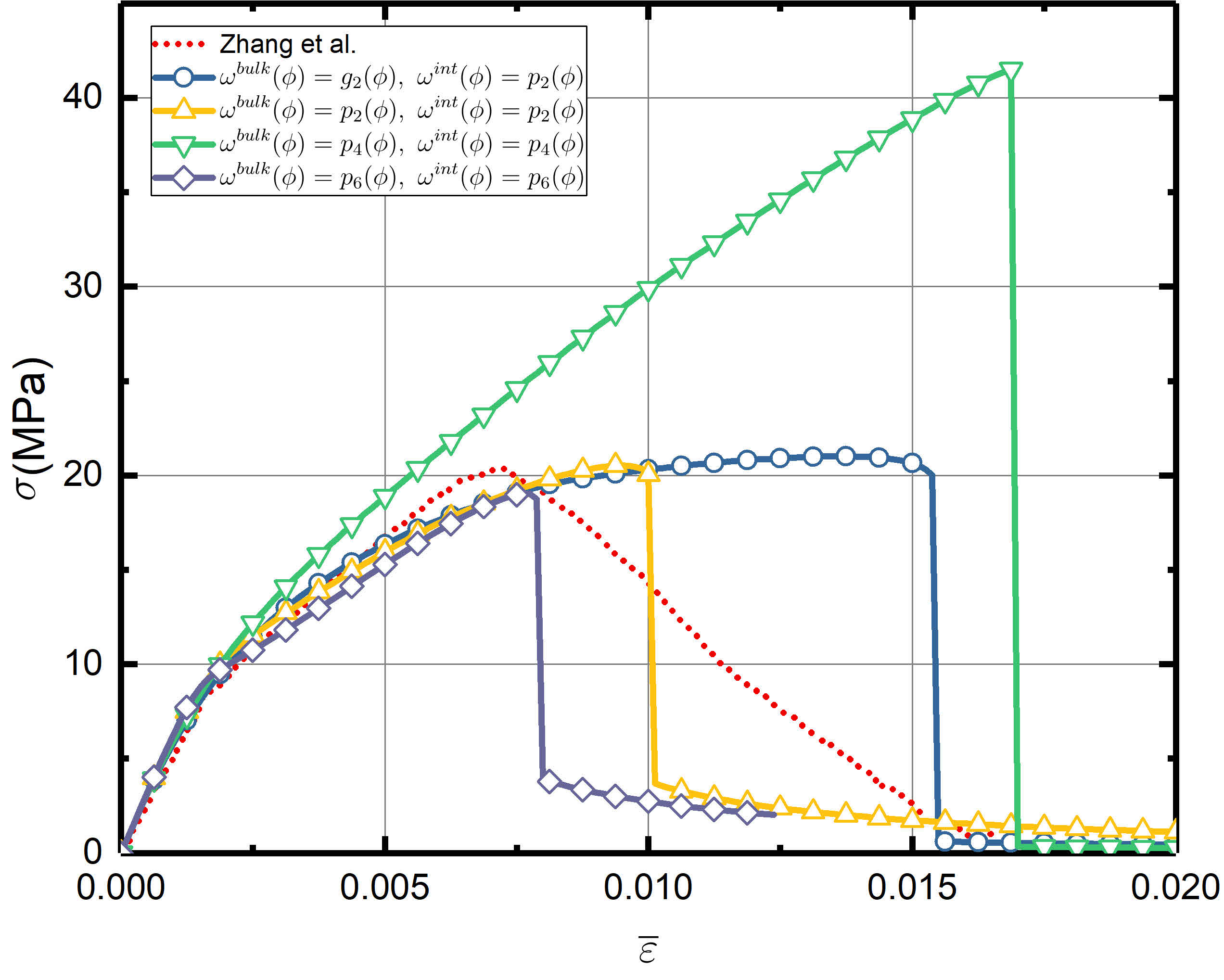}
        \caption{}
        \label{figSFSEcurve}
    \end{subfigure}
    \caption{Single fiber test: (a) detailed size of the RVE and (b) the displacement-traction curve of the models with different degradation function in bulk and interfacial material. }
\end{figure}

The averaged stress-strain curve of the composites with the identical material parameter and different degradation functions are illustrated in Fig. \ref{figSFSEcurve}. For the case that the $\omega^{\rm bulk}(\phi)=g_2(\phi)$, the model shows a much higher strength because the ultimate stress is determined with given $\mathcal{G}_c^{\rm bulk}$ and $\ell_0$. The other models have similar strength with the model of Zhang et al \cite{Zhang.2020}. In Zhang's work, the interface was modeled with a modified traditional cohesive element and the bulk matrix is modeled with a cohesive based phase-field method \cite{Wu.2018}. Nevertheless, obvious stress drop can be seen in the present model, which is caused by the crack kinking in the bulk materials. Moreover, the corresponding strain of the crack kinking increases with increasing $p$ of $\omega_p(\phi)$ in both bulk and interfacial regions. 

\begin{figure}[]
    \centering
    \begin{subfigure}[b]{0.24\textwidth}
        \centering
        \includegraphics[width=\textwidth]{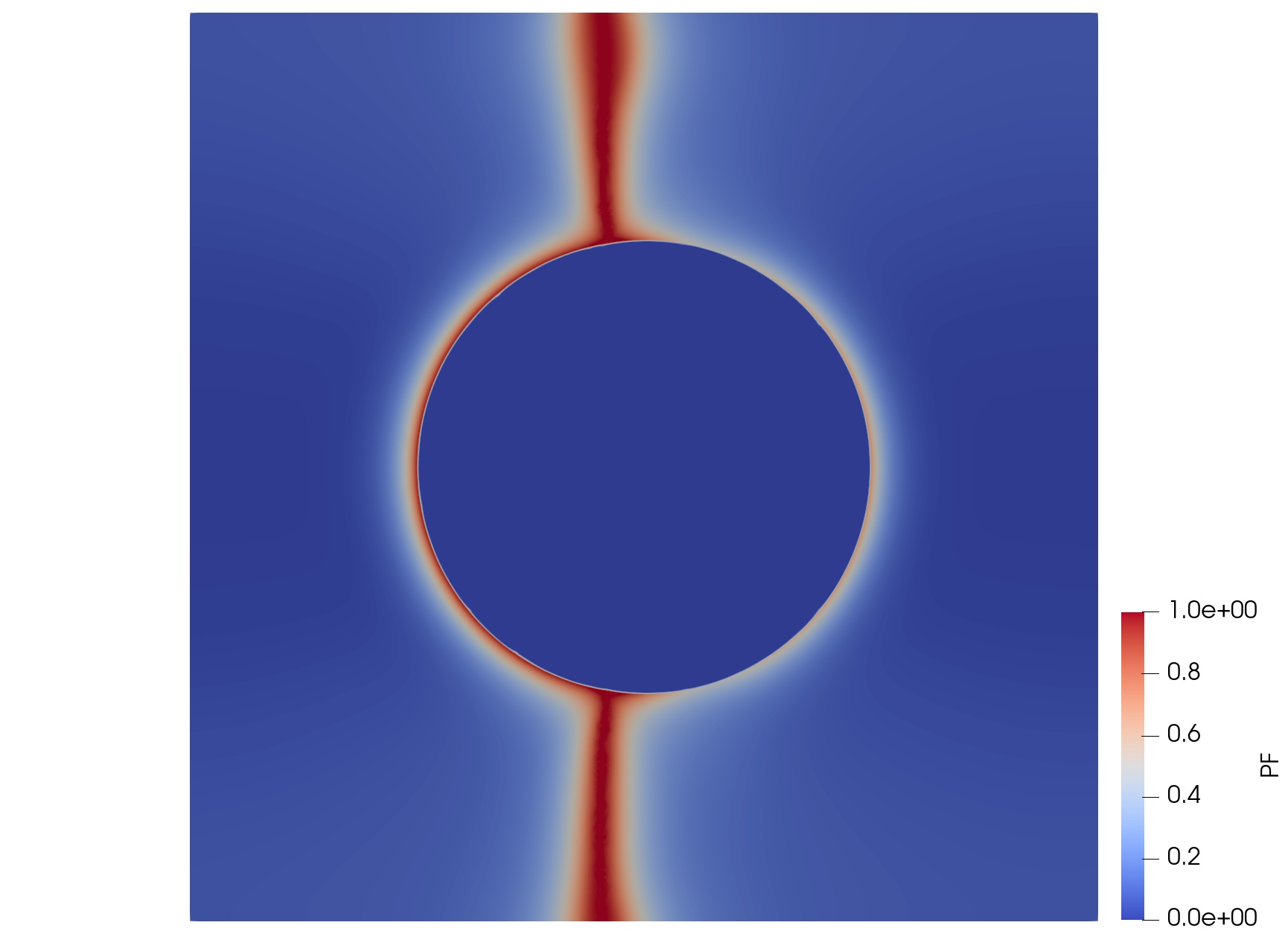}
        \caption{}
    \end{subfigure}
    \begin{subfigure}[b]{0.24\textwidth}
        \centering
        \includegraphics[width=\textwidth]{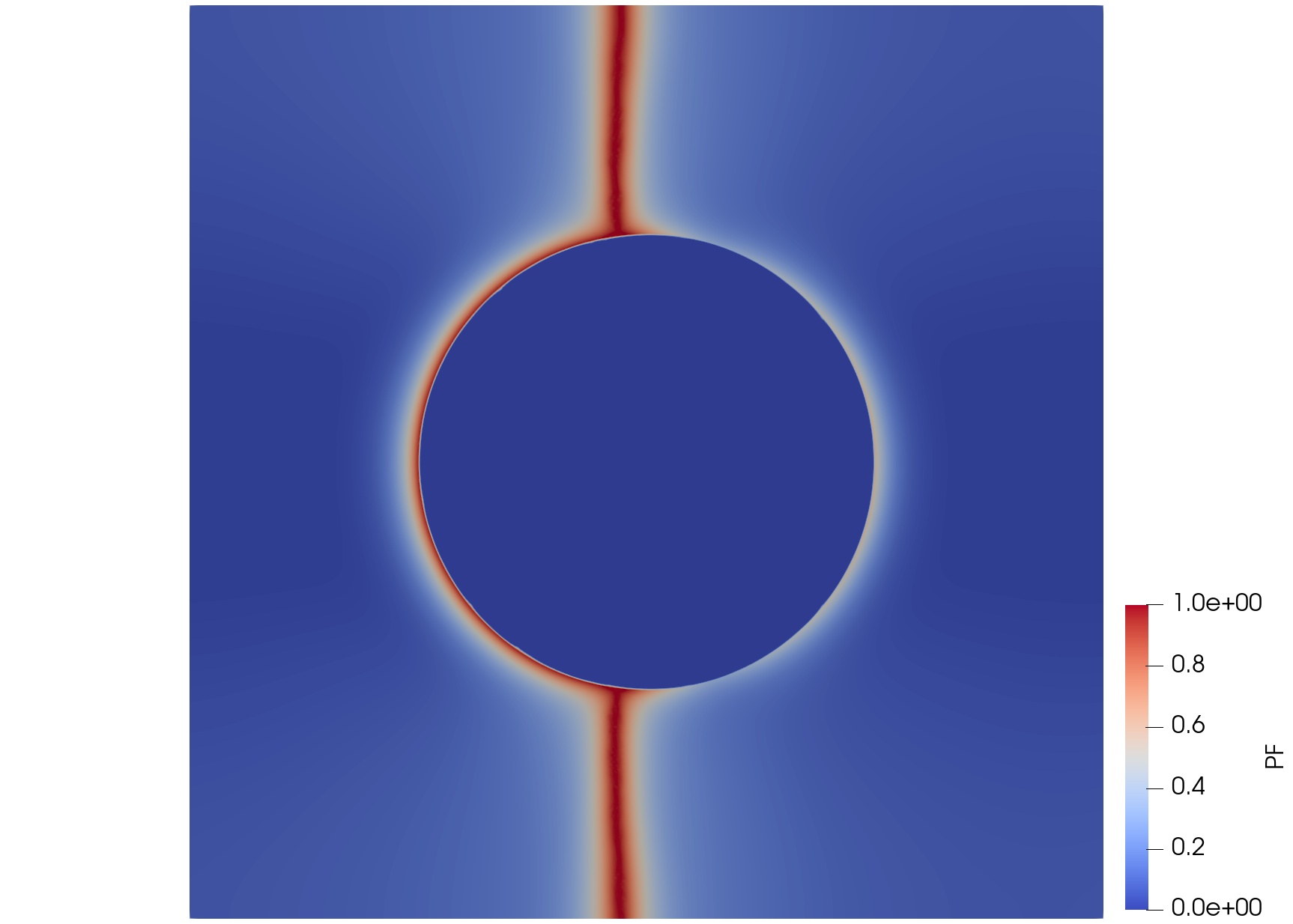}
        \caption{}
    \end{subfigure}
    \begin{subfigure}[b]{0.24\textwidth}
        \centering
        \includegraphics[width=\textwidth]{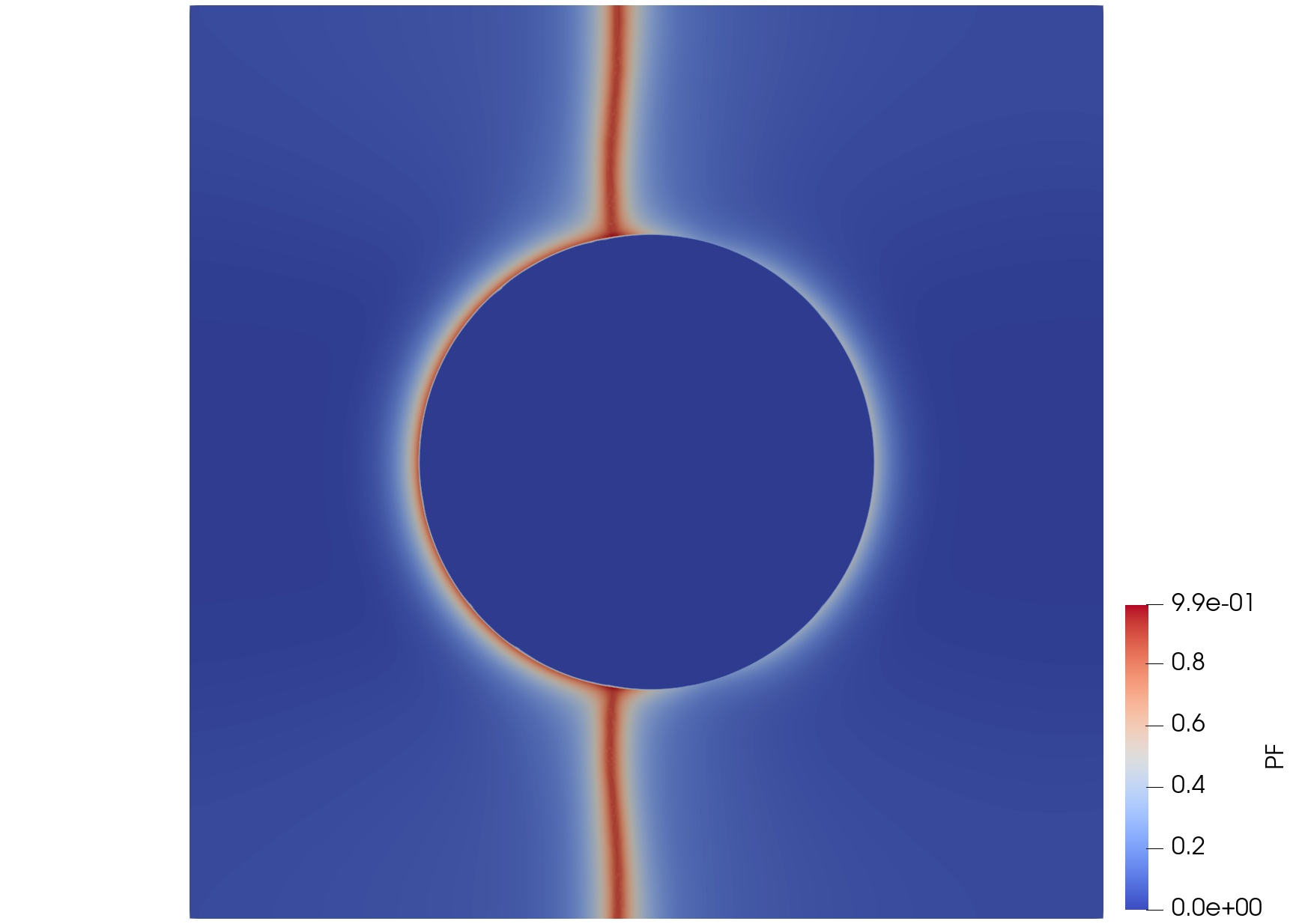}
        \caption{}
    \end{subfigure}
    \begin{subfigure}[b]{0.24\textwidth}
        \centering
        \includegraphics[width=\textwidth]{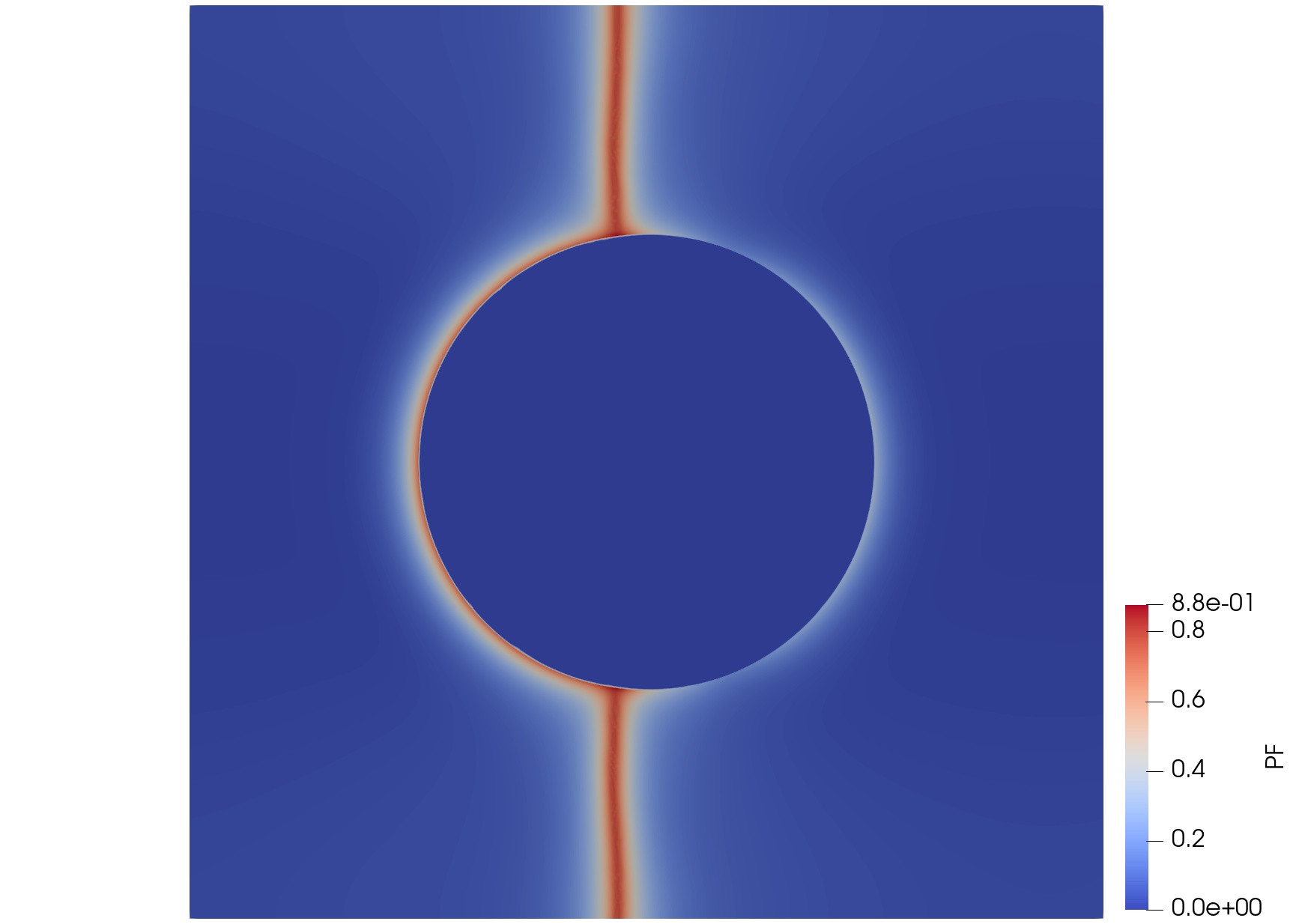}
        \caption{}
    \end{subfigure}
    \caption{Crack in the single fiber reinforced composites with different combination of degradation functions in the bulk and interfaces: (a) $\omega^{\rm bulk}(\phi)=g_2(\phi)$ and $\omega^{\rm int}(\phi)=\omega_2(\phi)$, (b) $\omega^{\rm bulk}(\phi)=\omega_2(\phi)$ and $\omega^{\rm int}(\phi)=\omega_2(\phi)$, (c) $\omega^{\rm bulk}(\phi)=\omega_4(\phi)$ and $\omega^{\rm int}(\phi)=\omega_4(\phi)$ and (d) $\omega^{\rm bulk}(\phi)=\omega_6(\phi)$ and $\omega^{\rm int}(\phi)=\omega_6(\phi)$.}
    \label{figCrackinSF}
\end{figure}

The cracks in different models are presented in Fig. \ref{figCrackinSF}. It can be seen that all of the kinking points are located at the same place. Meanwhile, the phase-field value of the crack after cracking decreases with increasing $p$ order, which agrees with the properties of degradation functions in Fig. \ref{figG2andWd}. For high order $\omega_p(\phi)$, there is a flat stage of $\omega_p(\phi)$ and its derivation $\partial\omega_p(\phi)/\partial\phi$ when $\phi$ approaches one, which can prevent the further increase of $\phi$ after brittle fracture occurring. As a result, the phase-field value can hardly reach one when high-order $\omega_p(\phi)$ is adopted. 

To figure out the relationship between the location of the crack kinking point and the properties of the bulk and interfacial material, a series of numerical experiments are carried out. 
To brief the discussion, just classical $g_2(\phi)$ is used as the degradation function in the bulk region. Meanwhile, the degradation function of interfacial region is fixed to $\omega_2({\phi})$. The elastic parameters of bulk and the interfacial region are the same values as in table \ref{tb:matSF}. The energy release rate of matrix in next discussion is also fixed to $\mathcal{G}_c^{\rm bulk} = 0.25\ \mathrm{N/mm}$.

\begin{figure}[]
    \centering
    \begin{subfigure}[b]{0.48\textwidth}
        \centering
        \includegraphics[width=\textwidth]{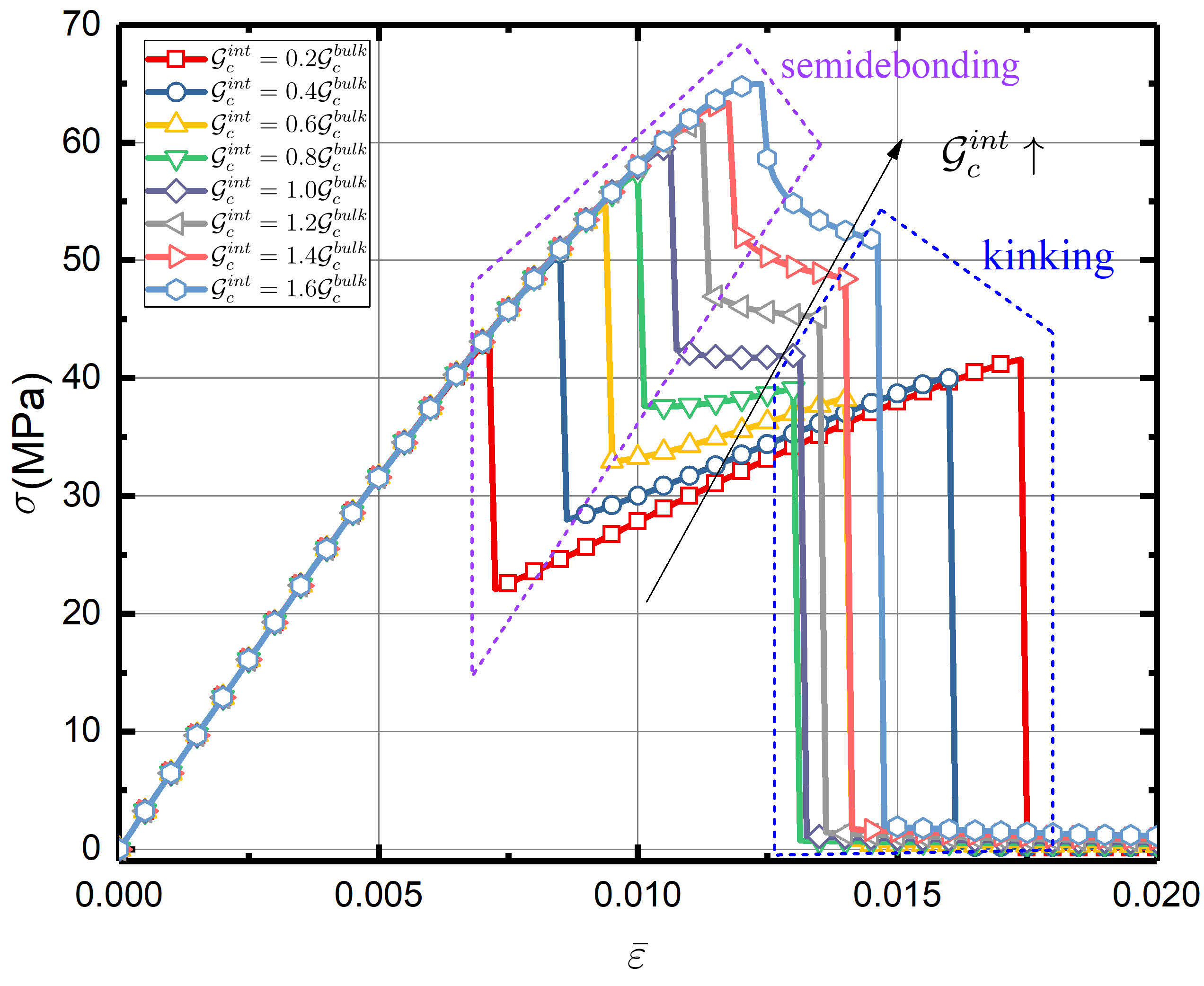}
        \caption{}
		\label{fig:SF_diffGc}
    \end{subfigure}
    \begin{subfigure}[b]{0.48\textwidth}
        \centering
        \includegraphics[width=\textwidth]{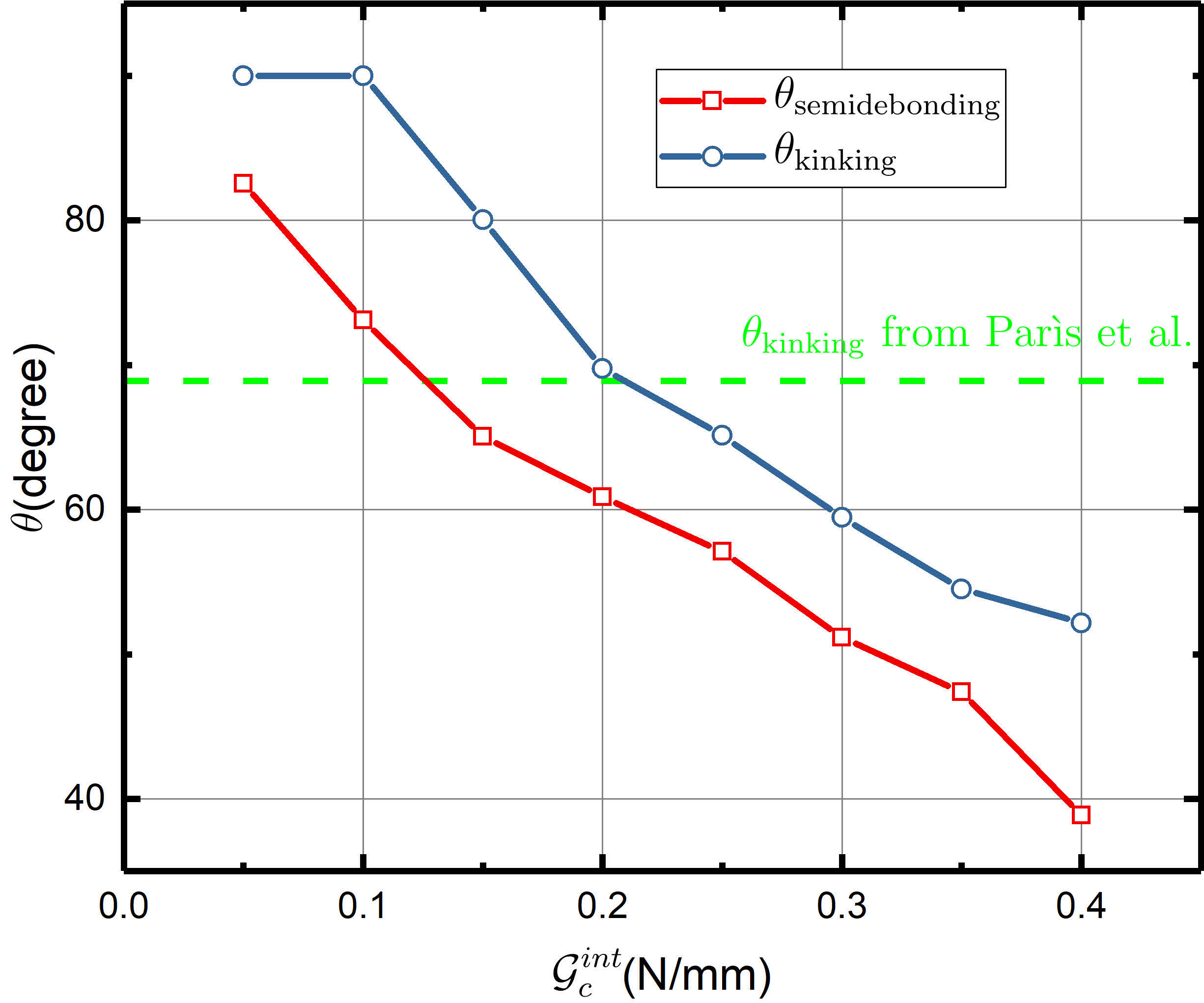}
        \caption{}
	\label{fig:SF_theta_diffGc}
    \end{subfigure}
    \caption{Tension test on the single-fiber reinforced composites with different $\mathcal{G}_c^{\rm int}$ and the identical $\sigma_{max}=80\ \mathrm{MPa}$: (a) stress-strain curves and (b) semidebonding and kinking angles.}
\end{figure}

Firstly, the relationship between $\mathcal{G}_c^{\rm int}$ and semi-debonding angle $\theta_{\mathrm{semideboning}}$ or kinking angle $\theta_{\mathrm{kinking}}$ is investigated. For convenience, the interfacial strength $\sigma_t=\sigma_n=\sigma_{\rm max}$ is fixed to be $80\ \mathrm{MPa}$. The average stress-strain curves of RVE with different $\mathcal{G}_c^{\rm int}$ are illustrated in Fig.\ref{fig:SF_diffGc}. The ultimate stress of RVE increases with increasing $\mathcal{G}_c^{\rm int}$ monotonically. Meanwhile, all these curves have a similar tendency, which shows that there are two obvious stress-drop points in each curve. The first stress-drop point is caused by the semi-debonding at the fiber-matrix interface. The second stress-drop point is caused by the crack kinking in the matrix. In addition,  when $\mathcal{G}_c^{\rm int}\le0.4\mathcal{G}_c^{\rm bulk}$, the ultimate strain corresponding to the crack kinking can be larger than those in the RVE with larger value $\mathcal{G}_c^{\rm int}$ at interfaces. 

The relationship between $\mathcal{G}_c^{\rm int}$ and $\theta_{\mathrm{semideboning}}$ or $\theta_{\mathrm{kinking}}$ is illustrated in Fig. \ref{fig:SF_theta_diffGc}.  Both of the $\theta_{\mathrm{semideboning}}$ and $\theta_{\mathrm{kinking}}$ decrease with increasing $\mathcal{G}_c^{\rm int}$. Nevertheless, the difference between $\theta_{\mathrm{semideboning}}$ and $\theta_{\mathrm{kinking}}$ are not sensitive to the $\mathcal{G}_c^{\rm int}$, which means that the interfacial crack propagates a similar path before penetrating into matrix. It also can be seen that the $\theta_{\mathrm{kinking}}=90^{\circ}$ when $\mathcal{G}_c^{\rm int}\le0.4\mathcal{G}_c^{\rm bulk}$. This phenomenon suggest that for a low value of $\mathcal{G}_c^{\rm int}$, the interfacial crack will not penetrate into the bulk region until stress in the bulk region reach its criterion. The criteria of crack kinking can be expressed as the competition between bulk and interfacial cracking \cite{He.1991}:
\begin{equation}
	\frac{\mathcal{G}^{\rm int}}{\mathcal{G}_c^{\rm int}}<\frac{\mathcal{G}^{\rm bulk}}{\mathcal{G}_c^{\rm bulk}}
\end{equation}
The $\mathcal{G}^{\rm int}$ and $\mathcal{G}^{\rm bulk}$ are the energy release rate of interface and bulk material at the crack tip. For increasing value of $\mathcal{G}_c^{\rm int}$ and fixed $\mathcal{G}_c^{\rm bulk}$, the value of $\mathcal{G}^{\rm int}/\mathcal{G}_c^{\rm int}$ will decrease, which suggests that crack kinking can happen at a earlier stage. The above-mentioned mechanism can explain the phenomenon that the increasing $\mathcal{G}_c^{\rm int}$ leads to decreasing $\theta_{\mathrm{kinking}}$. 

An analytical solution is also used to compare with the present result. According to Par\'{i}s et al., $\theta_{\mathrm{kinking}}$ evaluated with a maximum circumferential stress (MCS) criterion can be expressed as \cite{Paris.2007}:
\begin{equation}
	\theta_{\mathrm{kinking}} = -2 \mathrm{sgn}(\beta)\arccos\sqrt{\frac{2+|\beta|}{3+|\beta|}},\ \text{for} \ \beta = \frac{\mu_1(\kappa_2-1)-\mu_2(\kappa_1-1)}{\mu_1(\kappa_2+1)+\mu_2(\kappa_1+1)}
\end{equation}
where $\mu_k = E_k/2(1+\upsilon_k)$ is the shear modulus and $\kappa_k = 3-4v_k$ is the Kolosov's constant. For the present material parameters, $\theta_{\mathrm{kinking}}$ can be obtained as $68.88^{\circ}$. It can be seen in Fig. \ref{fig:SF_theta_diffGc} that the closet $\mathcal{G}_c^{\rm int}$ is $0.2\ \mathrm{N/mm}$, which is near the value of $\mathcal{G}_c^{\rm bulk}$.

\begin{figure}[]
    \centering
    \begin{subfigure}[b]{0.24\textwidth}
        \centering
        \includegraphics[width=\textwidth]{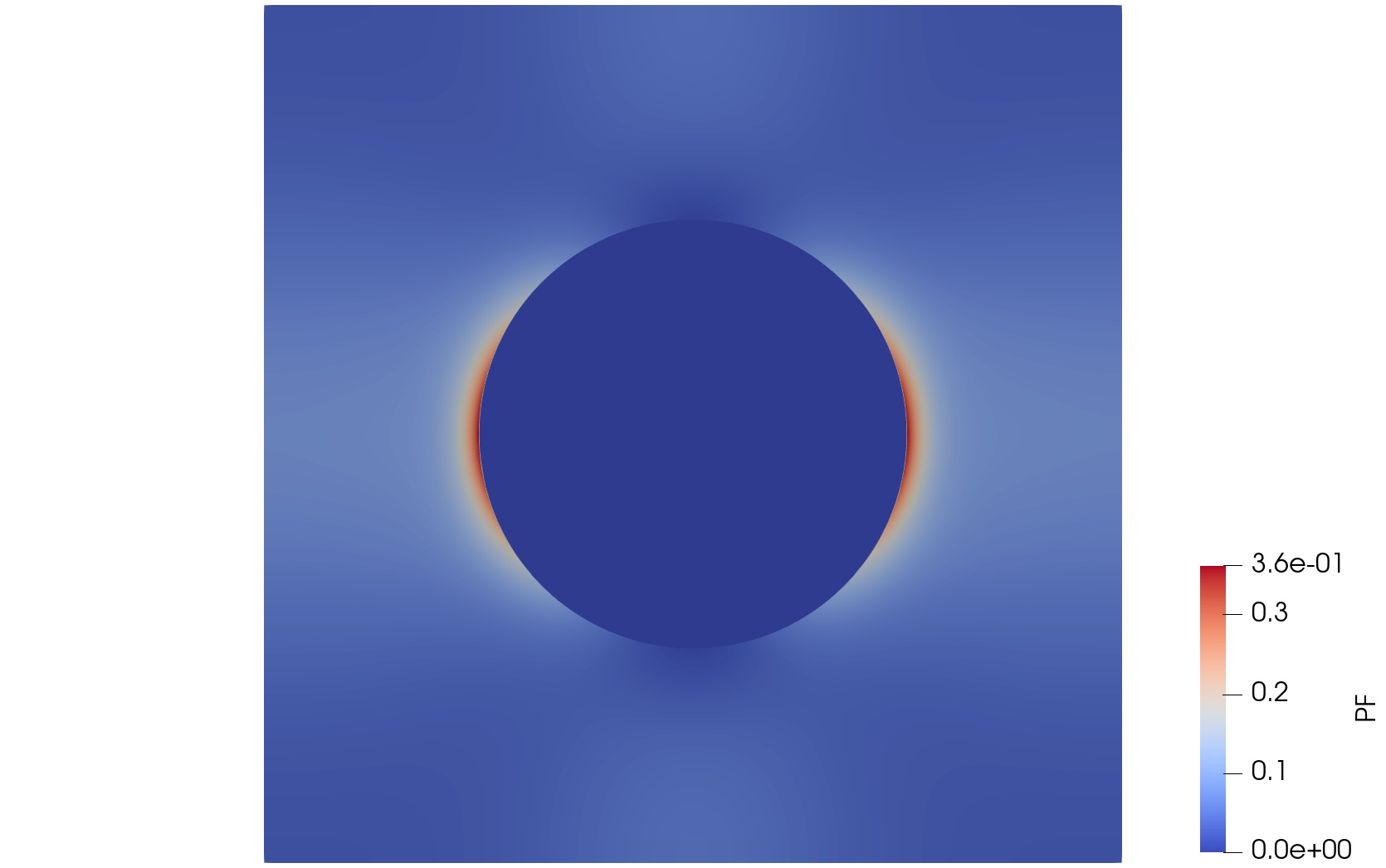}
        \caption{}
    \end{subfigure}
    \begin{subfigure}[b]{0.24\textwidth}
        \centering
        \includegraphics[width=\textwidth]{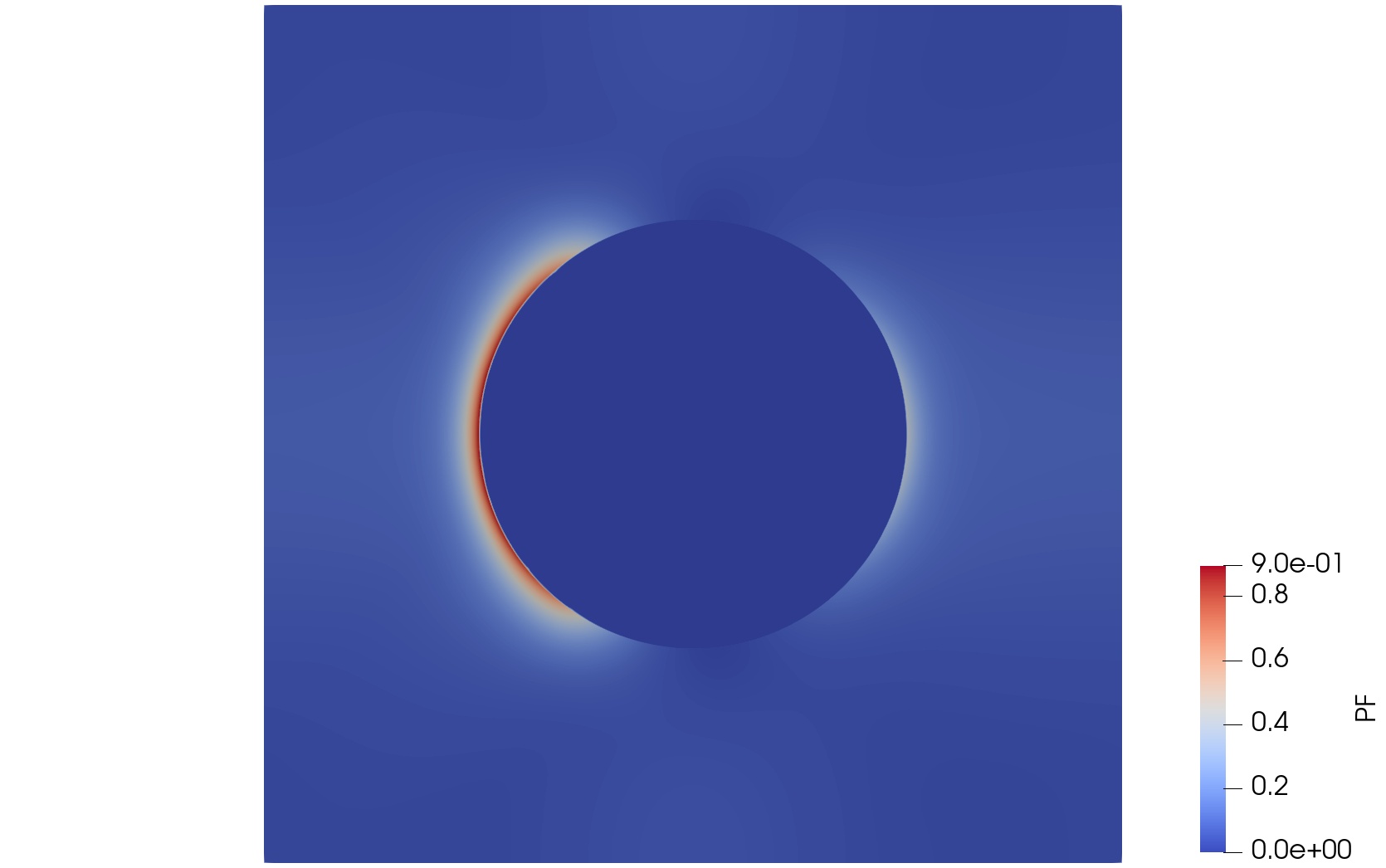}
        \caption{}
    \end{subfigure}
    \begin{subfigure}[b]{0.24\textwidth}
        \centering
        \includegraphics[width=\textwidth]{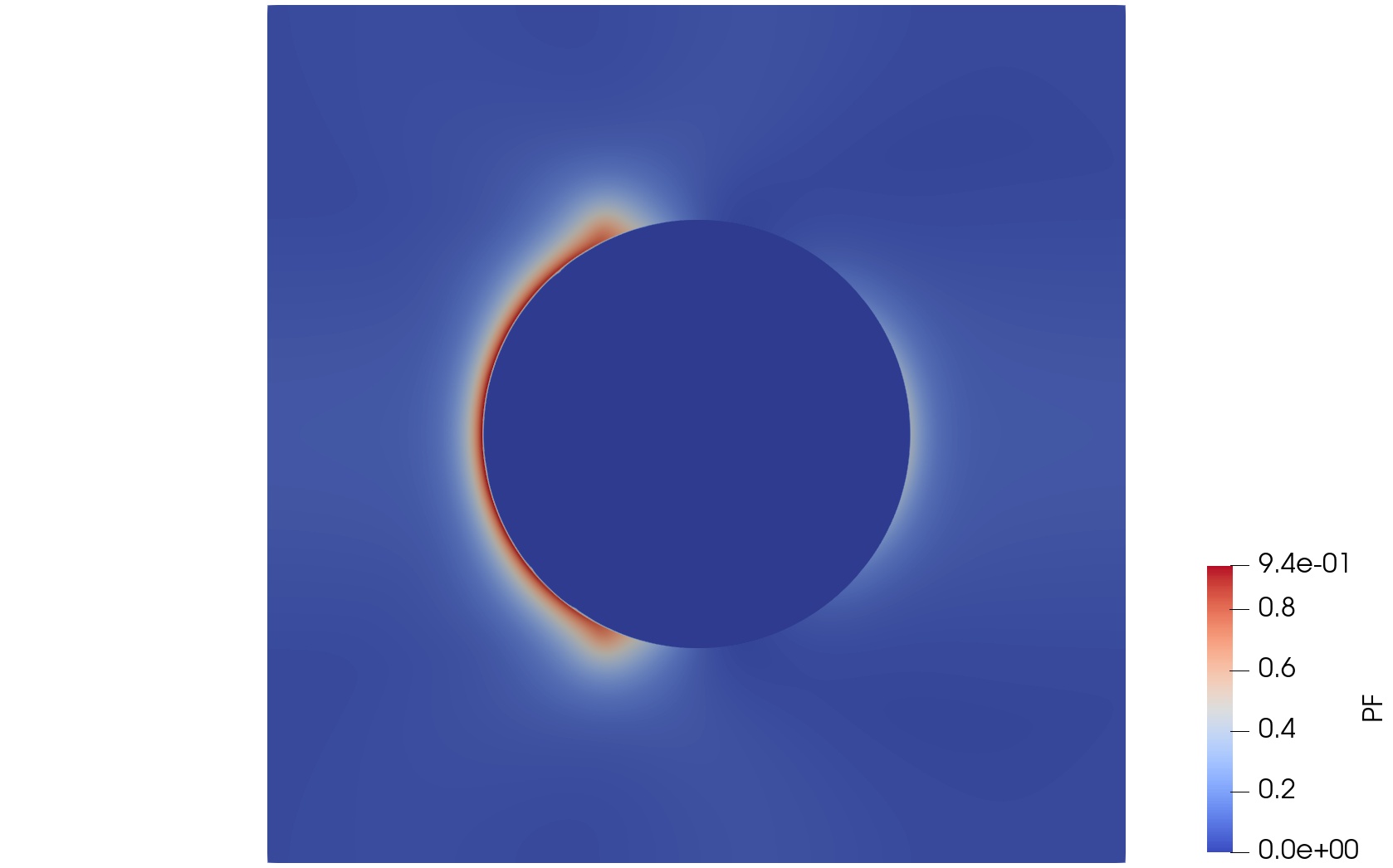}
        \caption{}
    \end{subfigure}
    \begin{subfigure}[b]{0.24\textwidth}
        \centering
        \includegraphics[width=\textwidth]{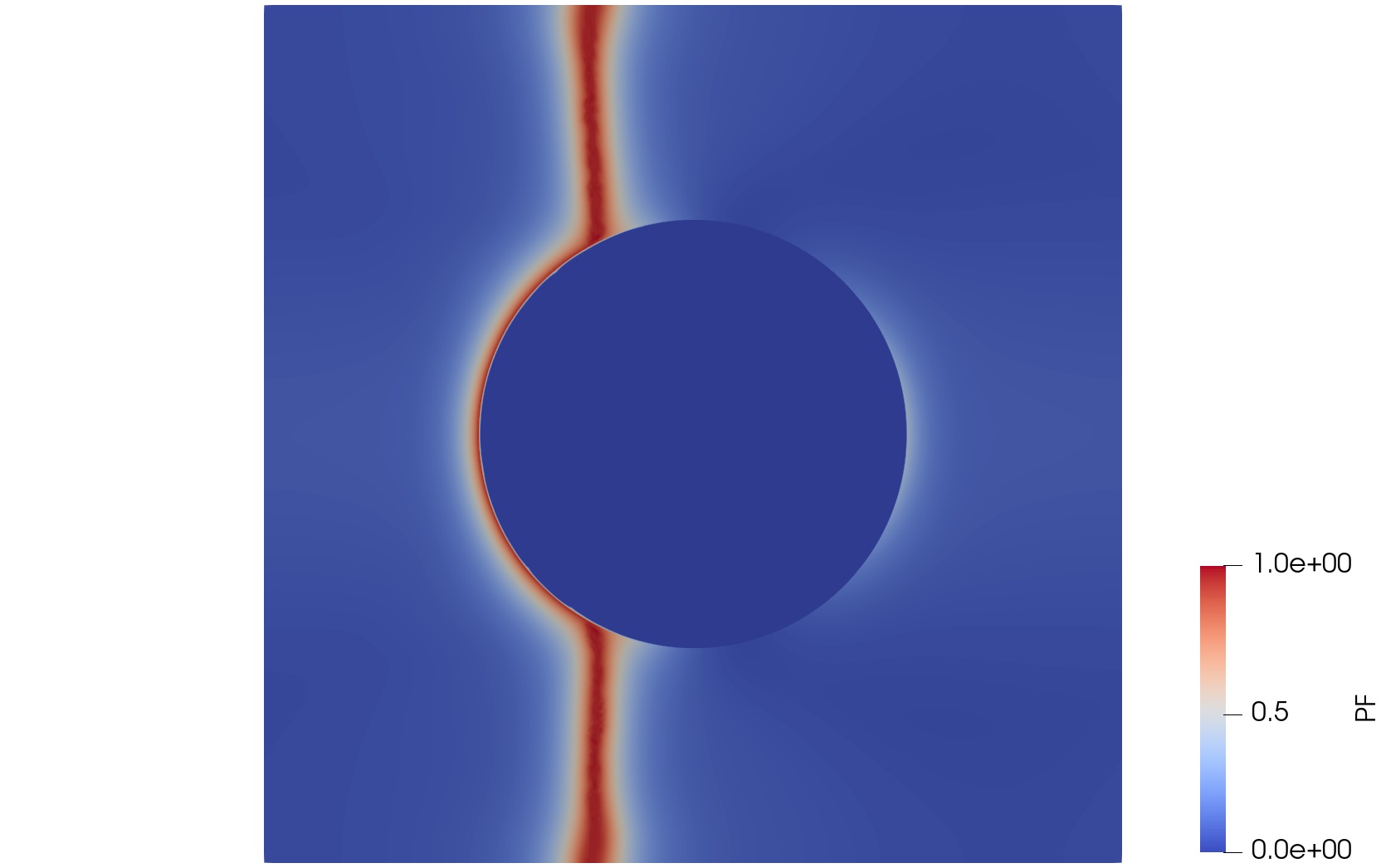}
        \caption{}
    \end{subfigure}
    \caption{The phase-field in the RVE with $\mathcal{G}_c^{\rm int}=\mathcal{G}_c^{\rm bulk}$: (a) the frame before semidebonding occurring at the interface; (b) semidebonding appearing; (c)  before crack kinking and (d) after crack kinking.}
	\label{fig:SF_4stage}
\end{figure}

\begin{figure}
	\centering
    \begin{subfigure}[b]{0.24\textwidth}
        \centering
        \includegraphics[width=\textwidth]{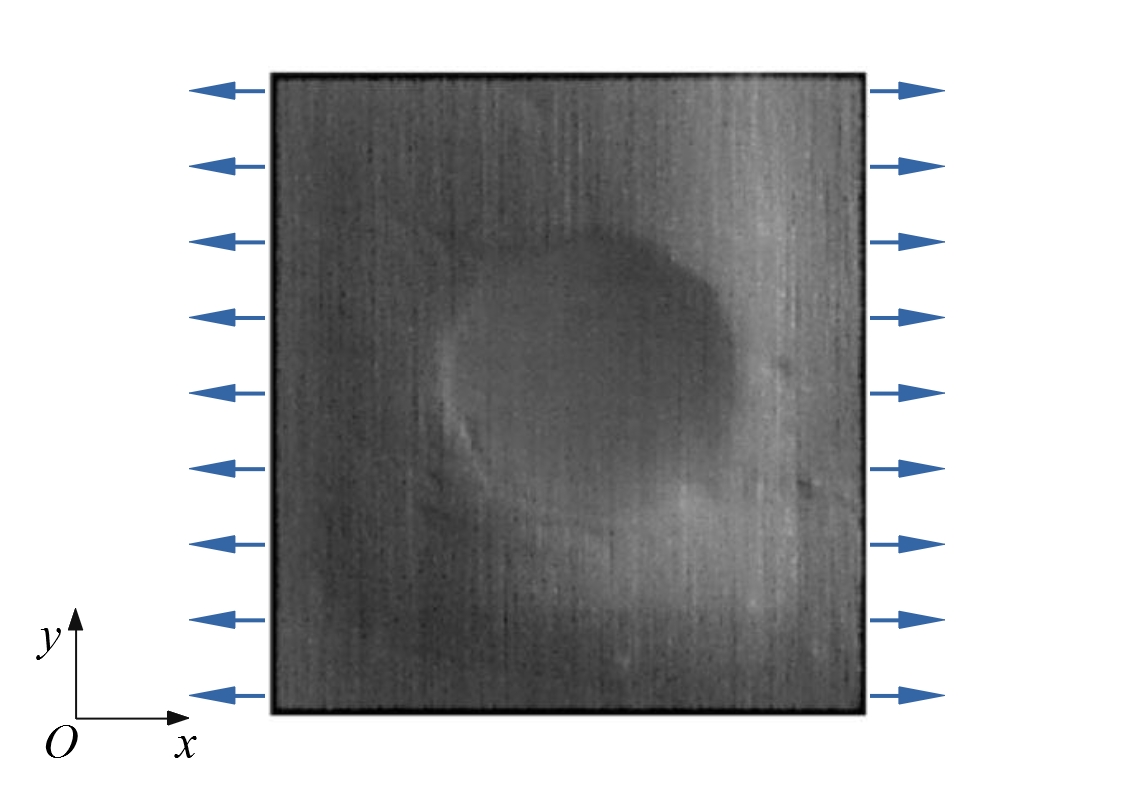}
        \caption{}
    \end{subfigure}
    \begin{subfigure}[b]{0.24\textwidth}
        \centering
        \includegraphics[width=\textwidth]{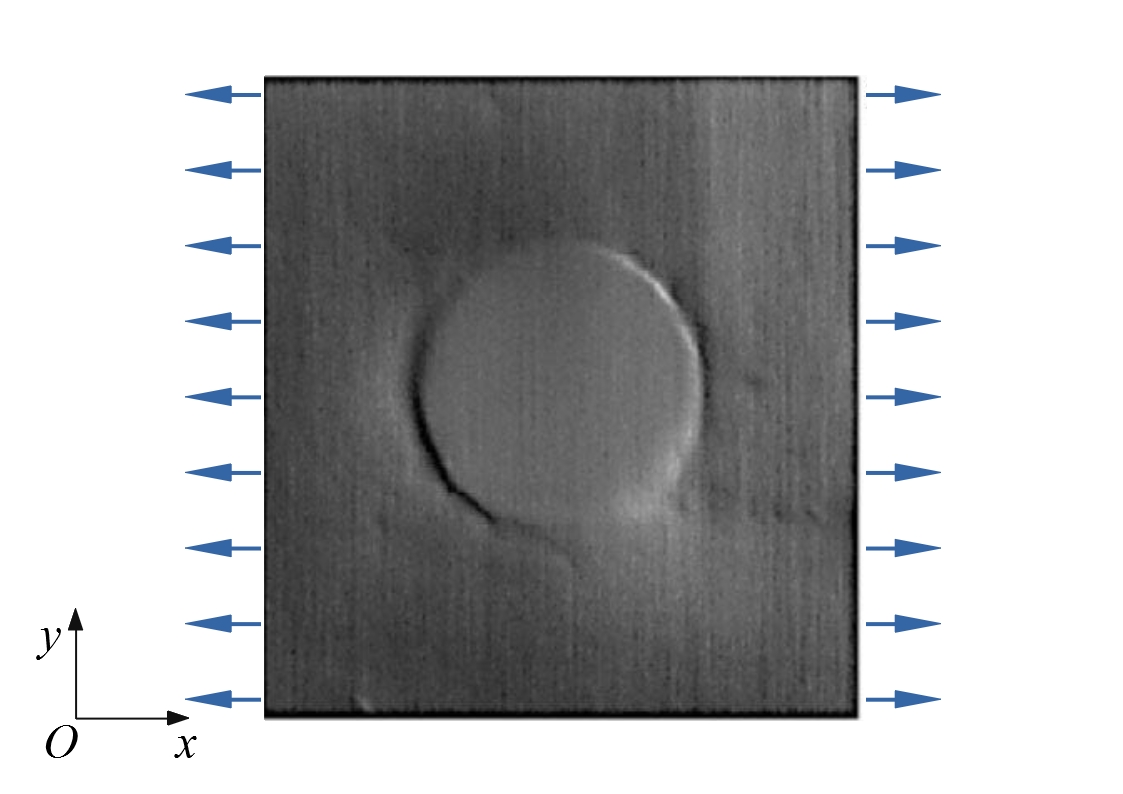}
        \caption{}
    \end{subfigure}
    \begin{subfigure}[b]{0.24\textwidth}
        \centering
        \includegraphics[width=\textwidth]{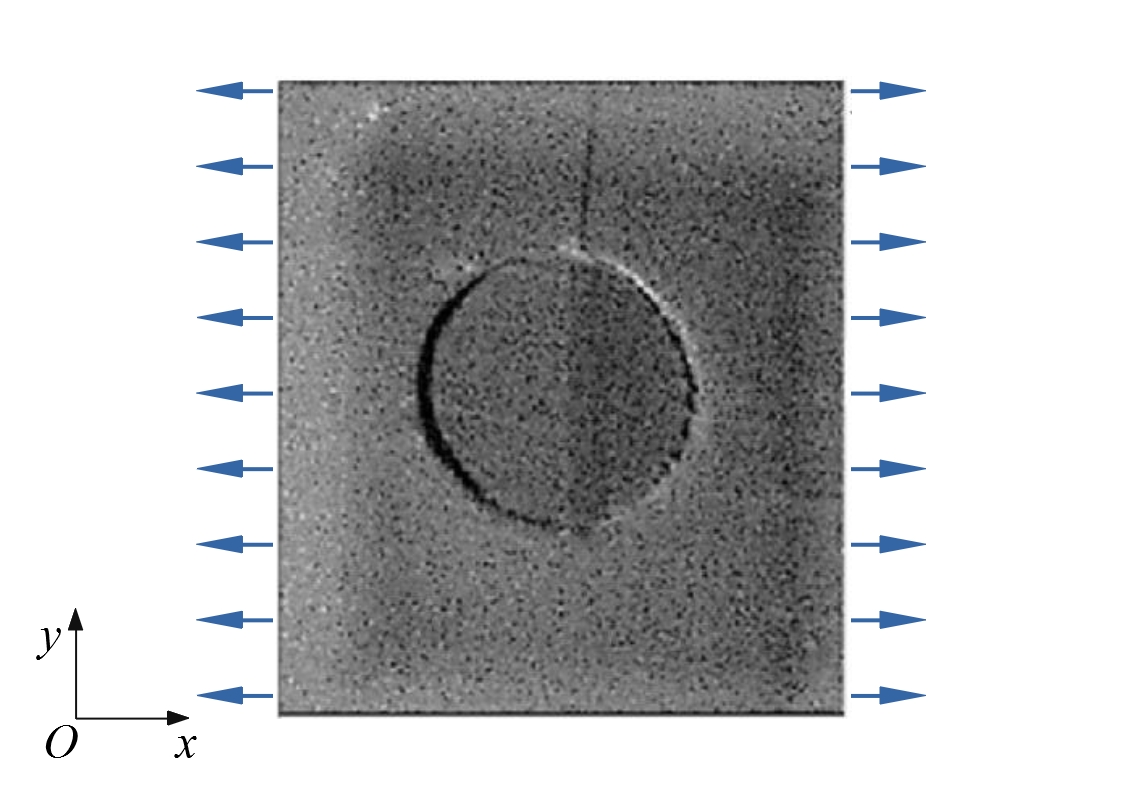}
        \caption{}
    \end{subfigure}
    \begin{subfigure}[b]{0.24\textwidth}
        \centering
        \includegraphics[width=\textwidth]{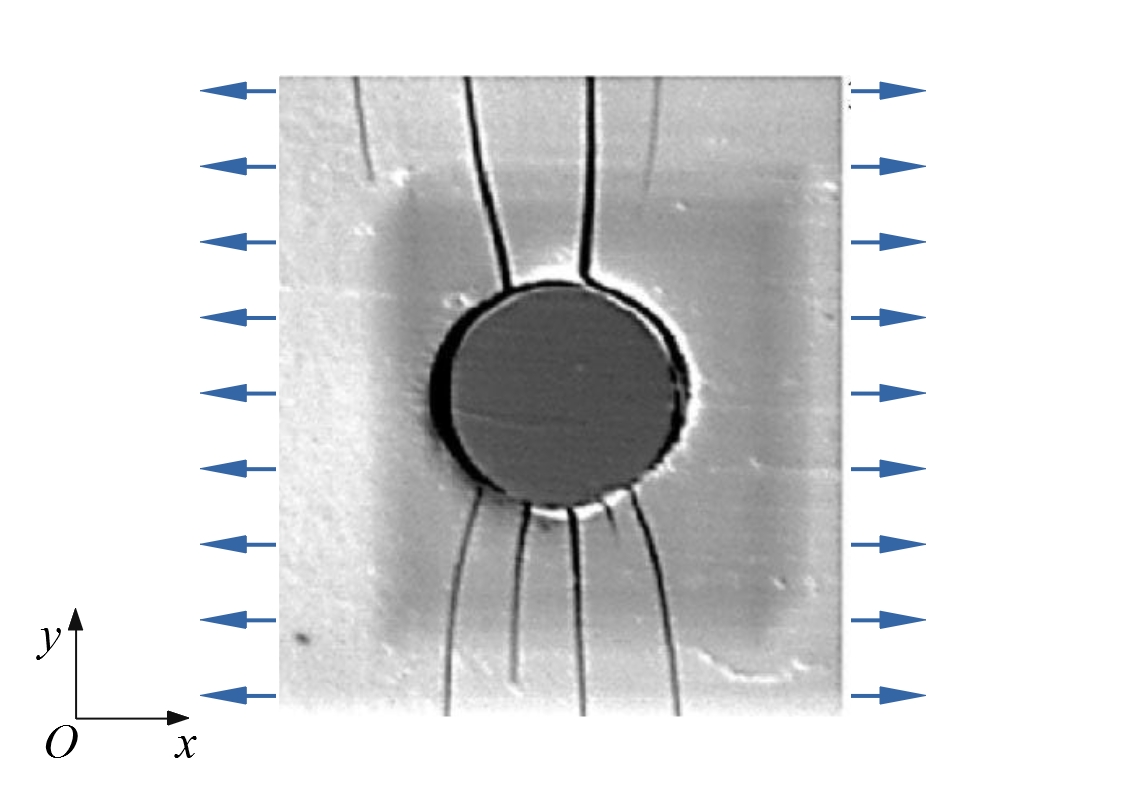}
        \caption{}
    \end{subfigure}
	\caption{SEM micrographs of different specimen of single fiber reinforced composites under transverse tension \cite{Totten.2016}: (a) before test; (b) initiation of interfacial crack; (c) semidebonding and (d) crack kinking.
	\label{fig:exp}
}

\end{figure}

The distribution of the phase-field in RVE with a typical material parameter ($\mathcal{G}_c^{\rm int}=\mathcal{G}_c^{\rm bulk}$) at different strain levels is illustrated in Fig. \ref{fig:SF_4stage}.  At first, the phase-field value in RVE increases slightly with the increase of load. Then semi-debonding occurs with an apparent drop in average stress. Next, the crack propagates along with the interface, and crack kinking happens at last. Scanning electron microscope (SEM) micrographs from Totten et al. \cite{Totten.2016} are illustrated in Fig.\ref{fig:exp}, in which a typical failure pattern is presented. It can be seen that result from the present CZM agrees with the realistic experimental result.

\begin{figure}[]
    \centering
    \begin{subfigure}[b]{0.48\textwidth}
        \centering
        \includegraphics[width=\textwidth]{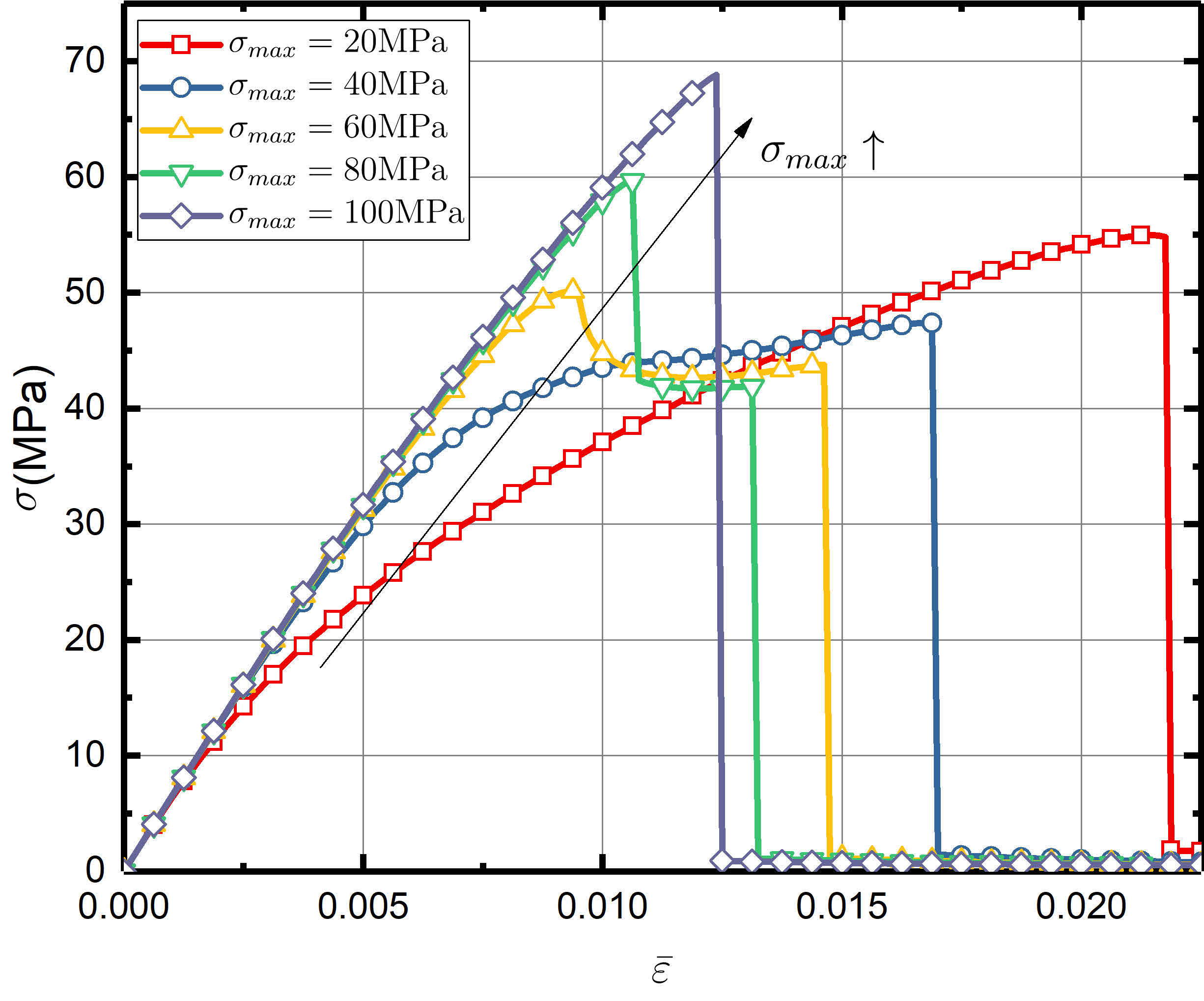}
        \caption{}
	\label{fig:SF_diffsmax}
    \end{subfigure}
    \begin{subfigure}[b]{0.48\textwidth}
        \centering
        \includegraphics[width=\textwidth]{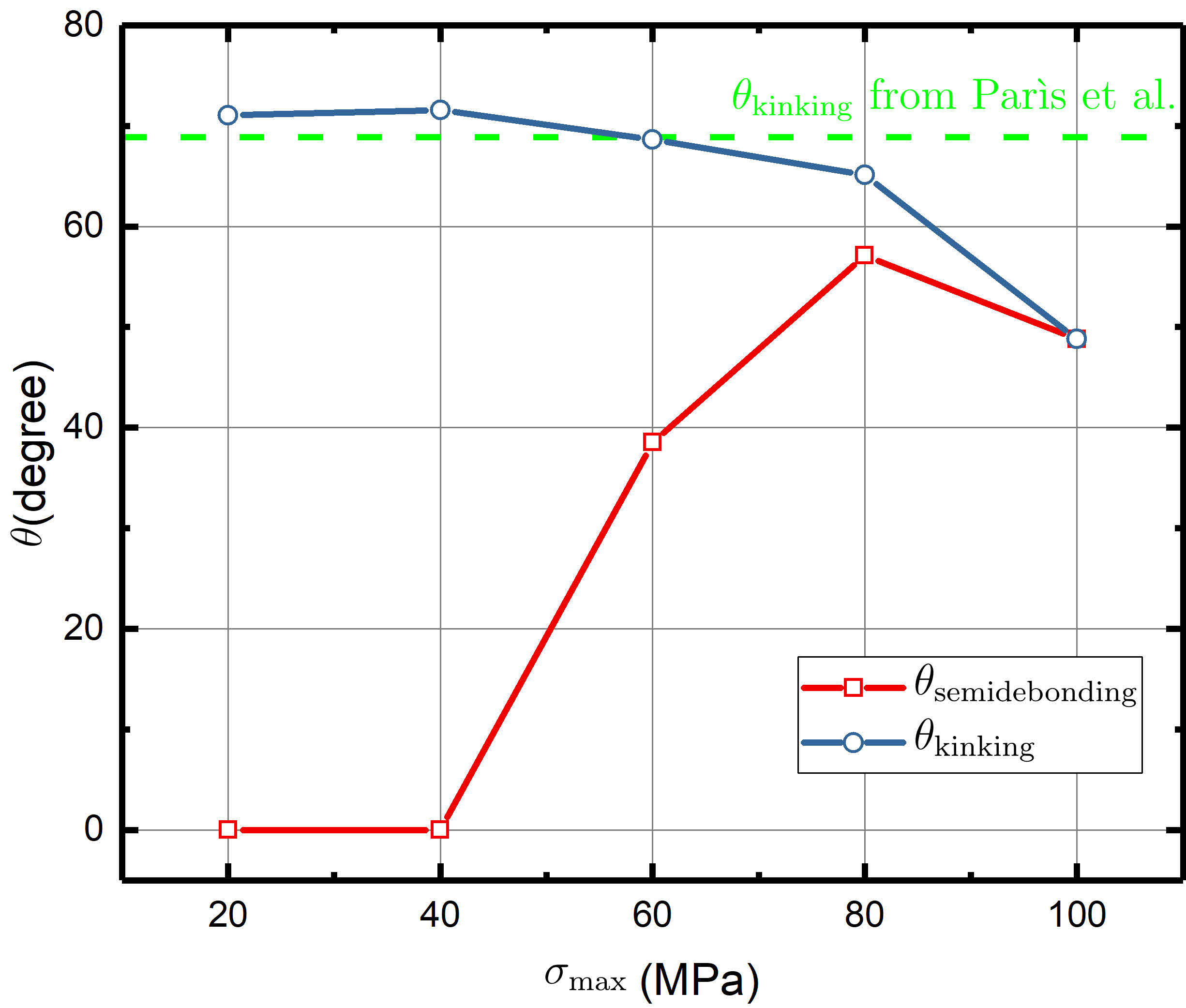}
        \caption{}
	\label{fig:SF_theta_diffsmax}
    \end{subfigure}
    \caption{Tension test on the single-fiber reinforced composites with different $\sigma_{\rm max}$: (a) stress-strain curves and (b) semidebonding and kinking angles.}
\end{figure}

On the other hand, the relationship between the interfacial strength $\sigma_t=\sigma_n=\sigma_{\rm max}$ and  $\theta_{\mathrm{semideboning}}$ or $\theta_{\mathrm{kinking}}$ is also investigated. For the convince, $\mathcal{G}_c^{\rm int}=\mathcal{G}_c^{\rm bulk}$ is fixed. The stress-strain curves of the RVE with different $\sigma_{\rm max}$ are illustrated in Fig. \ref{fig:SF_diffsmax}. For a low value of  $\sigma_{\rm max}$, i.e. $\sigma_{\rm max}\le40\ \mathrm{MPa}$, the semidebonding stage vanishes and the phase-field increases to one gently through the whole process of deformation. The semidebonding appears again when $60\ \mathrm{MPa}\le\sigma_{\rm max}\le80\ \mathrm{MPa}$. Nevertheless, the semidebonding stage disappears again when $\sigma_{\rm max}$ reaches $100\ \mathrm{MPa}$. The debonding and kinking happen at the same time point and only one stress-drop point can be seen at the stress-strain curve.

The relationship between the $\sigma_{\rm max}$ and $\theta_{\mathrm{semideboning}}$ or $\theta_{\mathrm{kinking}}$ is illustrated in Fig. \ref{fig:SF_theta_diffGc}. Compared to the $\theta_{\mathrm{kinking}}$, the $\theta_{\mathrm{semideboning}}$ is more sensitive to the $\sigma_{\rm max}$. Meanwhile, $\theta_{\mathrm{kinking}}$ is close to the MCS solution when $\sigma_{max}<80\ \mathrm{MPa}$. The difference between $\theta_{\mathrm{semideboning}}$ and $\theta_{\mathrm{kinking}}$ decreases with increasing $\sigma_{\rm max}$, which indicates that the interfacial crack becomes shorter and more brittle with increasing interfacial strength. 

\begin{figure}[]
    \centering
    \begin{subfigure}[b]{0.24\textwidth}
        \centering
        \includegraphics[width=\textwidth]{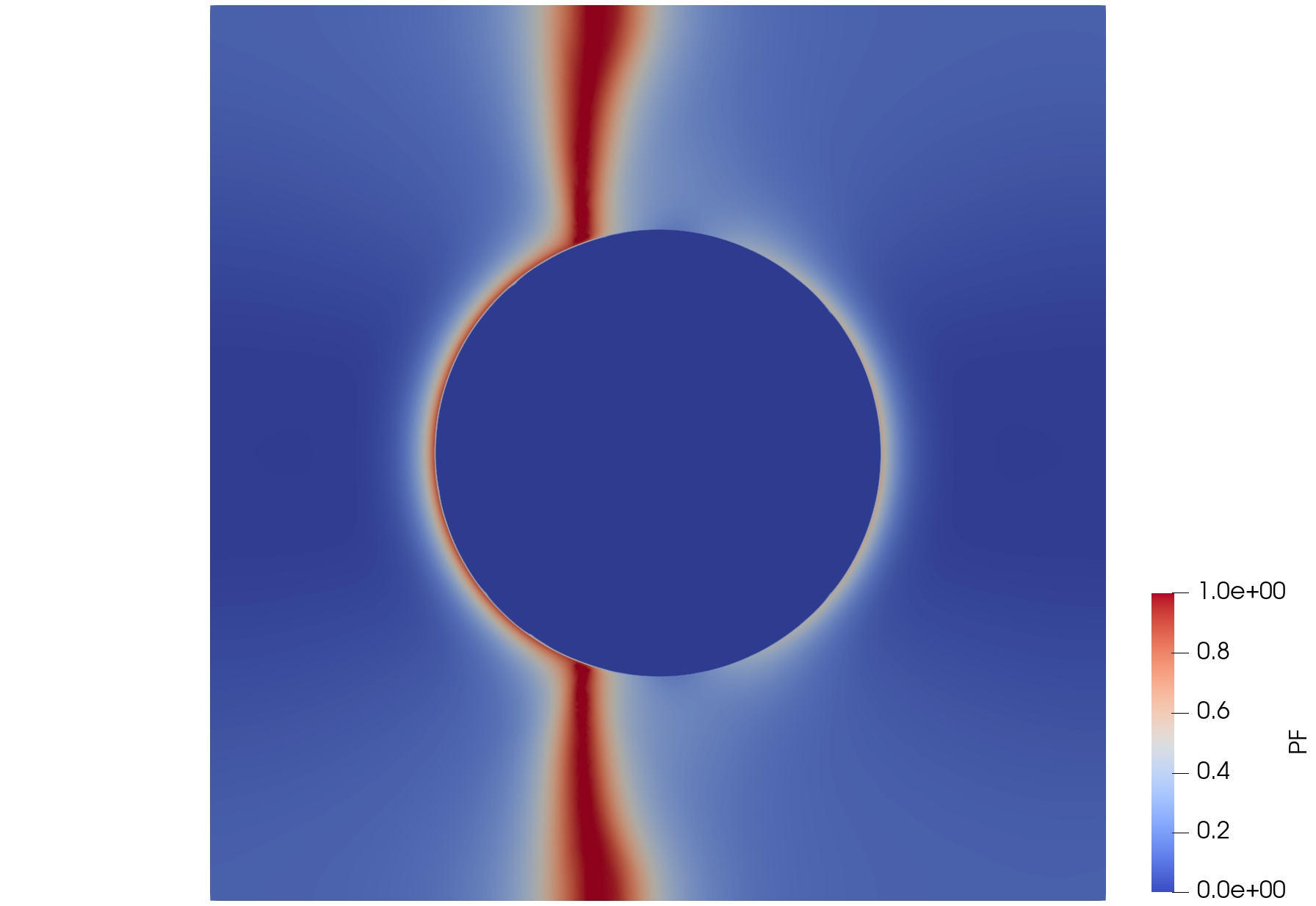}
        \caption{}
		\label{}
    \end{subfigure}
    \begin{subfigure}[b]{0.24\textwidth}
        \centering
        \includegraphics[width=\textwidth]{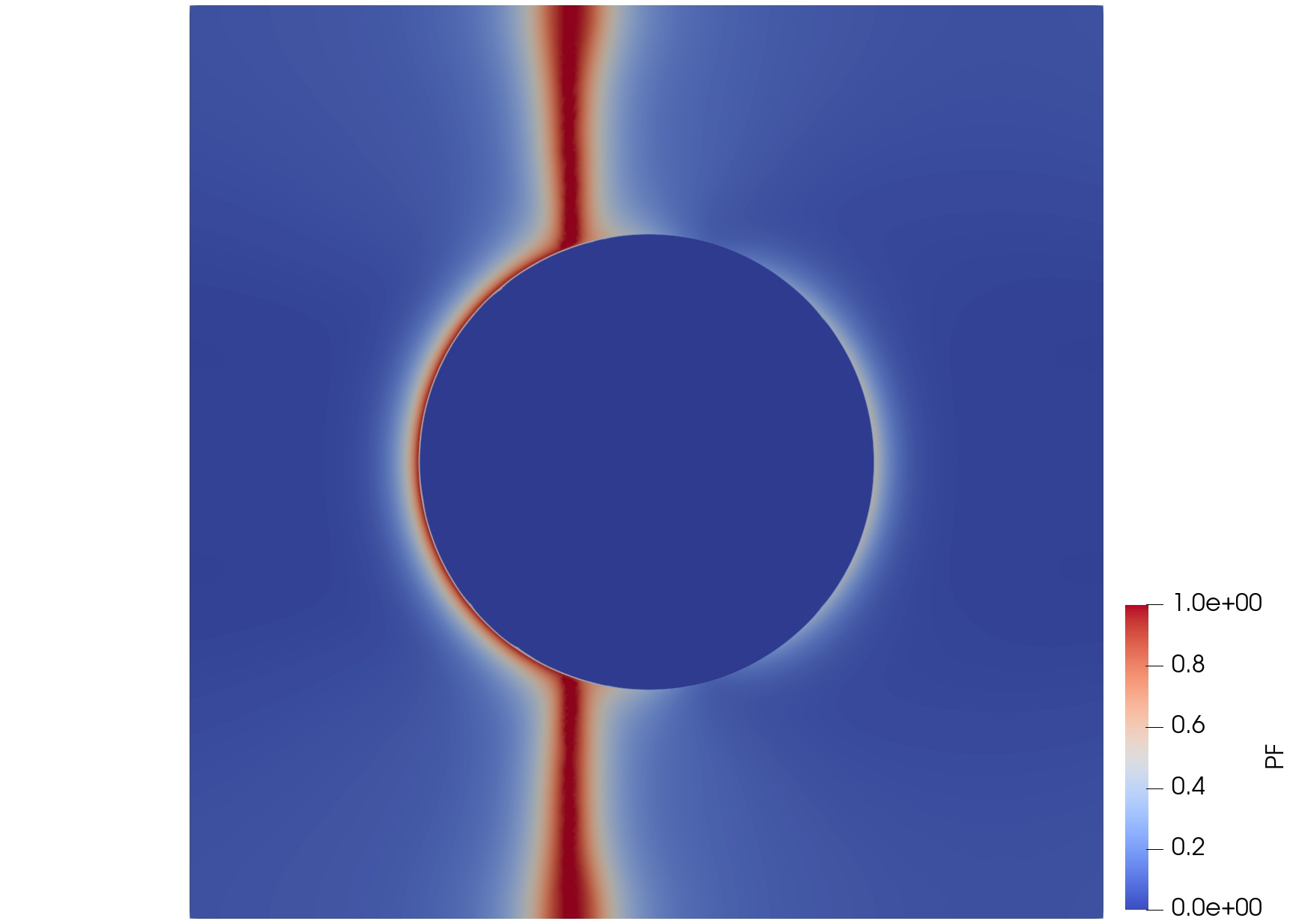}
        \caption{}
		\label{}
    \end{subfigure}
    \begin{subfigure}[b]{0.24\textwidth}
        \centering
        \includegraphics[width=\textwidth]{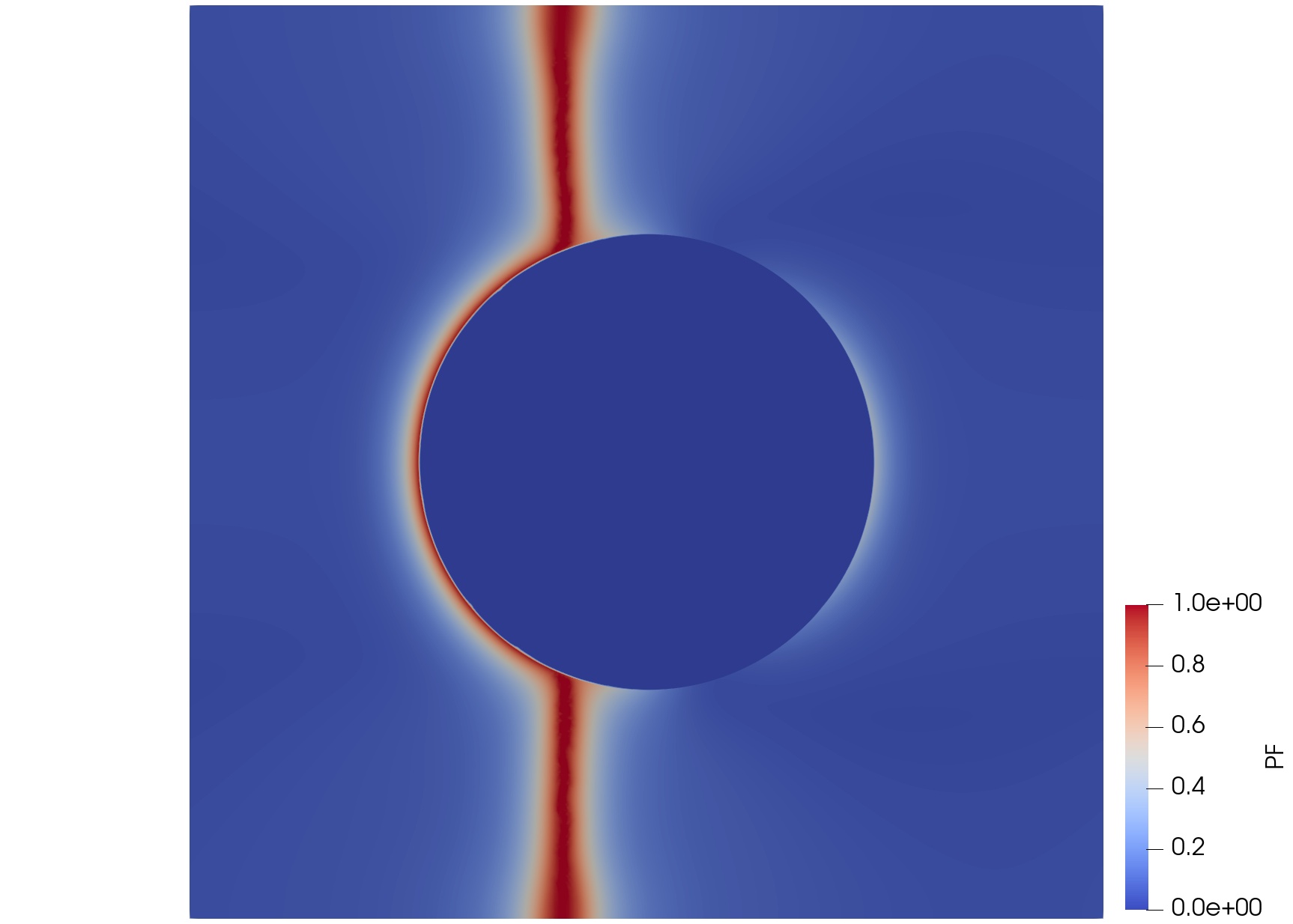}
        \caption{}
		\label{}
    \end{subfigure}
    \begin{subfigure}[b]{0.24\textwidth}
        \centering
        \includegraphics[width=\textwidth]{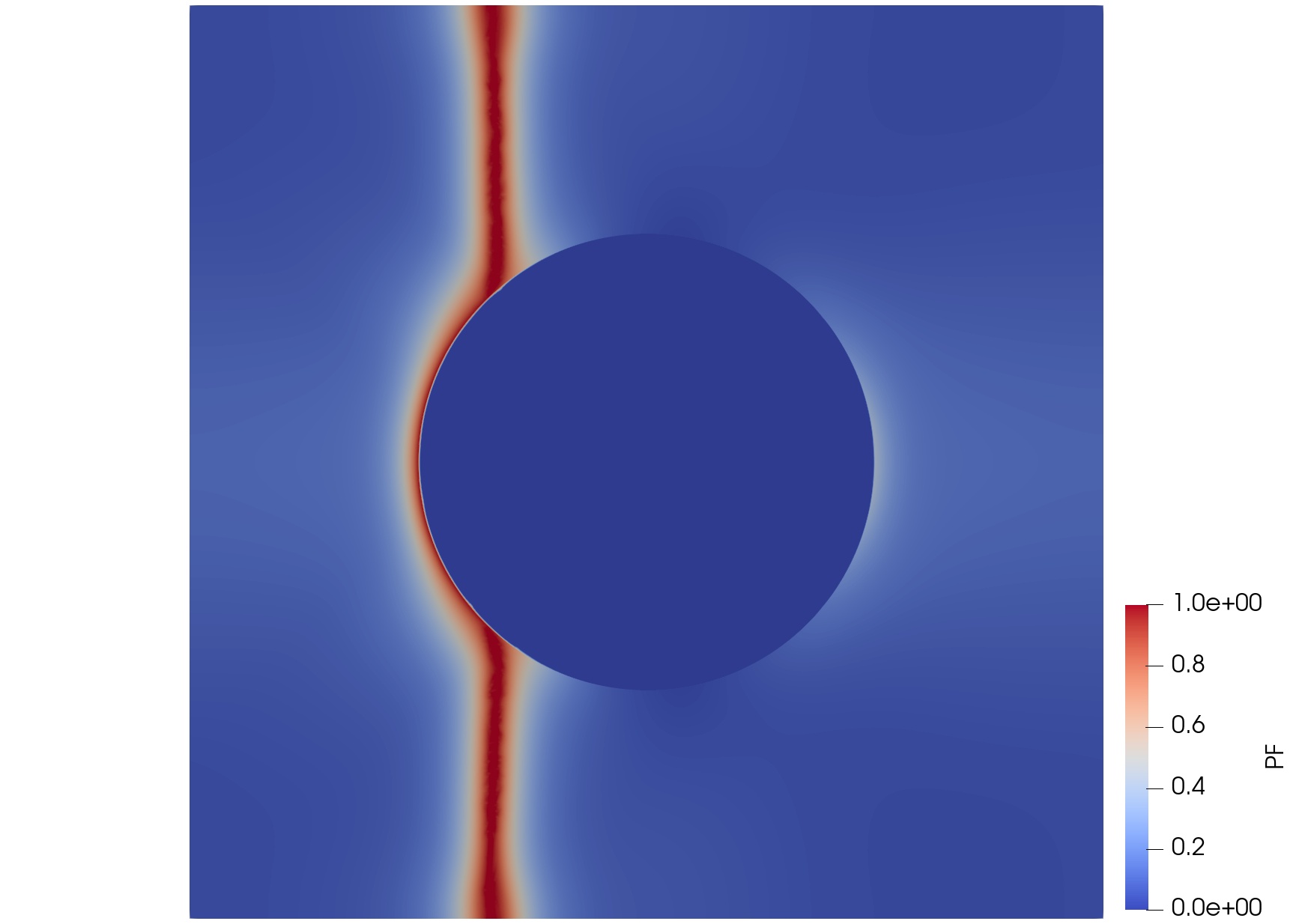}
    	\caption{}
		\label{}
    \end{subfigure}
    \caption{The phase-field in the RVE with different $\sigma_{\rm max}$: (a) $\sigma_{\rm max}=20\ \mathrm{MPa}$; (b)$\sigma_{\rm max}=40\ \mathrm{MPa}$; (c)$\sigma_{\rm max}=60\ \mathrm{MPa}$ and  (d) $\sigma_{\rm max}=100\ \mathrm{MPa}$.}
	\label{fig:SF_4strength}
\end{figure}

The distribution of phase-field in RVEs with different $\sigma_{\rm max}$ after crack kinking is illustrated in Fig. \ref{fig:SF_4strength}. $\theta_{\mathrm{semideboning}}$ decreases with increasing $\sigma_{\rm max}$. In addition, there is also debonding at the interface when $\sigma_{\rm max}=100\ \mathrm{MPa}$, which indicates that the interfacial crack can still appear even with a high value of $\sigma_{\rm max}$. The phenomenon can be explained as a result of the competition between elastic and free surface energy at the bulk region and interface.
%
%
\section{Conclusions}
In the present work, we proposed a novel type of cohesive element to deal with the failure in the interfaces. Under the framework, not only the bulk cracks but also the interfacial cracks are simulated with phase-field. Different from some existing models of the phase-field, the interfacial region is represented by a layer of quadrilateral elements, which is similar to the traditional cohesive element in the only displacement field. This character makes the present model be a better option to simulate composites with high volume fraction. The staggered algorithm is used to solve the coupled equations of the displacement and the phase-field. The interaction between the mode I and II interfacial fracture is also considered in the present work with a straightforward scheme. To describe the interfacial traction-separation law, a modified family of degradation functions are also presented here, in which the penalty stiffness, the ultimate traction, and the critical energy release rate can be taken into consideration.

Some numerical examples have been carried out to verify the present cohesive model. It can be seen that the interfacial mechanical properties are not affected by the bulk material properties. Meanwhile, the present model has a longer quasi-elastic stage compared to the existing models. In addition, the penalty stiffness here can prevent penetration when the cracks are closed. Besides, the competition between bulk and interfacial free surface energy can also be simulated simultaneously in the phase-field under the present framework. The present cohesive zone model shows a lot of potential in the simulations on composite. 

\section*{Acknowledgements}
The authors (Bian and Qing) is grateful for the support of the present work by the National Natural Science Foundation of China (11672131) and the Research Fund of State Key Laboratory of Mechanics and Control of Mechanical Structures (Nanjing University of Aeronautics and Astronautics, MCMS-I-0217G02) and the Priority Academic Program Development of Jiangsu Higher Education Institutions. The author (Bian) is also grateful for the scholarship provided by the China Scholarship Council for a one-year study at the University of Stuttgart (201806830018).

\appendix
\section{Details of finite-element implementation of the present CZM}
\label{sec:appendix}
\renewcommand{\theequation}{\thesection.\arabic{equation}}
\setcounter{equation}{0}  
In the appendix, we outline the process of finite-element discretization at interfaces. For a given element illustrated in Fig. \ref{fig:elementshape}, the nodal displacements and phase-field values are expressed as $\boldsymbol{u}$ and $\boldsymbol{\phi}$, respectively. For any point in the interface element, the value of the displacement jump vector $\boldsymbol{\delta}$ across the interface under local coordinate and phase-field $\phi(x)$ can be obtained by interpolating along the tangential direction:
\begin{equation}
    \delta_{\mathrm{local}}(\boldsymbol{x})=\mathbi{R}\mathbi{B}_{u}\boldsymbol{u},\ \phi(\boldsymbol{x}) = \mathbi{N}\mathbi{M}\boldsymbol{\phi} 
\end{equation}
The definition of the matrix $\mathbi{R}$ and $\mathbi{B}_{u}$ can be found in \cite{Paggi.2017}. The shape function matrix $\mathbi{N}$ and average matrix $\mathbi{M}$ are defined as:
\begin{subequations}
    \begin{align}
        \mathbi{N} = \left[\begin{array}{cc}
            \frac{1-\xi}{2} & \frac{1+\xi}{2} \\
        \end{array}\right]
    \end{align}
    \begin{align}
        \mathbi{M} = \left[\begin{array}{cccc}
            \frac{1}{2} & 0 & 0 & \frac{1}{2} \\
            0 & \frac{1}{2} & \frac{1}{2} & 0\\
        \end{array}\right]
    \end{align}
\end{subequations}
where $\xi$ is the parameter coordinate of the interface element. $\xi=-1$ represents the quadrature point one and $\xi=1$ represents the quadrature point two. The gradient of $\phi$ along the tangential and the phase-field jump along the normal directions can be expressed as:
\begin{subequations}
    \begin{align}
		\nabla_t\phi(\boldsymbol{x}) = \mathbi{B}_{\phi,s}\boldsymbol{\phi}= \mathbi{D}\mathbi{M}\boldsymbol{\phi}
	\end{align}
 	\begin{align}
		\nabla_n\phi(\boldsymbol{x}) = \mathbi{B}_{\phi,n}\boldsymbol{\phi}= \mathbi{N}\mathbi{L}\boldsymbol{\phi}
	\end{align}
\end{subequations}
The tangential gradient matrix $\mathbi{D}$ and difference matrix $\mathbi{L}$ are defined as:
\begin{subequations}
    \begin{align}
        \mathbi{D} = \left[\begin{array}{cc}
            \frac{1}{l_{\rm elem}} & -\frac{1}{l_{\rm elem}} \\
        \end{array}\right]    
    \end{align}
    \begin{align}
        \mathbi{L} = \left[\begin{array}{cccc}
            -1 & 0 & 0 & 1 \\
            0 & -1 & 1 & 0\\
        \end{array}\right]    
    \end{align}
\end{subequations}
where $l_{\rm elem}$ is the length of the interface element. Then the residual vectors in Eqs. (\ref{eq:ru}) and (\ref{eq:rp}) and Jacobian matrix in Eqs. (\ref{eq:Kuu}) and (\ref{eq:Kpp}) can be explicitly expressed. To guarantee the continuity of $\phi$ along the normal direction of the interface, the penalty function method can be used. A simple scheme to implement the penalty function method is used here by modifying the residual vector and Jacobian matrix of phase-field in Eqs. (\ref{eq:rp}) and (\ref{eq:Kpp}) with:
\begin{subequations}
    \begin{align}
        \overline{\mathbi{r}}_{\phi} = \mathbi{r}_{\phi} + \int_{\varGamma} \beta \mathbi{B}_{\phi,n}^{\mathrm{T}}\nabla\phi \mathrm{d}\varGamma    
    \end{align}
    \begin{align}
        \overline{\mathbi{K}}_{\phi\phi} = \mathbi{K}_{\phi\phi} + \int_{\varGamma} \beta \mathbi{B}_{\phi,n}^{\mathrm{T}}\mathbi{B}_{\phi,n} \mathrm{d}\varGamma    
    \end{align}
\end{subequations}
where $\beta$ is the penalty parameter, which should be large enough.




%
%

\bibliographystyle{elsarticle-num.bst}
\bibliography{PhaseFieldCohesiveZone}
\end{document}